\documentclass{egpubl}
\usepackage{pg2018}

 \SpecialIssuePaper         
 \electronicVersion 

\ifpdf \usepackage[pdftex]{graphicx} \pdfcompresslevel=9
\else \usepackage[dvips]{graphicx} \fi

\PrintedOrElectronic

\usepackage{t1enc,dfadobe}

\usepackage{egweblnk}
\usepackage{cite}
\usepackage{placeins}

\usepackage{overpic}
\usepackage{amsmath,amssymb}
\usepackage{todonotes}
\usepackage{color}
\usepackage{gensymb}
\usepackage[normalem]{ ulem }

\newcommand{\NB}[1]{{#1}}

\newcommand{\IP}[3]{\ensuremath{\langle{#2}{,\,}{#3}\rangle_{#1}}}

\newcommand{\dav}[1]{{#1}}
\newcommand{\jaco}[1]{{#1}}

\usepackage{xcolor,colortbl}

\title{Mumford-Shah Mesh Processing using the Ambrosio-Tortorelli Functional}

\author[N. Bonneel, D. Coeurjolly, P. Gueth, J.-O. Lachaud]{Nicolas Bonneel$^{*,1}$, David Coeurjolly$^{*,1}$, Pierre Gueth$^{*,2}$, Jacques-Olivier Lachaud$^{*,3}$\\$^*$joint first authors \hfill $^1$CNRS, Univ. Lyon \hfill $^2$Arskan \hfill $^3$Universit\'{e} Savoie Mont Blanc }

\begin{document}

\teaser{  \begin{tabular}{@{}c@{}c@{}c@{}c@{}c@{}}
 \includegraphics[width=0.2\linewidth]{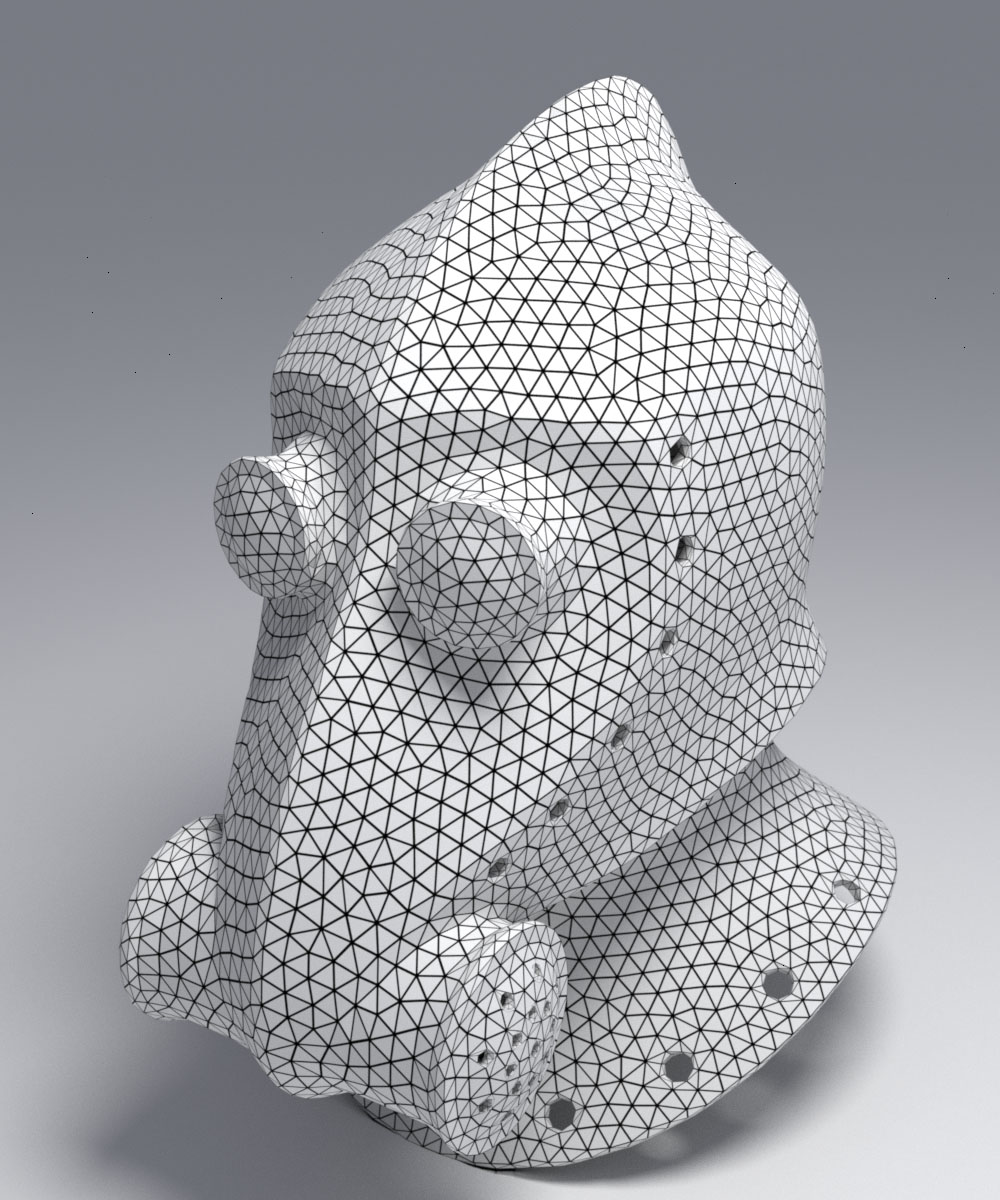}&
 \includegraphics[width=0.2\linewidth]{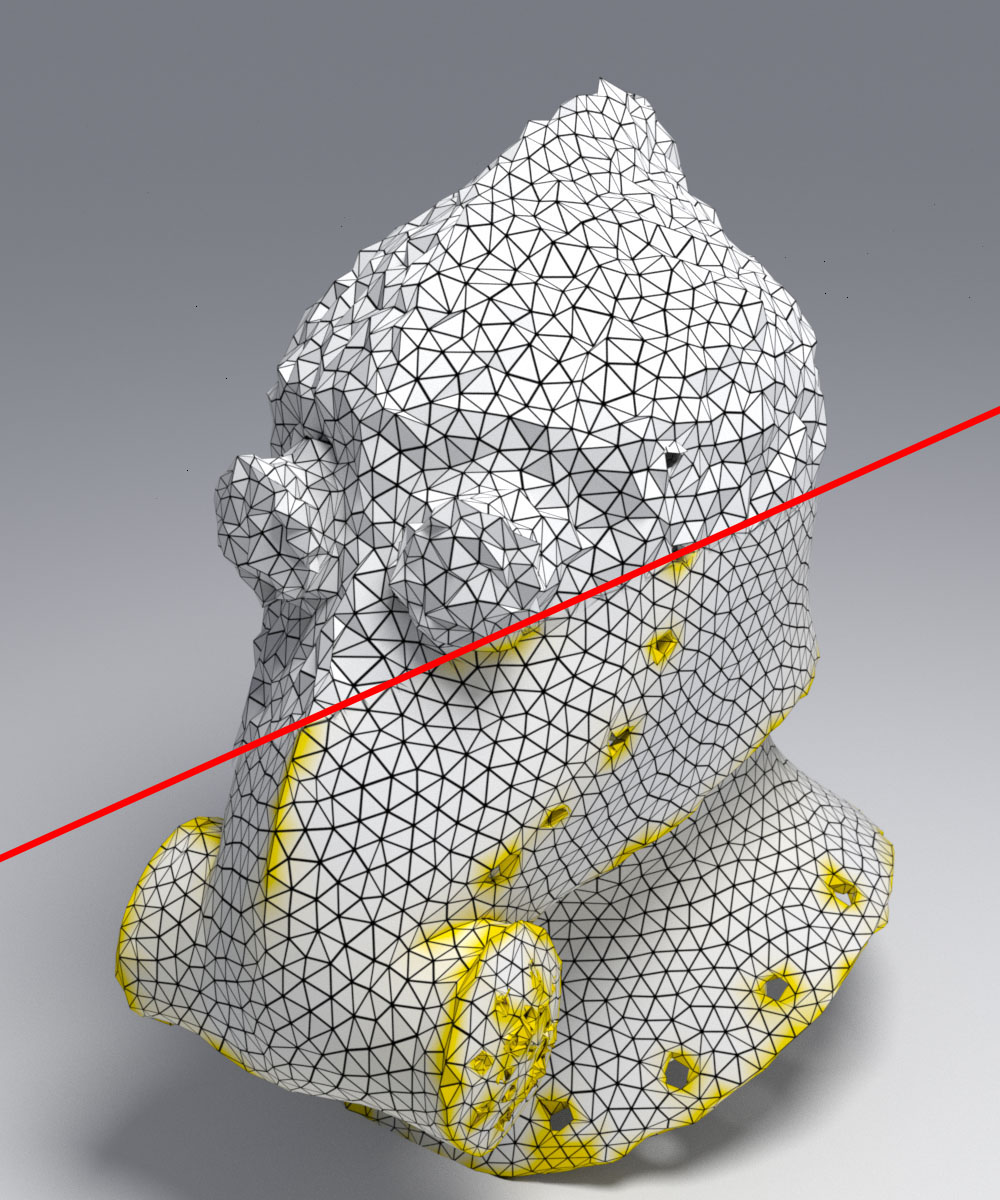}&
\includegraphics[width=0.2\linewidth]{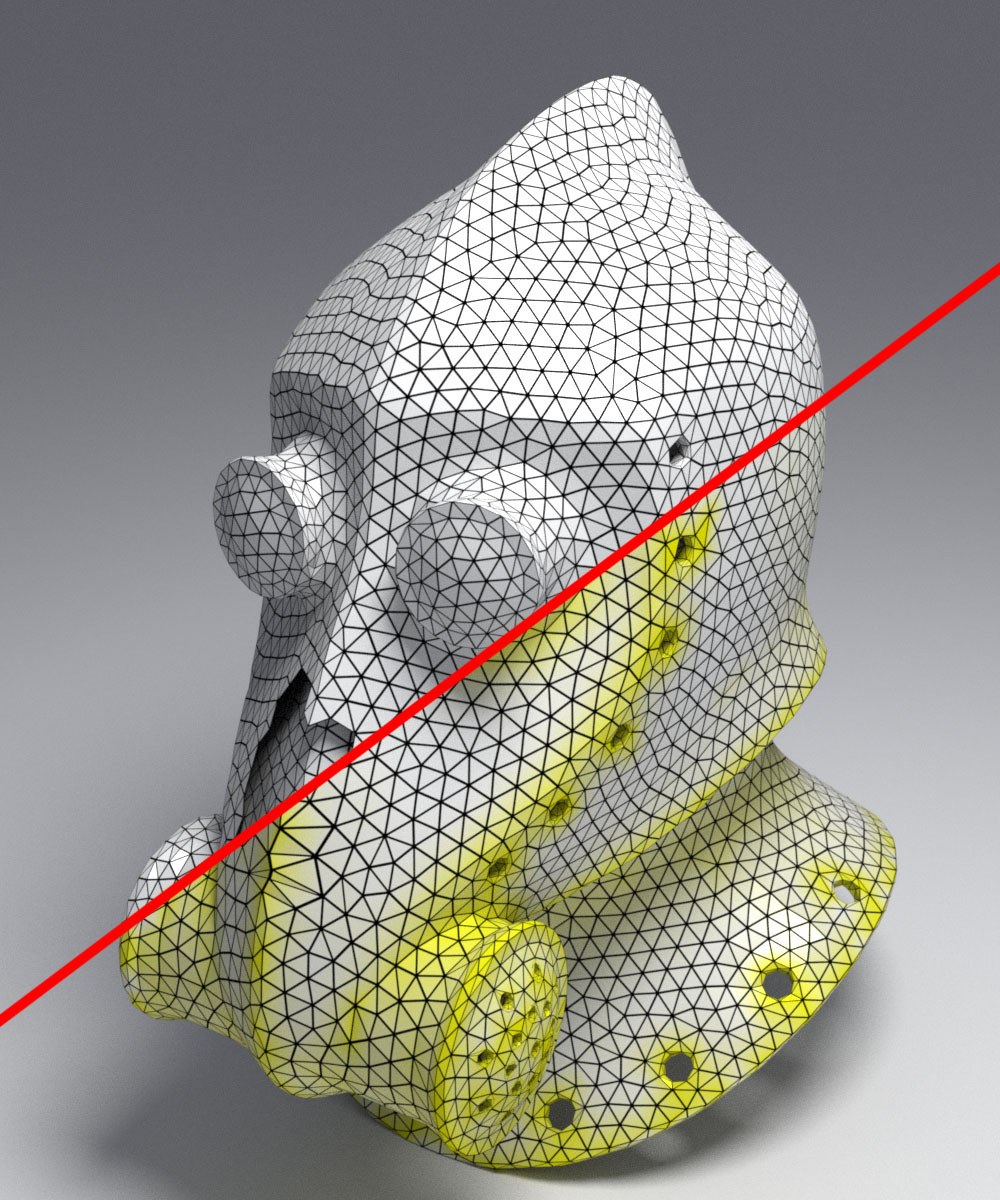}&
\includegraphics[width=0.2\linewidth]{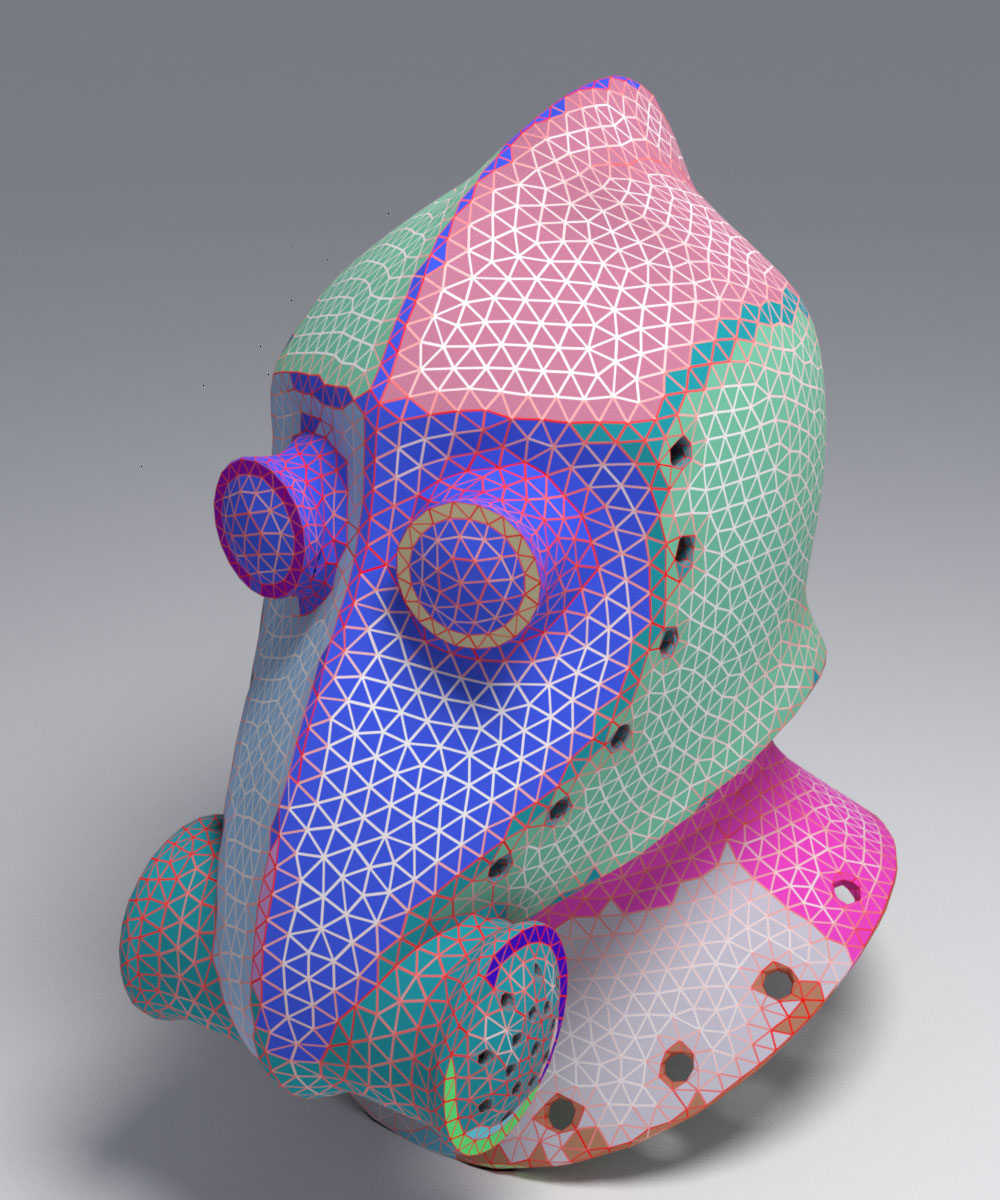} &
\includegraphics[width=0.2\linewidth]{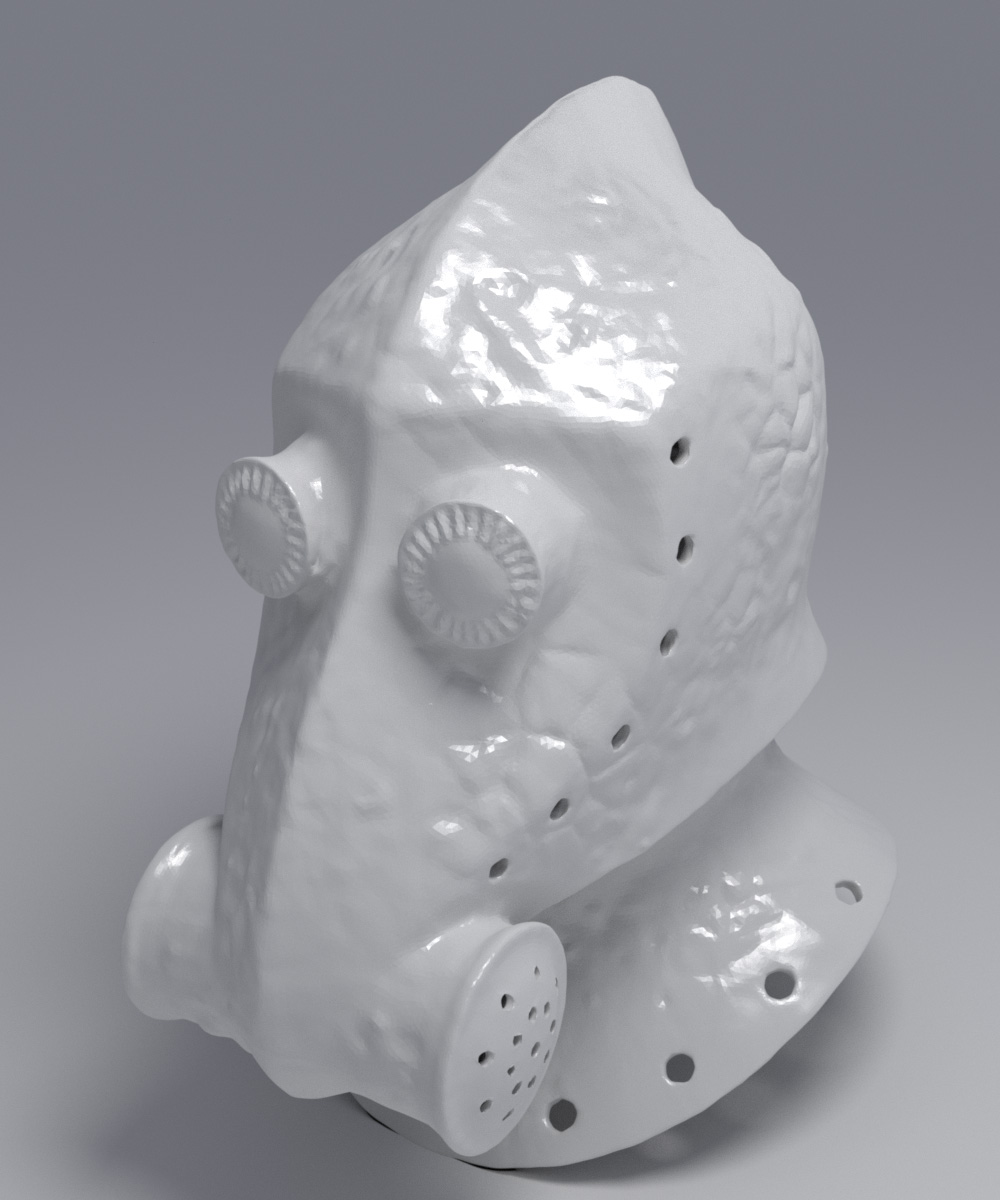} \\
 (a) Original & (b) Denoising  & (c)  Inpainting & (d) Segmentation  & (e) Embossing
 \end{tabular}
 \centering
  \caption{Our discretization of the MS functional along with our new
    vertex projection technique allows for applications such as mesh
    denoising, segmentation, inpainting and normal map
    embossing.
	(b) Our method removes heavy noise while preserving sharp features shown in yellow.  (c) We remove vertices highlighted in cyan from the original mesh (a): our method finds sharp features (shown in yellow) and sucessfully inpaints the missing area. (d) We decompose the original mesh (a) into piecewise smooth segments whose boundaries are characterized by sharp features shown on edges in red. (e) We emboss a normal map into the mesh vertices.}
    }

\maketitle

\begin{abstract}
The Mumford-Shah functional approximates a function by a piecewise
smooth function. Its versatility makes it
ideal for tasks such as image segmentation or restoration, and it is
now a widespread tool of image processing. Recent work has started to
investigate its use for mesh segmentation and feature lines detection, but we take the stance that
the power of this functional could reach far beyond these tasks and
integrate the everyday mesh processing toolbox. In this paper, we
discretize an Ambrosio-Tortorelli approximation via a Discrete Exterior Calculus formulation.
We show that, combined with a new shape optimization routine, several mesh
processing problems can be readily tackled within the same
framework.
In particular, we illustrate applications in mesh denoising, normal map embossing, mesh inpainting and mesh segmentation.
\end{abstract}

\section{Introduction}

The Mumford-Shah (MS) functional, originally intended for image
segmentation~\cite{Mumford89optimalapproximation}, is a well-known
functional which models an image as a piecewise-smooth
function. Minimizing this functional recovers both a piecewise-smooth
image that can be useful for applications such as image denoising, and
the set of discontinuities can be useful for edge
detection. Since the nineties, this model has seen a tremendous
success in image processing and has found many applications from image
segmentation~\cite{Tsai01,Vese2002} to denoising~\cite{Tsai01},
inpainting~\cite{Esedoglu2002}, magnification~\cite{Tsai01},
deblurring~\cite{Bar06} or registration~\cite{BenAri10}.

This functional has however attracted little attention in geometry
processing, where similar problems are encountered. Notable
exceptions include the work of Zhang et al.~\shortcite{Zhang:2012} that
makes use of a convexified version of the MS model for mesh
segmentation, and the finite element method of Tong and Tai~\shortcite{tong2016variational}
to construct feature lines on meshes. This lack of interest might have come from several
challenges that need to be overcome. First, the model itself is not
suited to a computational resolution. Hence, numerical methods only solve 
approximate MS models. Second their computational complexity
makes algorithms tuned for regular grids difficult to scale to large
unstructured meshes. Third, the MS formulation, once adapted to
manifolds, exhibits differential operators that need careful handling
when discretized over triangular meshes. Last, the MS model represents
a piecewise-smooth \textit{scalar function} over a domain; it is thus
not obvious what this function should be in the context of mesh
processing and for which applications.

Fortunately, recent progress on the first two issues allows to paint a
brighter picture. On the computational side, various approximations of
the MS functional have recently been proposed.
In this paper, we are interested with an
accurate approximation of the MS functional: 
the Ambrosio-Tortorelli's functional~\cite{Ambrosio_CommPureAppMath1990}.
Recent advances have
further made calculus on meshes practical and affordable.  In
particular, Discrete Exterior Calculus (DEC)~\cite{Hirani:2003} has
become popular for easily formulating and solving differential
equations on meshes. We will formulate the MS functional as a discrete
AT model in the language of DEC, in the spirit of the method of
Coeurjolly et al.~\shortcite{CoeurjollyFGL16}.

Regarding the choice of scalar function to use in the MS functional,
it obviously depends on the targeted applications. For our
applications, we take advantage of our DEC formulation to
consider each component of the normal vector at each point of the
surface as a 0-form stored on the faces of a dual mesh. This results
in a set of three scalar functions, generalizing MS to vector functions~\cite{focardi2014asymptotic}. 
We will jointly minimize MS over these three scalar functions.  
Overall, this essentially allows to smooth out the normals of an input mesh so they
better match the underlying continuous surface being approximated
while preserving mesh discontinuities.

Benefitting from these advances, we formulate a set of classical mesh
processing problems using the MS functional applied to the manifold's
normal vector field.  From the resulting regularized normal vector field, we further
introduce a shape optimization routine as an easy-to-implement
projection operator which deforms the geometry so that geometric
normals match the obtained regularized normals. We illustrate our
method on a number of applications, such as mesh denoising, normal map
embossing, mesh inpainting and mesh segmentation.


\section{Related work}

\textbf{The Mumford-Shah Model.}
Mumford and Shah described a functional representing an input image by
a piecewise-smooth
approximation~\cite{Mumford89optimalapproximation}. This functional
reads:
\begin{equation}
MS[u,C] = \alpha \int_\Omega (u - g)^2 \text{d}x +
\beta \int_{\Omega \setminus C} |\nabla u|^2 \text{d}x + \gamma
\int_{C} \text{d}s\,,
\end{equation}
with $g$ the input image on a two-dimensional planar domain $\Omega$,
$u$ its (unknown) approximation, and $C$ a set of (unknown) curves
describing the set of discontinuities. The parameters of the model
intuitively model the tightness $\alpha$ of the approximation to the
input image, the smoothness $\beta$ of the approximation, and the
length of discontinuities $\gamma$. While for segmentation purpose, it
is often assumed that $C$ forms a closed curve or is the boundary of
some partition of the space, this is not a requirement of the model
nor the optimal solution to it.  In addition, this model is not
restricted to images: in the general case, $\Omega$ need not be a
plane and can be an arbitrary surface, and $g$ a real-valued function
over this surface.

Unfortunately, except in limited scenarios, this functional is
non-convex and very difficult to optimize.

\begin{figure}[tbh]
\centering
\begin{tabular}{@{}c@{\hspace{2mm}}c@{}c@{}}
\rotatebox{90}{\hspace{1.5cm}Noise free} & 
\includegraphics[width=0.4\linewidth]{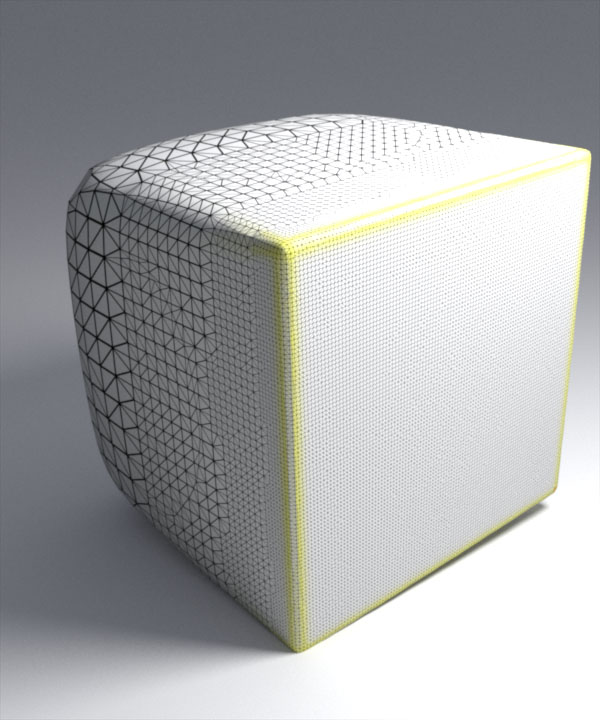}&
\includegraphics[width=0.4\linewidth]{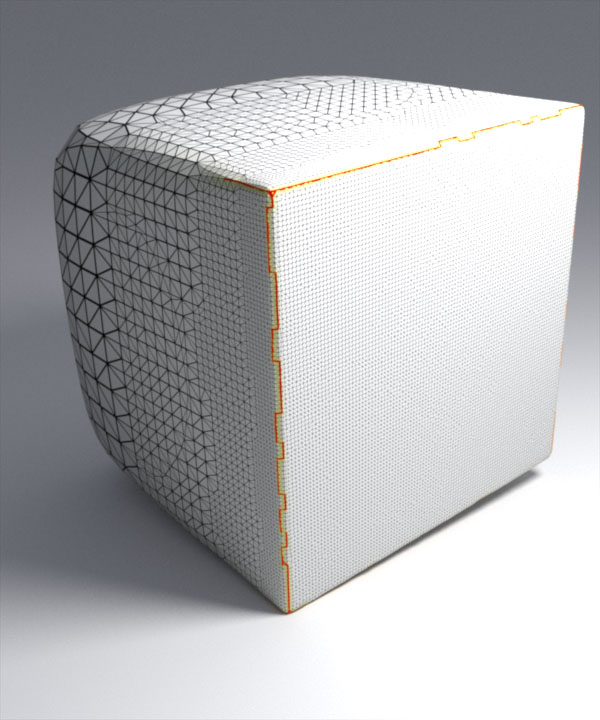}\\
\rotatebox{90}{\hspace{1.5cm}Noisy} & 
\includegraphics[width=0.4\linewidth]{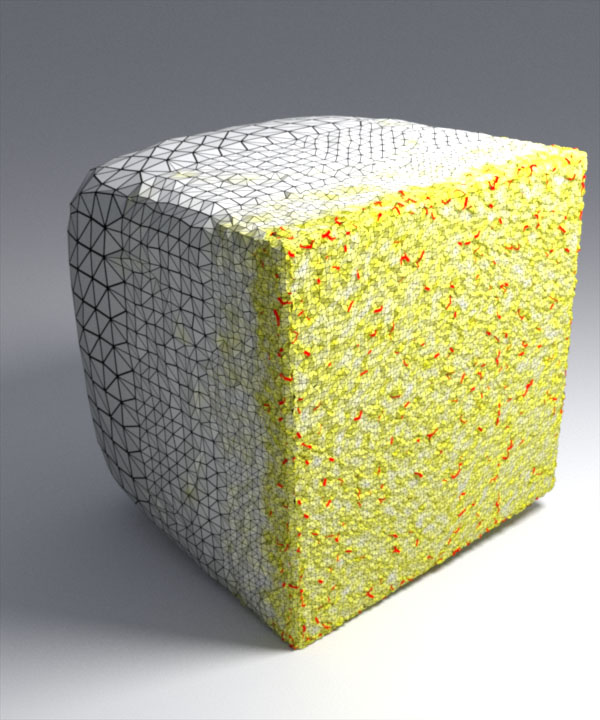}&
\includegraphics[width=0.4\linewidth]{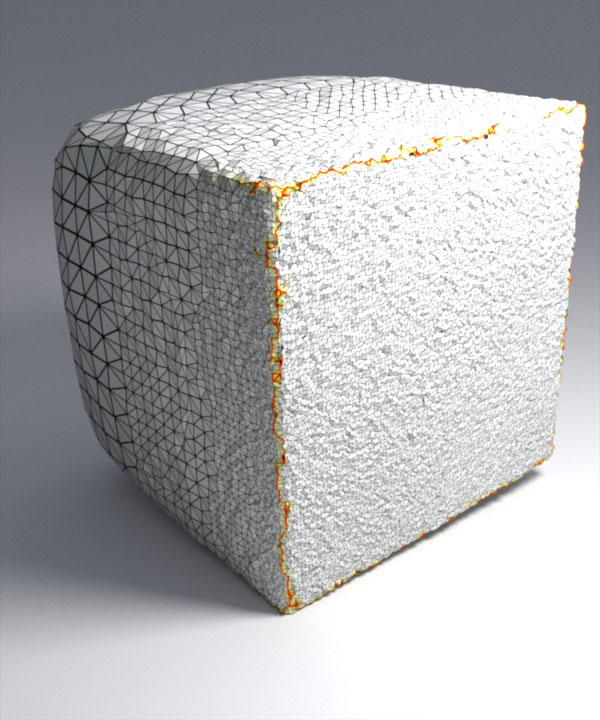}\\
&(a) \protect\cite{tong2016variational} & (b) Our method
\end{tabular}
\caption{We compare our features (b) to that extracted by the finite
  element discretization of Tong and
  Tai~\protect\cite{tong2016variational} (a) using the same parameters
  ($\lambda=0.06, \alpha=0.07$). The feature function is shown in
  yellow, while extracted feature lines are shown in red. Our method
  is able to detect features even at relatively high noise
  levels.}
\label{fig:femcomparison}
\end{figure}

\textbf{Optimizing the MS functional.}
To avoid these difficulties, various approximations to MS have been
proposed, and among them, the most natural ones are convex {envelopes}
or convex approximations to MS functional. When restricting the
problem to foreground / background extraction, the MS functional can
be convexified by replacing the term accounting for the length of
segment boundaries by the total variation of the gradient of a segment
binary indicator function~\protect\cite{Chan06}. Instead of a
piecewise-smooth assumption, this assumes a piecewise constant
function and precludes open segment boundaries or isolated
discontinuities. This can be extended to multiple segments but still
necessarily produces disjoint sets. While this assumption can be
appropriate for certain segmentation applications and has indeed been
sucessfully used for mesh segmentation~\cite{Zhang:2012}, this may not
be suitable in our context where our function of interest is the
normal field that varies relatively smoothly across the
surface~\cite{Zhang:2012} and may exhibit internal discontinuities
(see Fig.~\ref{fig:denoising}, first row). Similarly, Tsai et
al.~\shortcite{Tsai01} directly optimize the MS functional using level
sets, but also requires closed segments boundaries. These approaches
thus lack the generality we are targeting.  We instead resort to an
early approximation from Ambrosio and
Tortorelli~\cite{Ambrosio_CommPureAppMath1990} that relies on
$\Gamma$-convergence results, and provably converges towards the MS
functional.  Its numerical optimization is nonetheless far from being
trivial. For instance, Chambolle and Dal
Maso~\shortcite{Chambolle:1999-mmna} and Bourdin and
Chambolle~\shortcite{Bourdin:2000-nm} employ a Finite Element Method
with adaptive mesh refinement and edge alignment is required for its
optimization. Even with such advanced technique, these numerical
methods are very sensitive to noise~\cite{foare2016image}.
Fortunately, new discretization schemes for the Ambrosio-Tortorelli
(AT) model have been designed on
grids~\cite{foare2016image,CoeurjollyFGL16} and do not require FEM
anymore nor adaptive mesh to get piecewise smooth solutions.  This
discretization has seen applications for image
restoration~\cite{foare2016image} and feature extraction on
voxel-based digital geometries~\cite{CoeurjollyFGL16}.  \jaco{A first
  discrete differential calculus version of AT has been employed by
  Pokrass {\em et al.} \cite{pokrass2011} to solve the partial
  matching of non-rigid 3D shapes. Their discretization is different
  from ours since discontinuities and values live on vertices, while
  the cross-term is evaluated on faces (see below), resulting in smoother
  features. This is fine for their specific objective but limits its
  range of applications. In contrast our approach allows
  coarse-to-fine detection of features for a variety of geometry
  processing tasks.}  An alternative finite element discretization of
AT has been recently proposed for triangle
meshes~\cite{tong2016variational}.  \dav{This approach is the closest
  one to our formulation but it is not able to recover piecewise
  smooth patchs on noisy data as illustrated in Figure
  \ref{fig:femcomparison}. } \jaco{We propose a specific Discrete Exterior
Calculus discretization of this functional, which achieves sharper and
more robust features, as we show in our experiments.}


\textbf{Applications in image processing.} The MS functional
originated from the image processing community as a way to represent
an image as a piecewise-smooth function. In the last 40 years, it has
led to numerous applications, even if most of the time it is only
simpler variants of the MS functional that are solved. Image
segmentation~\cite{Tsai01,Vese2002} and denoising~\cite{Tsai01} are
the two main applications that directly follow from the variables
being optimized for (that is, discontinuity boundaries and piecewise
smooth approximations).  Image inpainting~\cite{Esedoglu2002} is
rendered possible by removing the data attachment term inside the
inpainted area.  A 3x image magnification~\cite{Tsai01} is performed
by interlacing the input image in a 3x upsampled grid, and inpainting
the missing pixel values.  Bar et al. extend a deconvolution
functional by summing it with the MS functional~\cite{Bar06}, hence
leading to a new image deconvolution method with piecewise smoothness
prior.  Similarly, Ben-Ari and Sochen integrate the MS model in a
stereo-matching functional, resulting in {piecewise} smooth disparity
maps~\cite{BenAri10}.  We believe the MS model has the potential to
produce a similar number of applications in geometry processing, and
this paper presents an initial set of applications.

\textbf{Variational methods in mesh processing.} Variational methods
for mesh processing have long been studied. A variational method
optimizes a given functional over a space of functions. The most
common examples of such functionals used in mesh processing include
Lloyd's functional~{\cite{Cohen-Steiner:2004}}, Total
Variation~{\cite{zhang2015variational}}, the variational problem
associated with the Poisson equation~{\cite{Yu:2004}}, the curvature
energy~{\cite{Zorin05}} or the MS
functional~{\cite{CoeurjollyFGL16,Zhang:2012}}.  Individual mesh
processing applications have their own set of priors, hypotheses and
constraints, and it is illusory to search for an all-purpose
functional. However, we show that the MS functional appropriately
solves a number of problems and is of value to the geometry processing
community.

\section{Discretizing Mumford-Shah over Surfaces}

This section describes our model for mesh processing, that extends the model of 
Coeurjolly et al.~\cite{CoeurjollyFGL16} tailored for voxel-based geometries. In particular,
we briefly discuss the AT approximation, and expose
our discretization over triangular meshes within the DEC framework.

\subsection{Ambrosio-Tortorelli's Approximation}

Ambrosio and Tortorelli reformulates the MS
functional~\cite{Ambrosio_CommPureAppMath1990} by working on a
smoothness indicator function rather than the set of discontinuities
$C$. This indicator function, denoted by $v$ in the following, is
optimized such that it takes the value $v(x) = 1$ wherever $g(x)$ is
smooth, and $v(x) = 0$ on discontinuities. Therefore we call it the
\textit{feature} function. In their model, this function $v$ is smooth and kept \textit{close to} $1$ where $g$ is
smooth, and its oscillations are prevented by keeping $\nabla v$
close to $0$. Their functional can be written as follows:
\begin{equation}
AT_\epsilon[u, v] = \int_\Omega \alpha (u - g)^2 +  |v \nabla u|^2 + \lambda\,\epsilon|\nabla v|^2 + \frac{\lambda}{\epsilon} \frac{(1-v)^2}{4} \text{d}x\,.
\label{eq:at-energy}
\end{equation}

This functional now depends on the same parameters $\alpha$ and $\lambda$ ($\beta$ is set to 1 without loss of generality), 
but also requires a parameter $\epsilon$. Parameter $\lambda$ is still such that $1/\lambda$
represents the length of discontinuities, and parameter $\epsilon$
controls the smoothness of the feature $v$. Ambrosio and Tortorelli
proved that when $\epsilon$ tends to zero, minimizers of this
functional exactly corresponds to those of the $MS$
functional. However, the problem is now quasi convex, and the
integration domain is fixed and does not involve unknown curves.

\subsection{Minimizing the Energy}

For a fixed $\epsilon$, since the energy (\ref{eq:at-energy}) is quasi
convex, we can minimize it by alterning minimizations over $u$ and
$v$. We follow the AT energy formulation proposed in
  \cite{CoeurjollyFGL16}. The feature scalar field $v$ is discretized
  as a primal $0-$form, that is a collection of scalar values
  associated with each vertex of the mesh.  The normal vector fields
  (both $u$ and $g$) are each discretized as three dual $0-$forms, one
  for each coordinate of the $\mathbb{R}^3$ embedding space.  Note
  that Focardi and Iurlano showed that AT in the vectorial case also
  converges towards an approximation of
  MS~\cite{focardi2014asymptotic}.  Each of those three dual $0-$forms
  can be seen as a scalar value associated with each
  primal face (or dual vertex) of the mesh.  Denoting $\text{d}$ and $\bar{\text{d}}$ the
  primal and dual exterior derivatives, it follows that
  $\text{d}v$ is a primal $1-$form and $\bar{\text{d}}u$ corresponds to
  three dual $1-$forms.  Norms are induced by the natural inner
  products between $k-$forms. We rewrite then the AT functional as a
  sum of inner products on primal and dual $0$ and $1$-forms:
\begin{align}
   AT_\epsilon[u, v] = & \alpha \IP{\bar{0}}{u-g}{u-g}  +
   \IP{\bar{1}}{v \bar{\text{d}}u}{v \bar{\text{d}}u} \nonumber \\ &+
   \lambda\,\epsilon \IP{1}{\text{d}v}{\text{d}v} +  \frac{\lambda}{4\epsilon} {\IP{0}{1-v}{1-v}}\,.
   \label{eq:discrete-at-energy}
\end{align}

Note that we have discretized (\ref{eq:at-energy}) such that our
feature function $v$ does not  become identically one when epsilon
tends to 0, and remains zero around feature edges. To achieve such
discontinuities, the function $v$ acts as a rescaling of the
derivatives of $u$ and not as a rescaling of the squared gradient
norm. Hence it operates directly on edges in
(\ref{eq:discrete-at-energy}) and it does not operate on vertices as
would give a standard DEC discretization of (\ref{eq:at-energy}). Our
discretization of AT is thus more related to discontinuous FEM than to
continuous and linear basis elements. For instance, the FEM approach
of Tong and Tai~\shortcite{tong2016variational} oversmoothes features
(function $v$ becomes almost 1 everywhere) and requires
post-processing to find its valleys.

\begin{figure}
  \definecolor{myblue}{RGB}{0,104,170}
\definecolor{myred}{RGB}{201,25,35}
\definecolor{mygreen}{RGB}{0, 185 40}
\definecolor{myorange}{RGB}{255,103,37}\centering
\begin{overpic}[width=5cm]{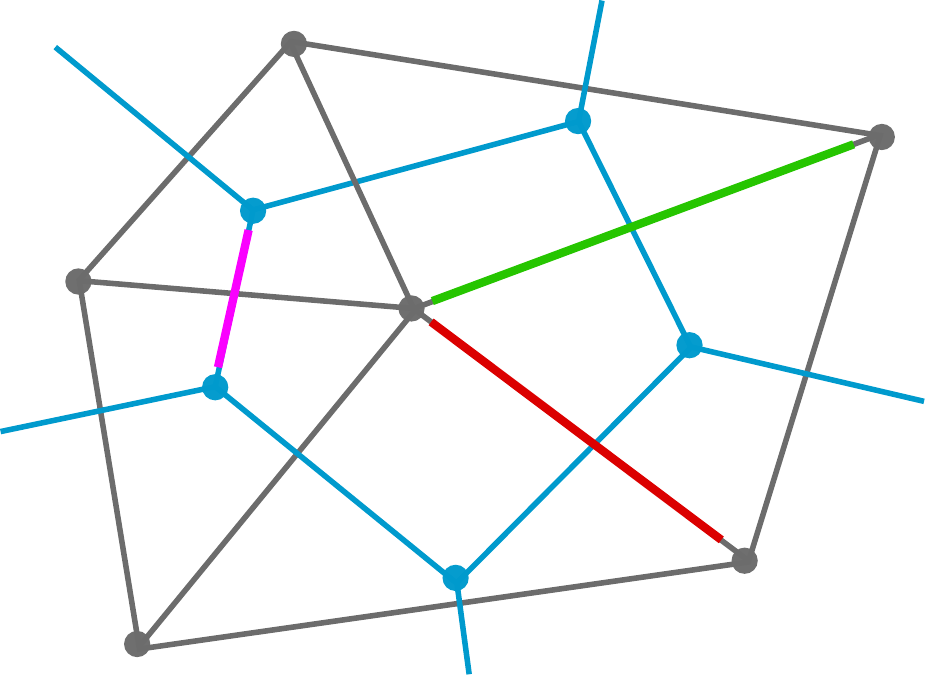}
  \put(45,44){$0.5$}
   \put(82,7){$0.6$}
   \put(92,62){$0.2$}
  \put(60,30){\color{myred}$0.1$}
  \put(62,50){\color{mygreen}$0.35$}\small
    \put(10,20){\color{myblue}$(0.3,0.3,0.3)^T$}
    \put(-20,36){\color{magenta}$(0.3,-0.2,-0.2)^T$}
    \put(15,55){\color{myblue}$(0.0,0.5,0.5)^T$}
\end{overpic}
  \caption{Notations on a primal and dual mesh for the energy
    optimization (edge orientation not shown for the sake of clarity):
    $v$ is a $0-$form on vertices. In red (resp. green), we have the
    $Av$ values (resp. $Mv$ values) on
    edges. In blue, the triple of dual $\bar{0}-$forms, $u$ on dual
    vertices and the $Bu$ values in magenta on dual edges.}
  \label{fig:notation}
\end{figure}
\jaco{We write the $AT_\epsilon$ energy in matrix form, using the
  following notations \dav{using thec classical discrete exterior calculus (DEC)
  framework where the dual vertices are circumcircle of primal
      triangles~\cite{Crane:2013:DGP}: } the Hodge star $\star_i$ is
  written as the diagonal matrix $S_i$ , the vector diagonalization
  operator is denoted by $\operatorname{Diag}$, differential operators
  $\text{d}_0$ and $\bar{\text{d}}_1$ are respectively denoted by
  matrices $A$ and $B$, and the matrix $M$ is the operator averaging
  incident 0-form values at each 1-cell ($M := \text{abs}(A)/2$).}  We
thus have: \jaco{
\begin{align}
  AT_\epsilon[u, v] = & \alpha (u-g)^T S_{\bar{0}} (u-g) + u^T B^T \operatorname{Diag}(M v) S_{\bar{1}} \operatorname{Diag}(M v) B  u  \nonumber \\
  & + \lambda\,\epsilon\, v^T A^T S_1 A v + \frac{\lambda}{4\epsilon} (1-v)^T S_0 (1-v).
\end{align}
}
 \jaco{The
  cross term ($v\nabla u$ in Eq.~(\ref{eq:at-energy}) and its discretizations) can indifferently be written with the vectors $v$ or $u$
  inside or outside the inner product.}
Observe that the energy gradient with respect to $u$ (\ref{eq:solve-u}) (resp. $v$ (\ref{eq:solve-v}))
is a linear operator with respect to $u$ (resp. $v$). Thus the alternating
minimization method amounts to alternating between the resolution
of the following linear systems:
\begin{align}
  \nabla_u AT_\epsilon[u, v] = 0 \Leftrightarrow \qquad \qquad \qquad
   && \nonumber \\
  \left[ \alpha\,S_{\bar{0}} - B^T \operatorname{Diag}(M v) S_{\bar{1}} \operatorname{Diag}(M v) B \right] u  = \alpha\,S_{0}\,g, \label{eq:solve-u}
\end{align}
\begin{align}
  \nabla_v AT_\epsilon[u, v] = 0 \Leftrightarrow \qquad \qquad \qquad \qquad \qquad
  && \nonumber \\
  \left[ \frac{\lambda}{4\epsilon} S_0 + \lambda\,\epsilon\,A^T S_1 A  + M^T \operatorname{Diag}(B u) S_{\bar{1}} \operatorname{Diag}(B u) M \right] v = \frac{\lambda}{4\,\epsilon} S_0, \label{eq:solve-v}
\end{align}




\jaco{As long as the diagonal Hodge stars $S_0$ and $S_{\bar{0}}$ are
  non-degenerate, the left-hand sides are full rank, yielding a unique
  minimum at each iteration. It is a known result in convex analysis
  linked to block coordinate descent algorithms
  \cite[Prop. 2.7.1]{bertsekas1999nonlinear}, that these iterations
  must converge to a stationary point of $AT_\epsilon$.} 
  
Convergence and stability are improved by progressively reducing
$\epsilon$ rather than directly solving the problem with a small
$\epsilon$. We hence perform the above minimization repeatedly,
dividing $\epsilon$ by 2 at each iteration (we typically start at
$\epsilon_0 = 2$ and decrease until $\epsilon=0.25$). This allows to
better infer the general position of features at a coarse scale (large
$\epsilon$), and then to better precisely delineate them at a fine
scale (small $\epsilon$). \jaco{This coarse-to-fine optimization is
  especially important in the inpainting case, where the value of
  $\epsilon_0$ should be around the radius size of the inpainted
  region.}

\section{Deforming a Mesh to a Prescribed Normal Field}
\label{sec:projection}

Given a regularized normal field $u$ over an input mesh (e.g., given
by a first AT minimization), we wish to deform the mesh by moving its
vertices such that the triangles normal vectors match the prescribed
normal field $u$. We perform this operation by minimizing an energy
$E$ consisting of three terms: a normal matching term $E_m$, a
fairness term $E_f$, and a data attachment term $E_d$, described in
this section.  Note that some terms may appear in various forms in
other works
\cite{taubin2001linear,fleishman2003bilateral,pottman2007architectural,sun2007fast,bouaziz2012shape}. We
describe them for self-containedness \dav{and propose a fairness energy
  to take into account the feature field}.  Our method then proceeds by
alternately solving for the unknown positions of the deformed mesh
with a fixed feature field $v$ and solving the AT functional to update
the feature field $v$ and piecewise smooth normals $u$.

\textbf{\dav{Matching normals.}} Let $p$ be the vertices positions of
the deformed mesh, and $F = \{f_i\}_{i=1..N}$ the set of triangles
where $f_i^j$ denotes the $j^{th}$ vertex of the $i^{th}$ triangle,
with $j \in \{0,1,2\}$. As we are deforming the mesh, the topology
(and hence the set of triangles) of the deformed mesh is the same as
that of the input mesh. We denote by $|p|$ the number of vertices.
Our goal is to find $p$ such that per-triangle geometric normals match
the prescribed normal field $u$. We formulate this condition by
imposing that each edge of each triangle be orthogonal to $u$:
\begin{align*}
E_m &= \sum_{f_i \in F} ((p_{f_i^1} - p_{f_i^0})\cdot u)^2 + ((p_{f_i^2} - p_{f_i^1})\cdot u)^2 + ((p_{f_i^0} - p_{f_i^2})\cdot u)^2\\
    &= \sum_{f_i \in F} E_m(f_i)\,.
\end{align*}

Note that the summation is performed over all triangles and not over
edges since $u$ is known per triangle. This induces sharper
discontinuities.  Its gradient with respect to each triangle vertex
position is given by:
\begin{eqnarray*}
\nabla_{p_{f_i^j}} E_m(f_i) &=&  2 ((2 p_{f_i^j} - \sum_{k\neq j} p_{f_i^k})\cdot u)  u\\
							&=& {C}(f_i) p\,,
\end{eqnarray*}
where ${C}(f_i)$ is a $3 \times 9$ matrix, that can be assembled in a $3|p| \times 3|p|$ matrix ${C}$ for all vertices of all triangles of the mesh.

\begin{figure}[tbh]
\centering
\begin{tabular}{ccc}
\includegraphics[height=2.8cm]{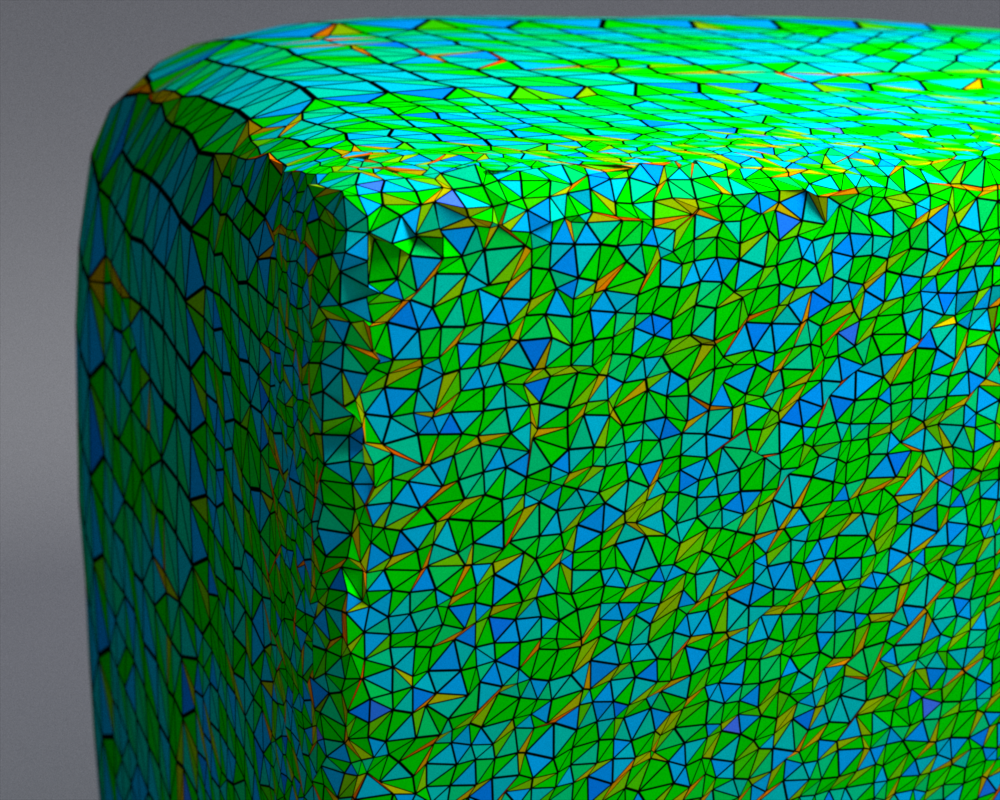} &
                                                 \includegraphics[height=2.8cm]{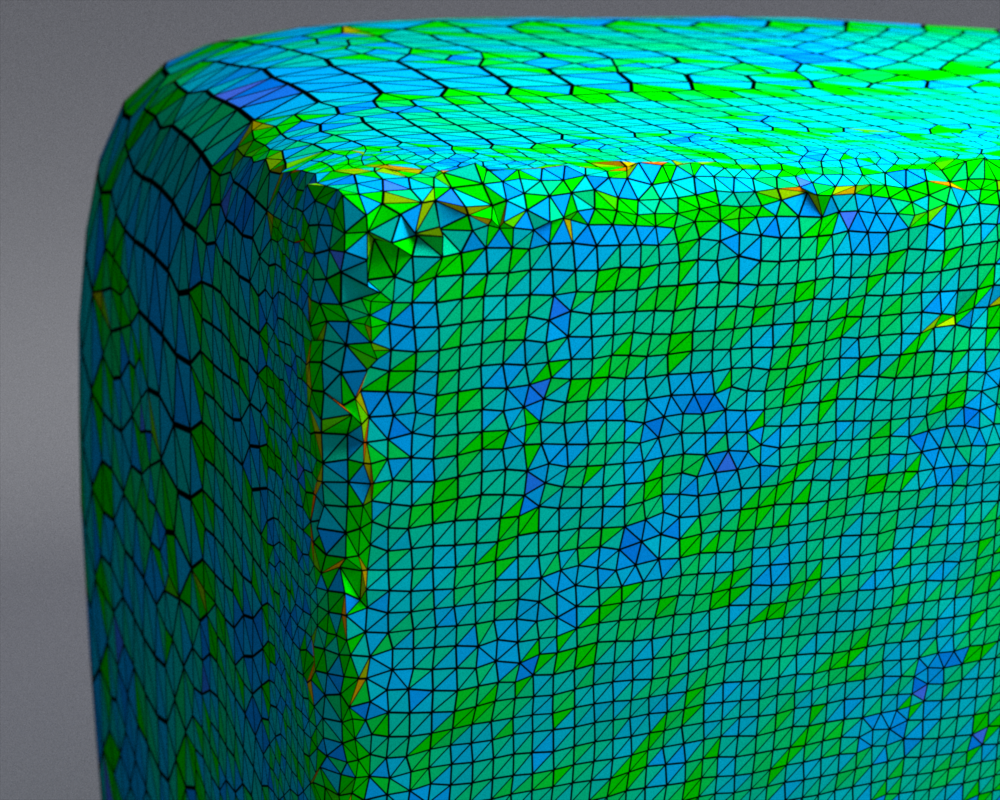} &
                                                                                                  \begin{overpic}[height=2.8cm,width=0.2cm]{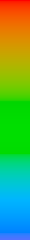}\put(10,0){$0$}\put(10,97){$1$}
                                                                                                \end{overpic}
  \\
(a) no fairness & (b) with fairness
\end{tabular}
\caption{Denoising the model of Fig.~\protect\ref{fig:femcomparison}
  without (a) and with (b) a small fairness term $E_f$ ($w_1 =
  0.5$). The fairness term favors more regular near-Delaunay
  meshes. \dav{The colormap indicates the aspect ratio of each
    triangle ($1- \frac{\alpha_T}{\pi/3}$ with $\alpha_T$ the
    minimum angle of triangle $T$) }.\label{fig:fairness}}
\end{figure}

\textbf{\dav{Fairness.}} 
We additionally enforce the piecewise smoothness of the deformed mesh
by adding a fairness term. This term forces neighboring triangles to
share similar geometric normals, at least wherever the feature $v = 1$
and corresponds to a thin plate energy. Its effect can be appreciated in Fig.~\ref{fig:fairness}. 
For an edge $e = (p_{i_1},
p_{i_2})$ between triangles $t_1 = (p_{i_1}, p_{i_2}, p_{i_3})$ and
$t_2 = (p_{i_1}, p_{i_2}, p_{i_4})$, the fairness energy reads:
$$E_f(e) = \left(\frac{v(p_{i_1}) + v(p_{i_2)}}{2} \right)^2
\|p_{i_1}+p_{i_2}-p_{i_3}-p_{i_4}\|^2\,,$$ and we obtain $E_f$ by
summing over all edges $E_f = \sum_e E_f(e)$.  Fixing $v$, the
gradient of $E_f(e)$ with respect to $p_{i_1}$ (resp. $p_{i_2}$) is:
\begin{eqnarray*}
\nabla_{p_{i_1}} E_f(e) &=& 2 \left(\frac{v(p_{i_1}) + v(p_{i_2})}{2} \right)^2  (p_{i_1}+p_{i_2}-p_{i_3}-p_{i_4})\\
						&=& {D}(e_i) p\,,
\end{eqnarray*}
where ${D}(e_i)$ is a $3 \times 12$ matrix. These matrices again be
assembled in a $3|p| \times 3|p|$ matrix ${D}$ by summing contributions
of all edges.

\textbf{\dav{Data attachment.}}
We finally prevent vertices to depart too much from their original
position, and add a data attachment term. Denoting $q$ the original
position of the vertices, this reads:
$$E_d = \sum_i \|p_i-q_i\|^2\,.$$
Its gradient is $\nabla_p E_d = 2 (p-q)$.

\textbf{\dav{{Solving for positions.}}}
Denoting $E = E_m + w_1 E_f + w_2 E_d $ the weighted sum of the
energies, we can now obtain $p$ by solving $\nabla_p E = 0$. This
amounts to solving the linear system
$$ ({C} + w_1 {D} + w_2 \text{Id} ) p = w_2 q \,,$$ where Id is the $3|p|
\times 3|p|$ identity matrix. {One can easily show that by definitions, operators $C$
  and $D$ are positive-semidefinite. As long as  $w_2>0$, the overall left-hand
  linear operator is thus
  positive-definite and efficient
  inversion algorithms can be used to solve the system.}
 Except for inpainting which requires large values of $w_1$ as discussed next, 
 most results were obtained with $w_1 = 1$, though decreasing $w_1$ can be useful for large noise or 
 unreliable vertex positions. We have kept $w_2$ fixed and small throughout all our experiments, 
 with $w_2 = 0.05$.
 Note that meshes are uniformly scaled to fit the unit ball, so that parameters are comparable across meshes.

\section{Applications}

Inspired by image processing, our method serves various geometry processing applications illustrated in this section. This includes restoring noisy meshes, segmenting meshes, inpainting missing data, or embossing a normal map into the mesh vertices.

\subsection{Denoising}

\begin{figure*}[tbh]
\centering
\begin{tabular}{@{}c@{}c@{}c@{}c@{}c@{}c@{}c@{}c@{}}
\includegraphics[width=0.122\linewidth]{images/helmet_original.jpg}&
\includegraphics[width=0.122\linewidth]{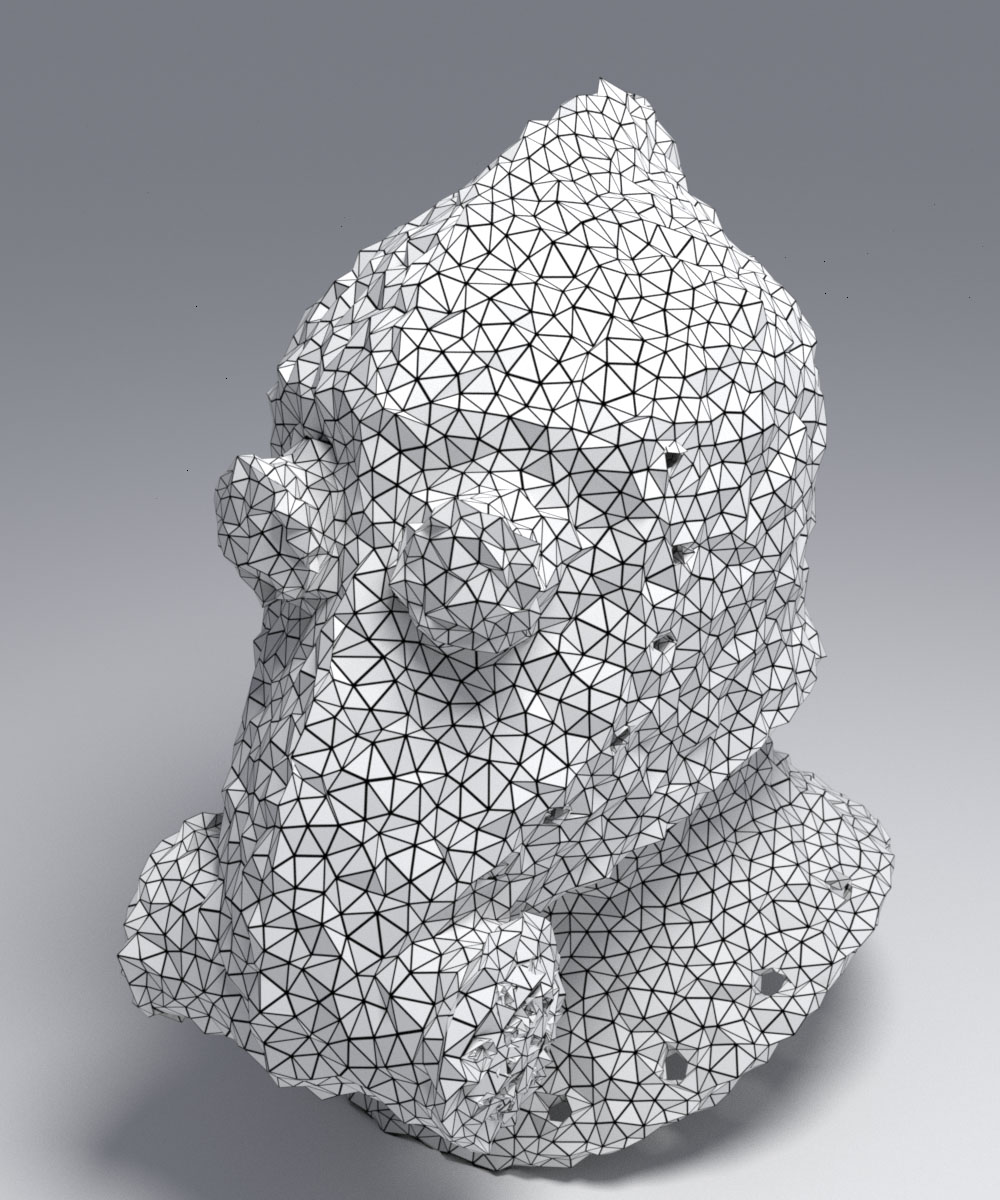}&
\includegraphics[width=0.122\linewidth]{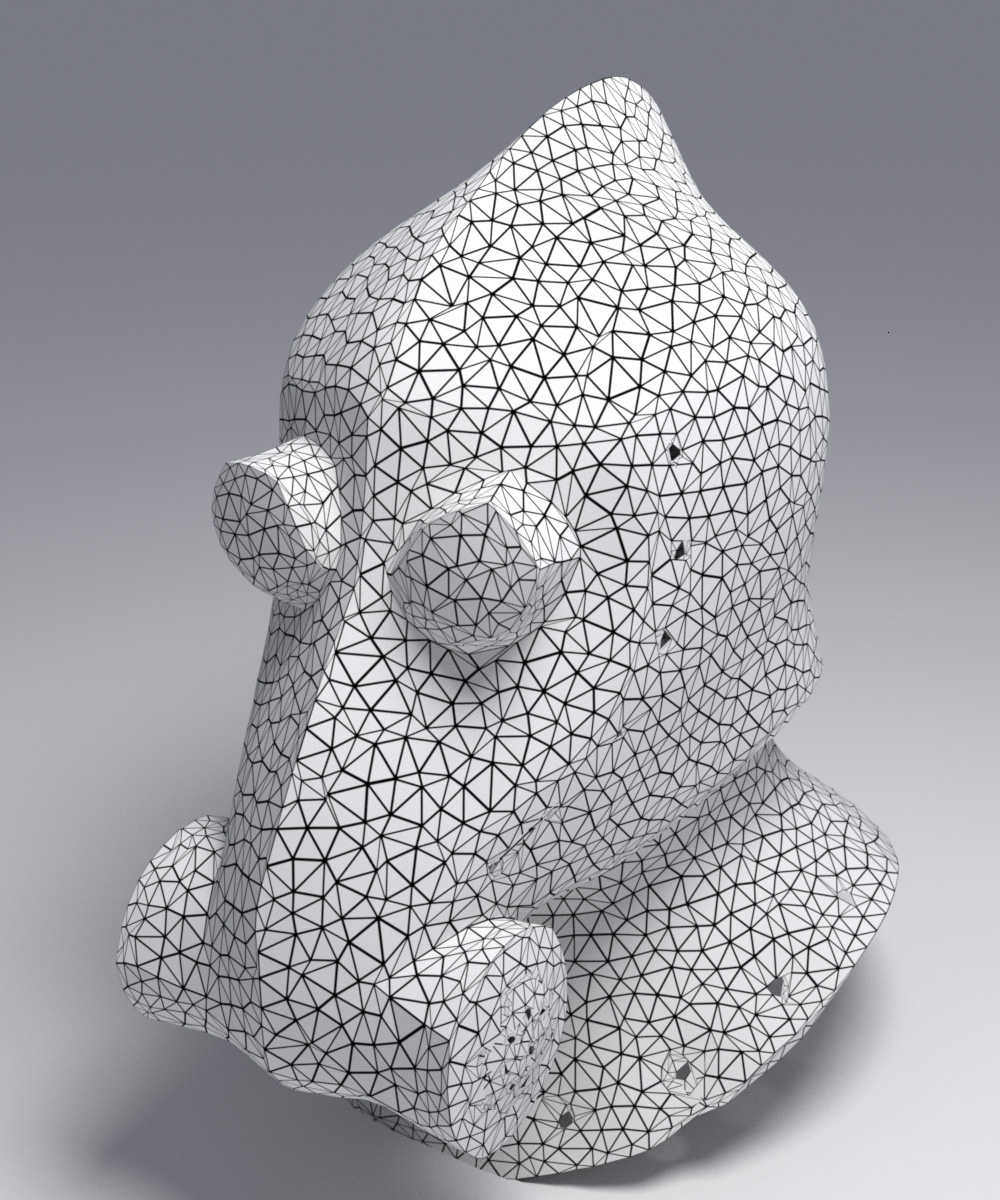}&
\includegraphics[width=0.122\linewidth]{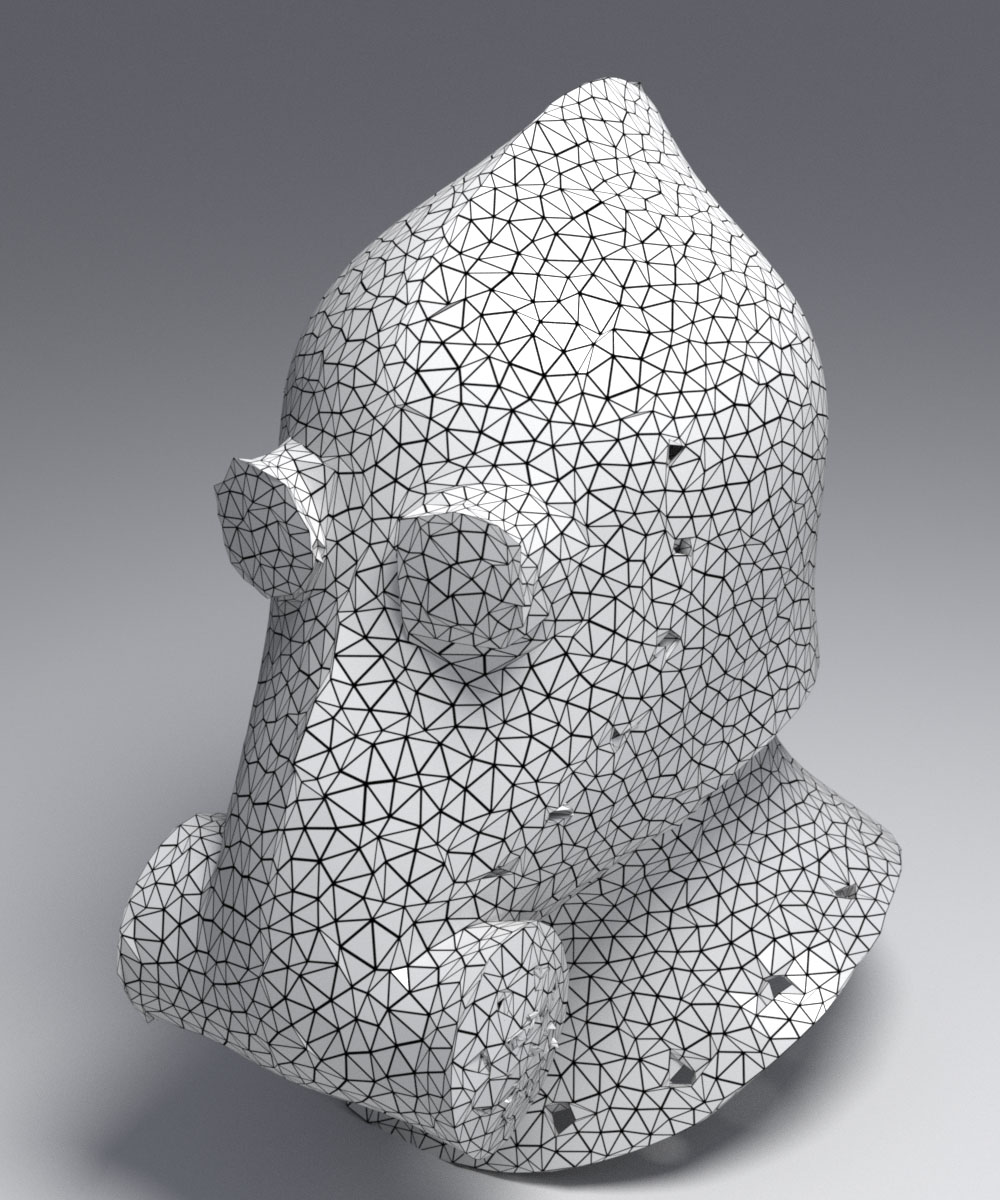}&
\includegraphics[width=0.122\linewidth]{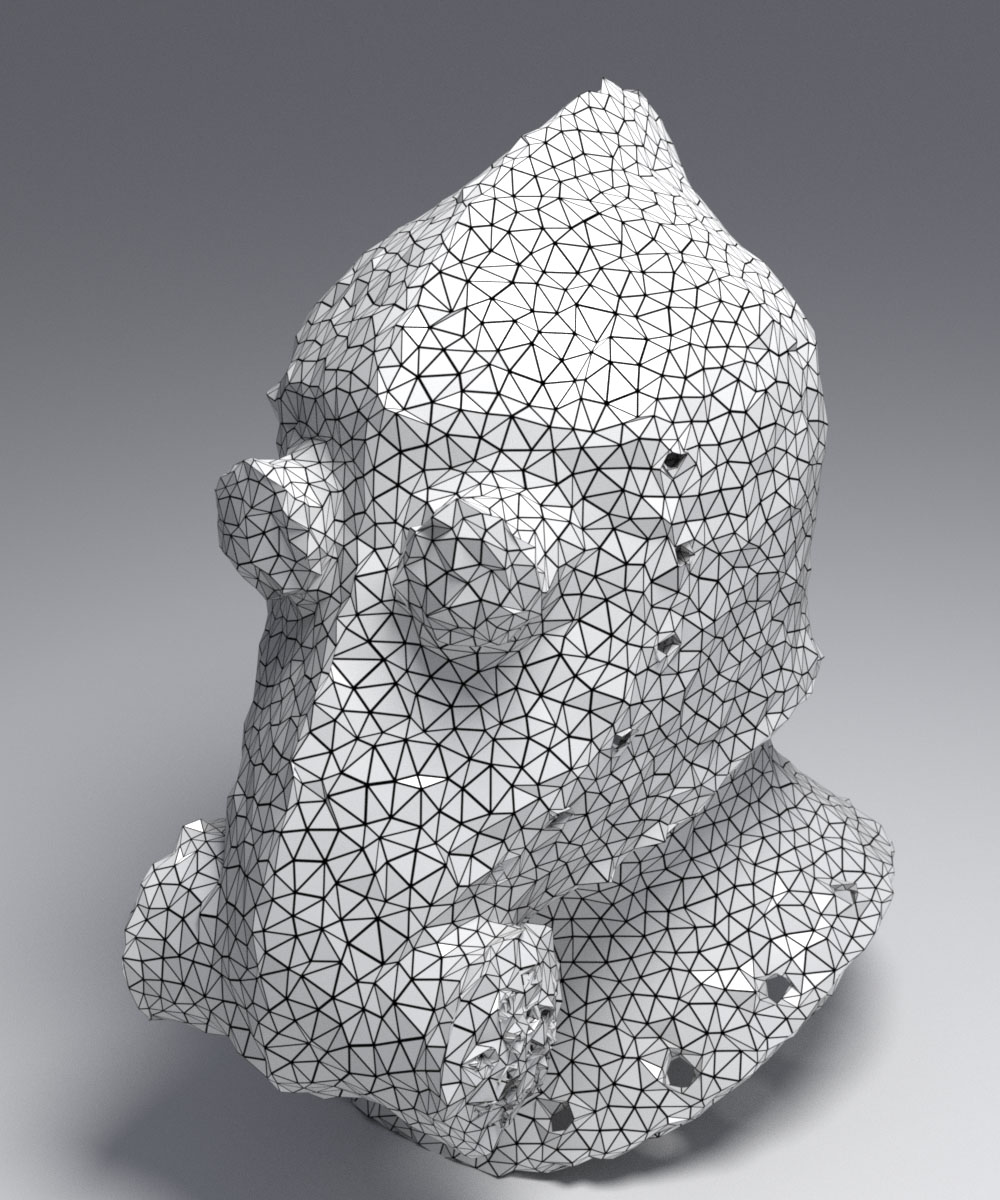}&
\includegraphics[width=0.122\linewidth]{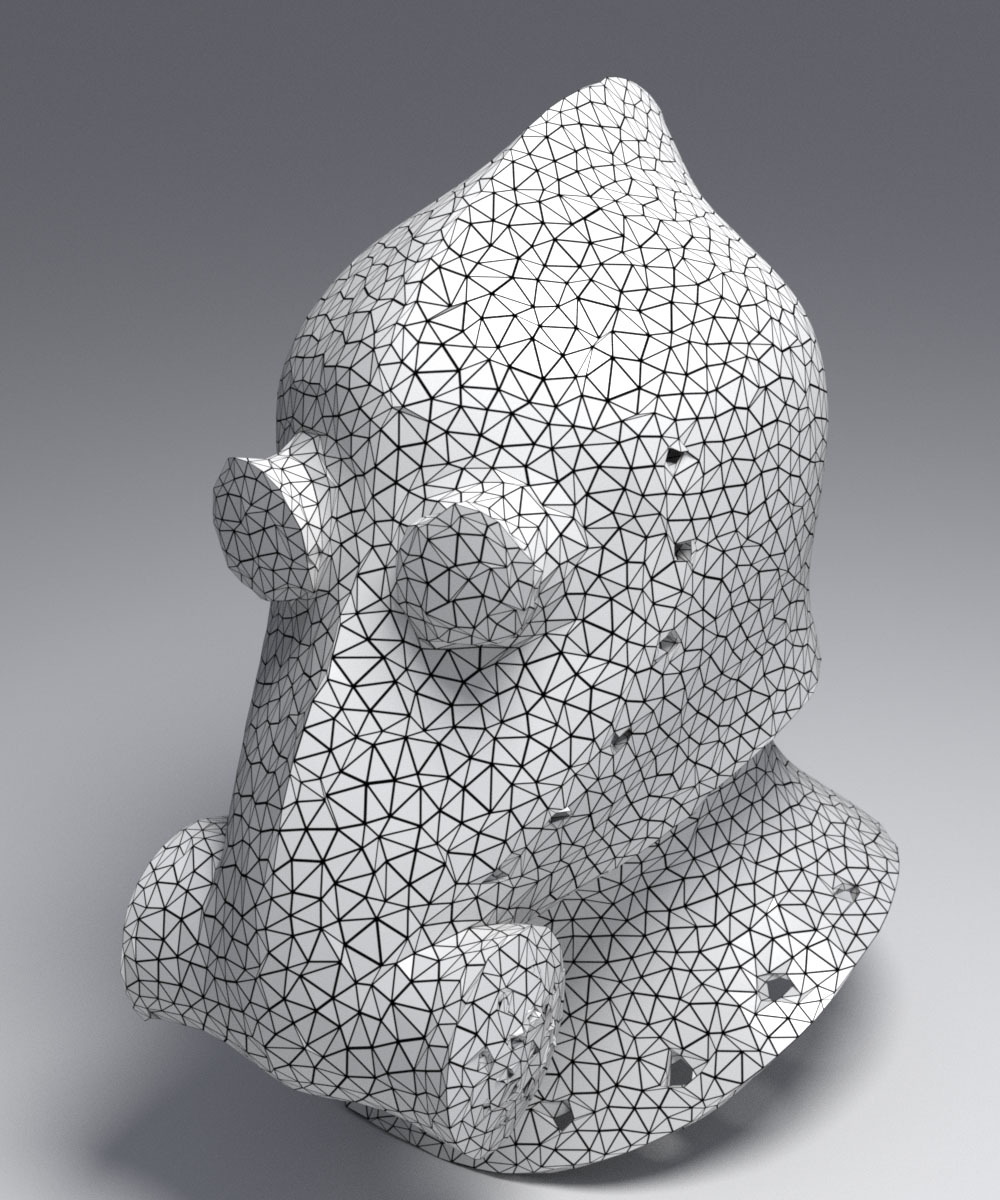}&
\includegraphics[width=0.122\linewidth]{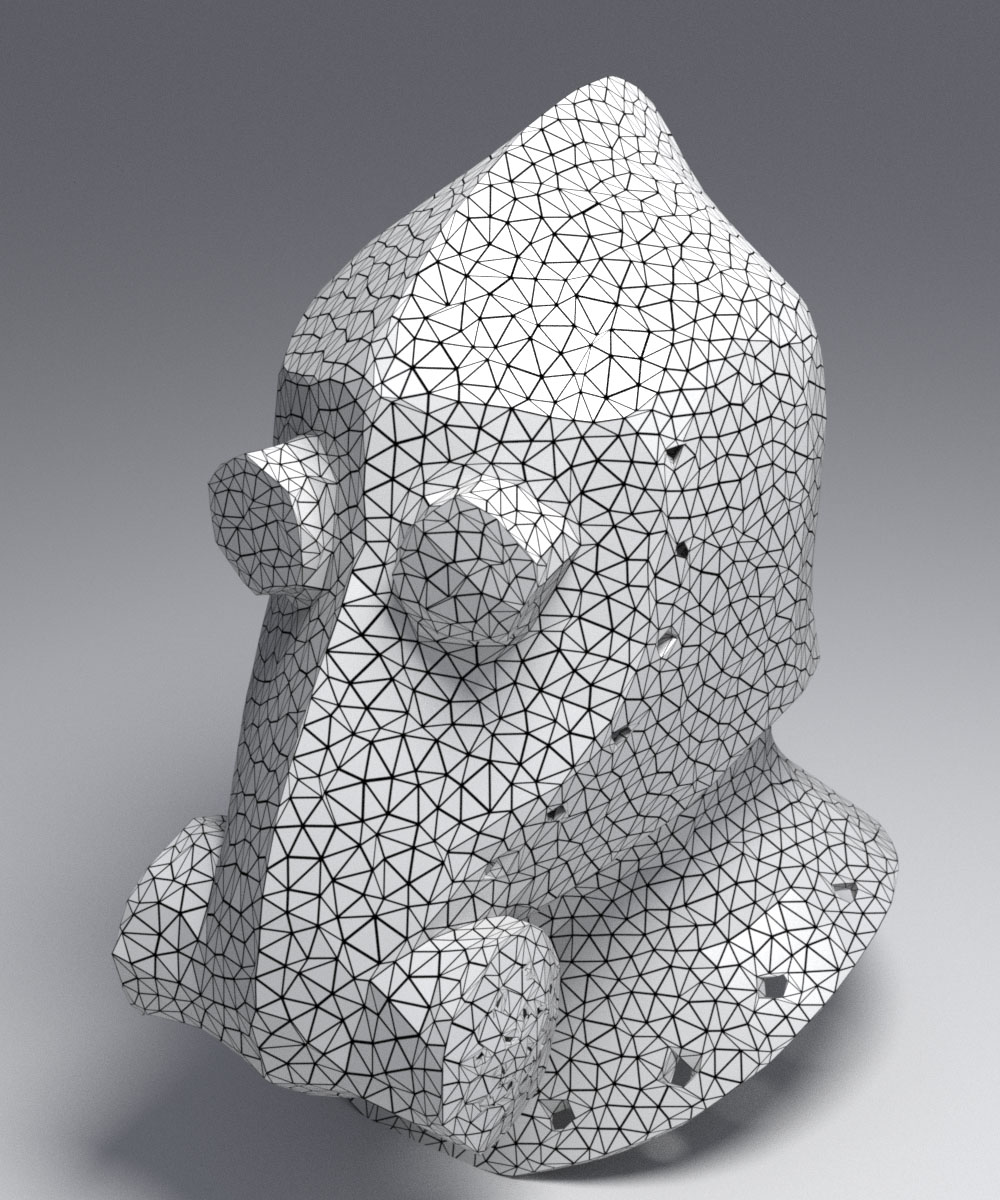}&
\includegraphics[width=0.122\linewidth]{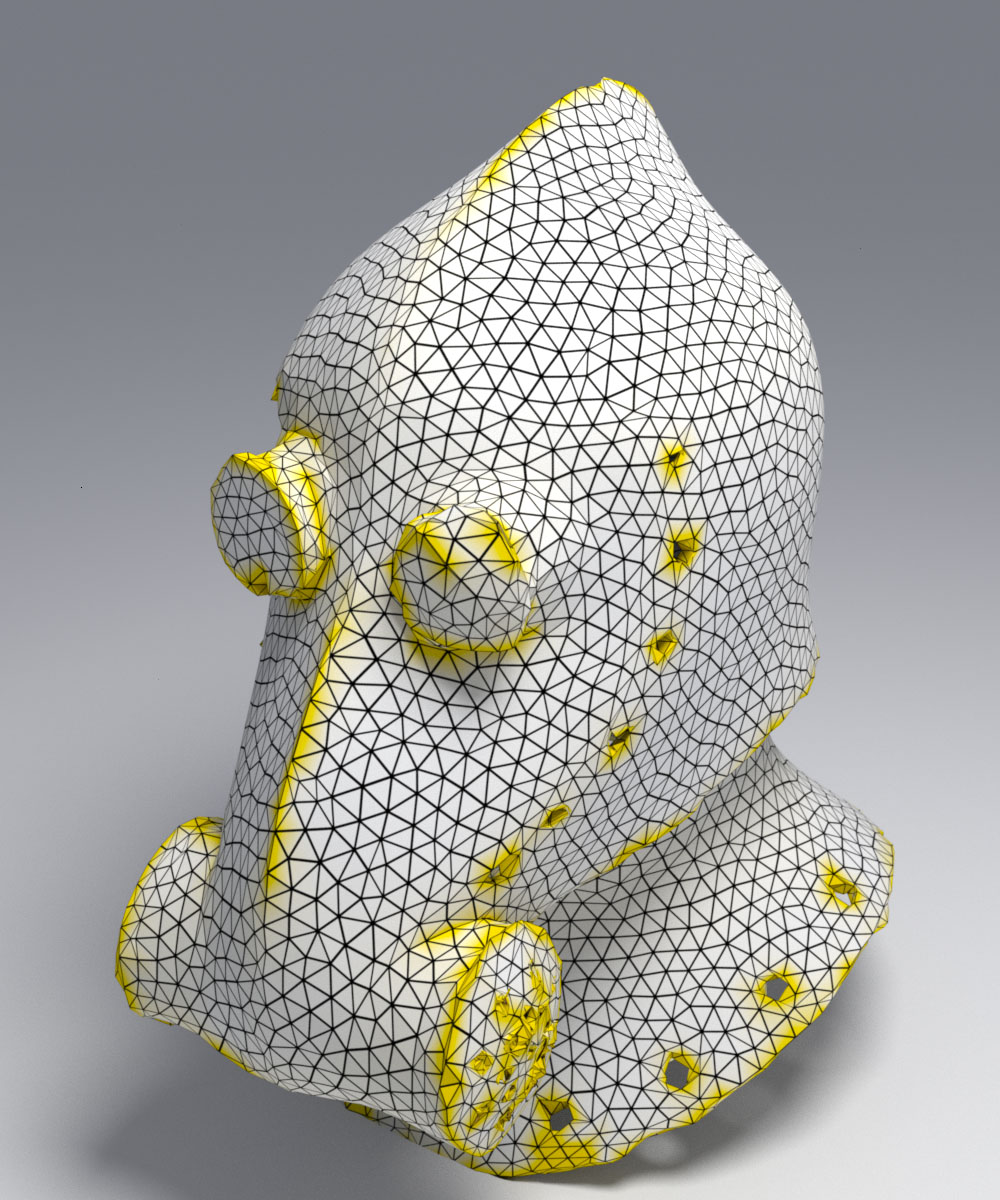} \\
\includegraphics[width=0.122\linewidth]{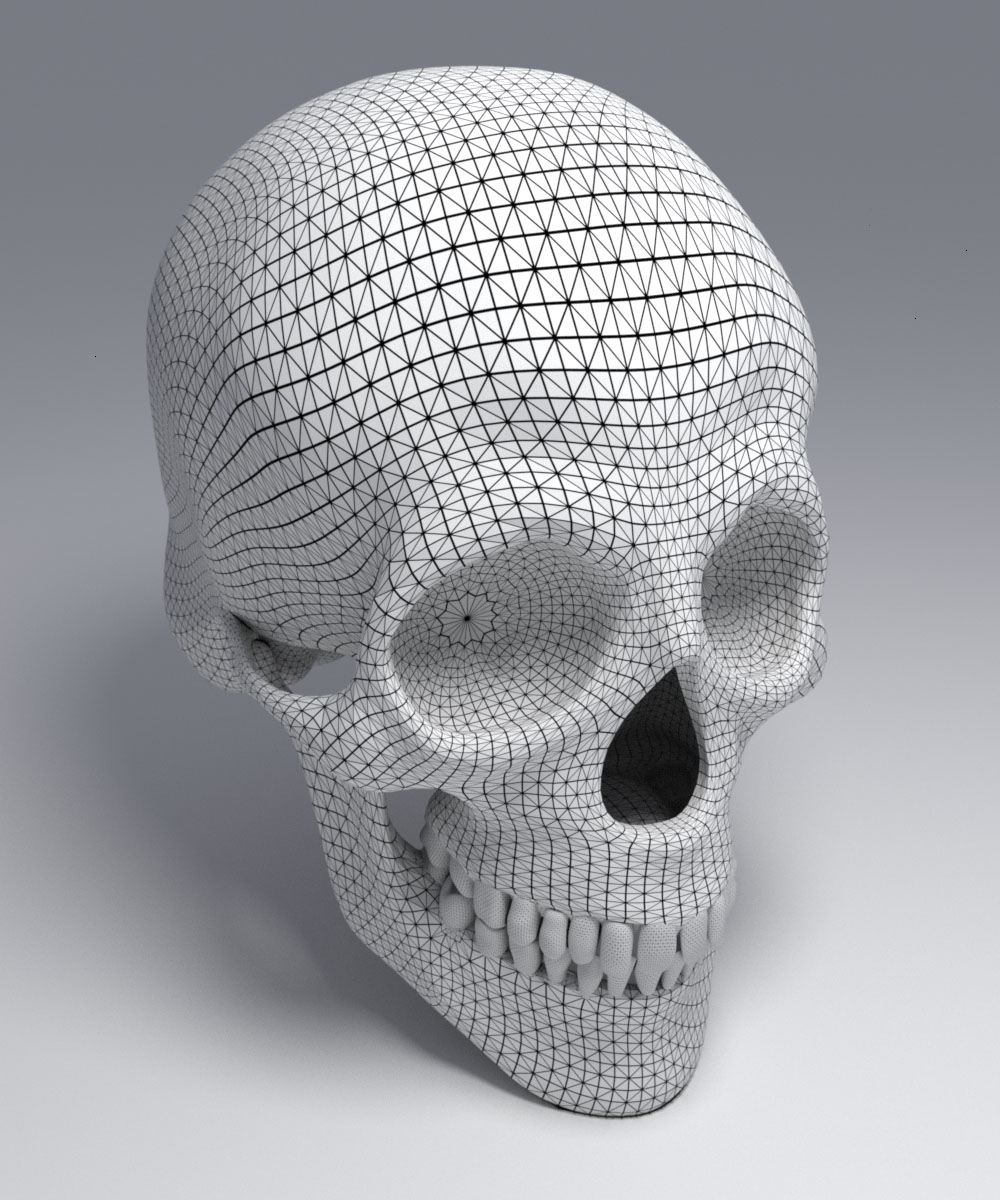}&
\includegraphics[width=0.122\linewidth]{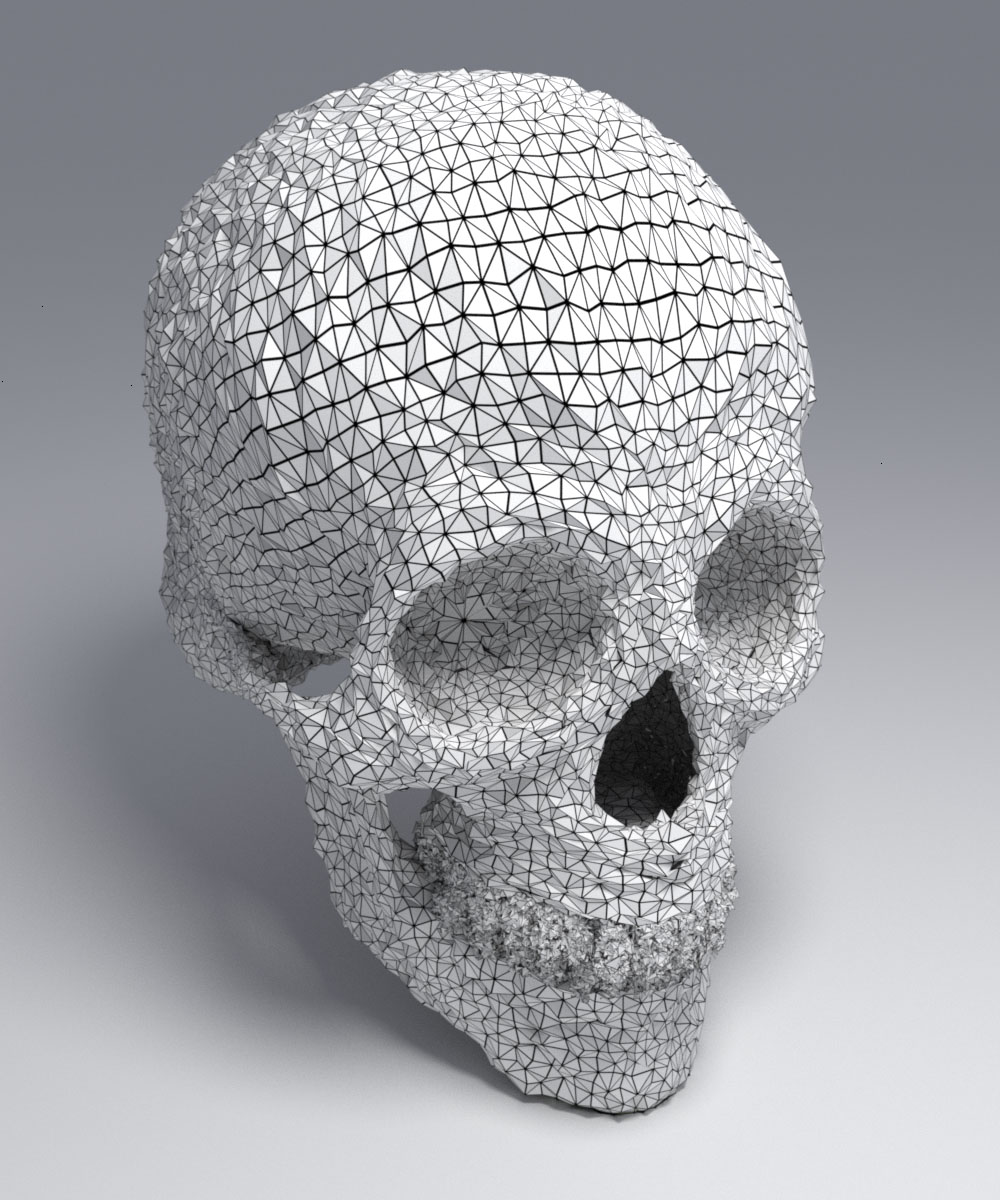}&
\includegraphics[width=0.122\linewidth]{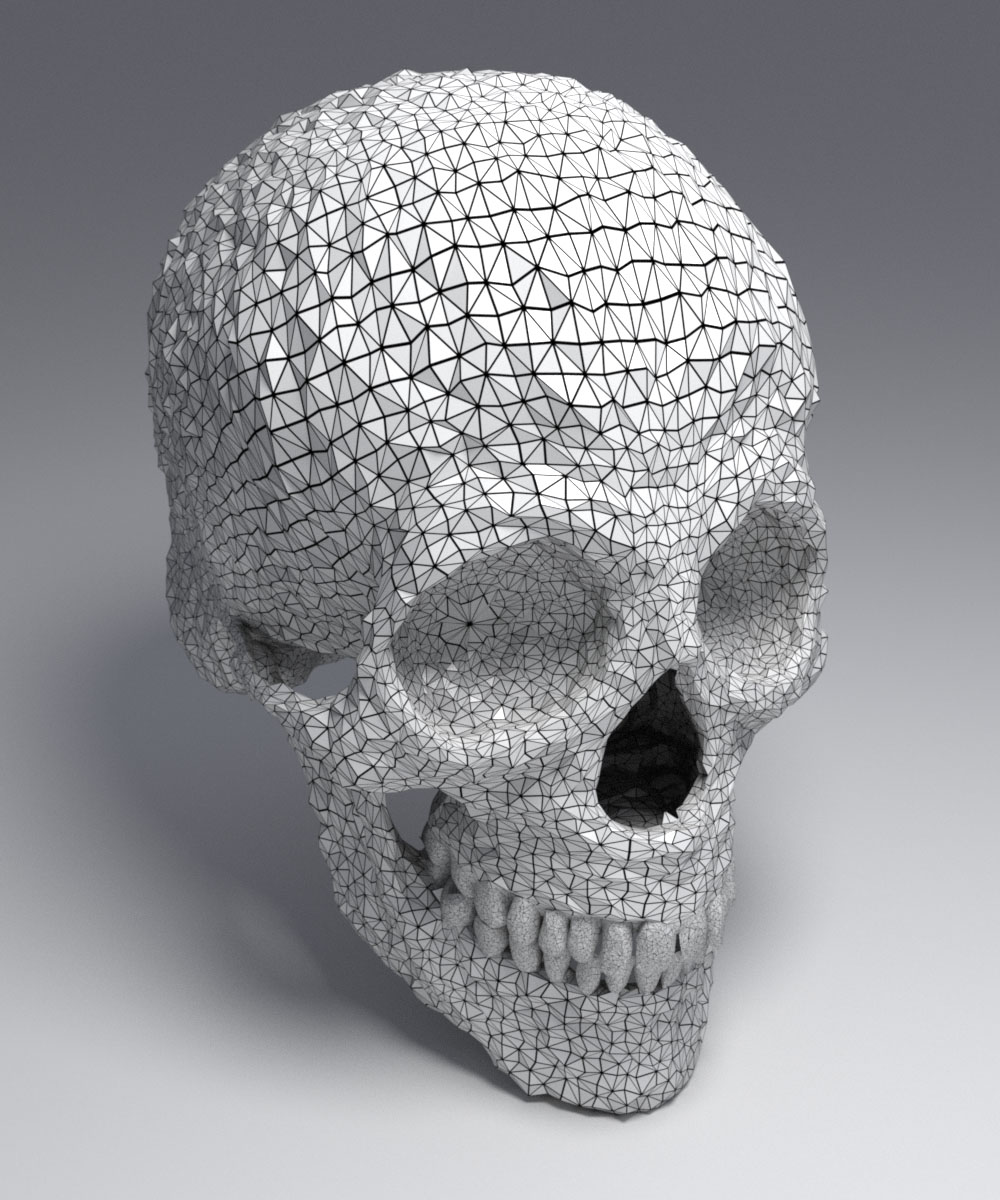}&
\includegraphics[width=0.122\linewidth]{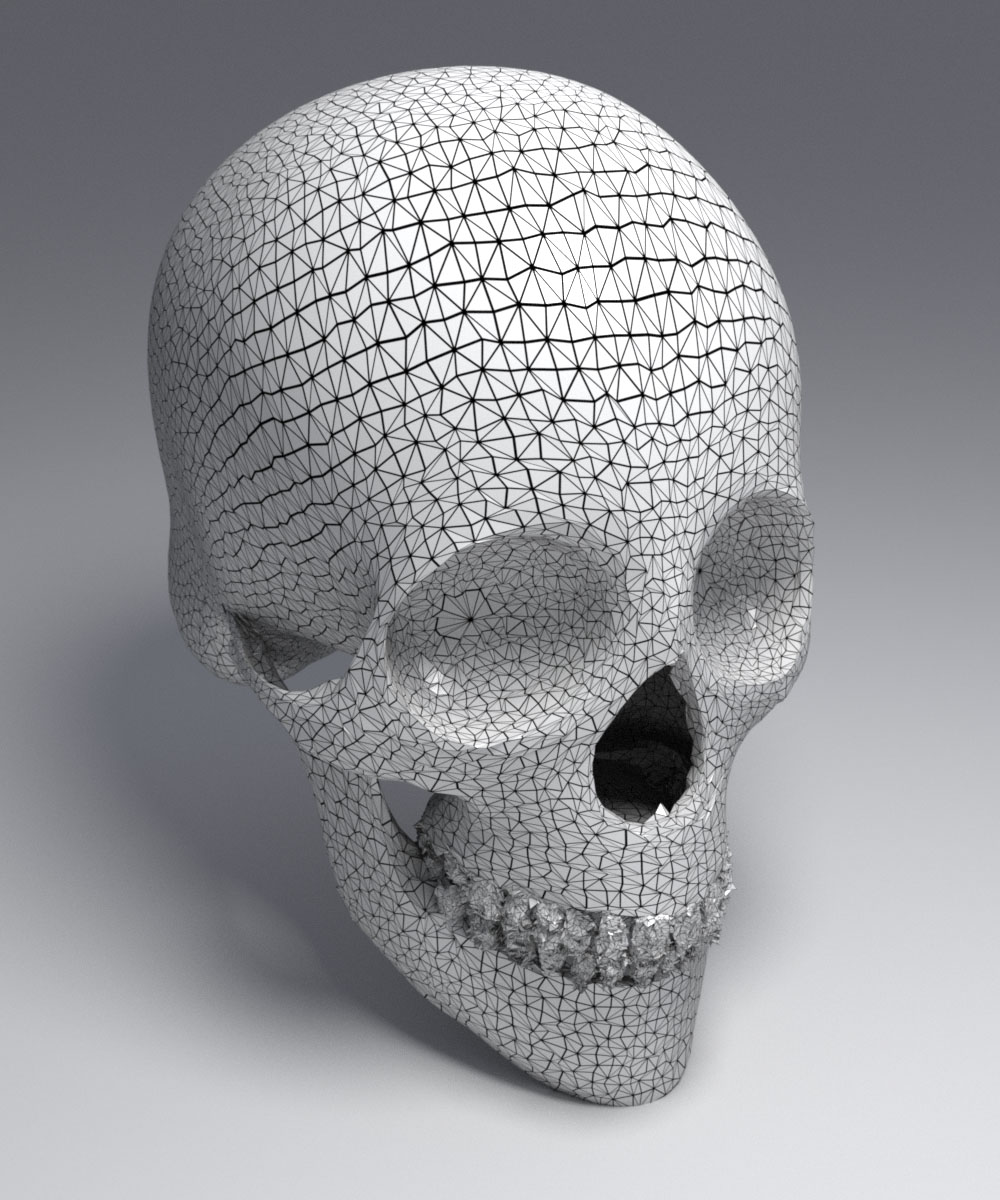}&
\includegraphics[width=0.122\linewidth]{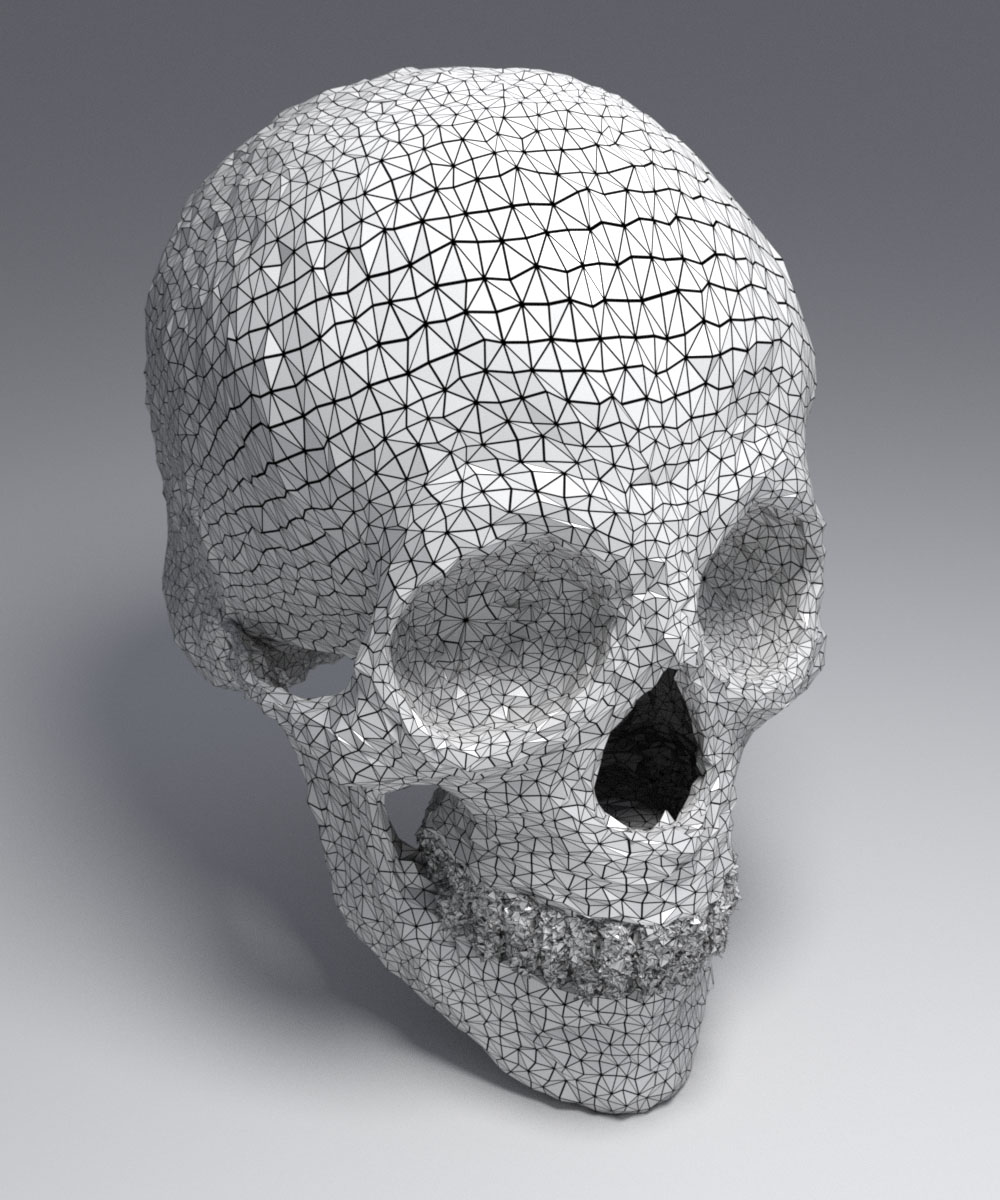}&
\includegraphics[width=0.122\linewidth]{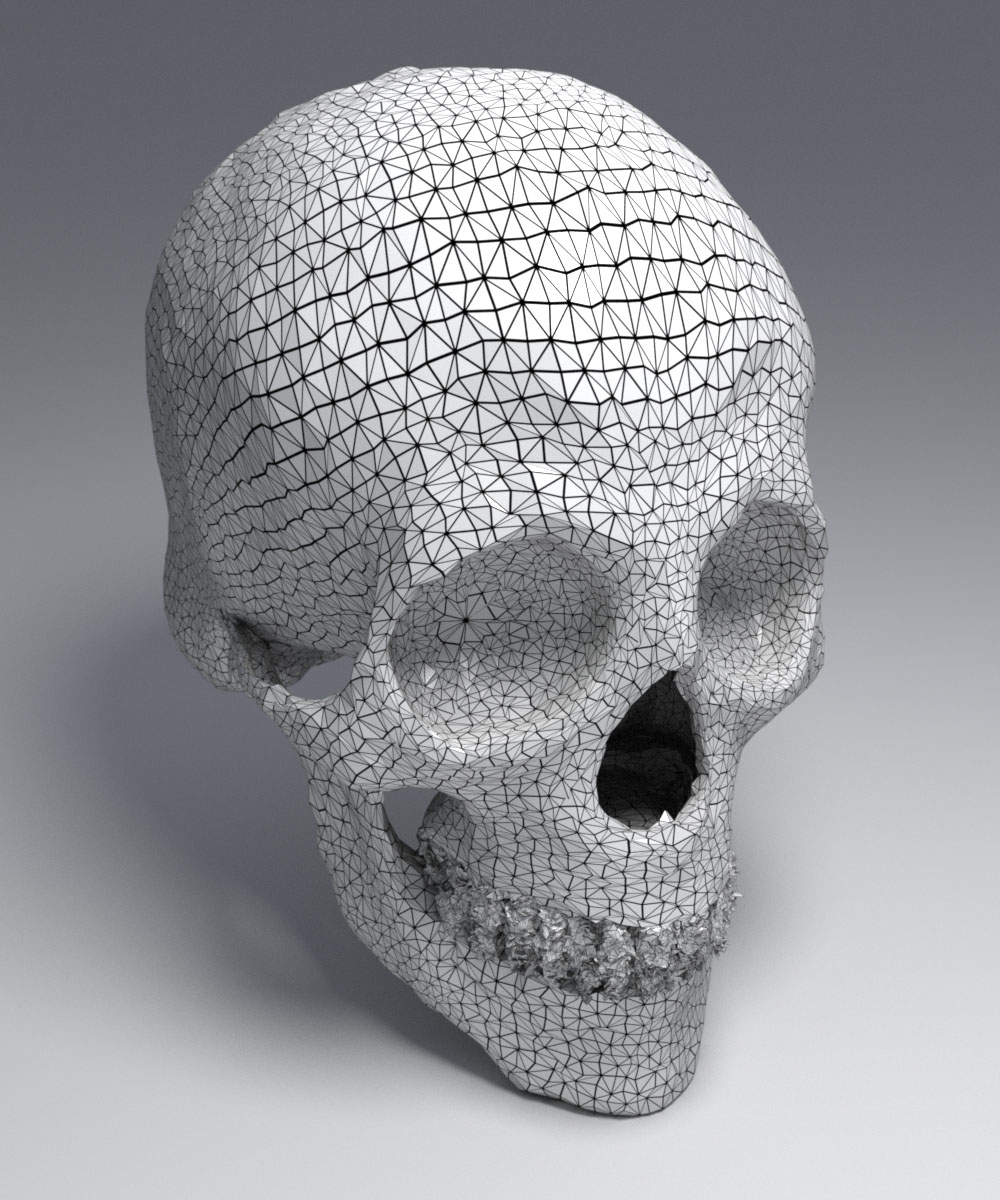}&
\includegraphics[width=0.122\linewidth]{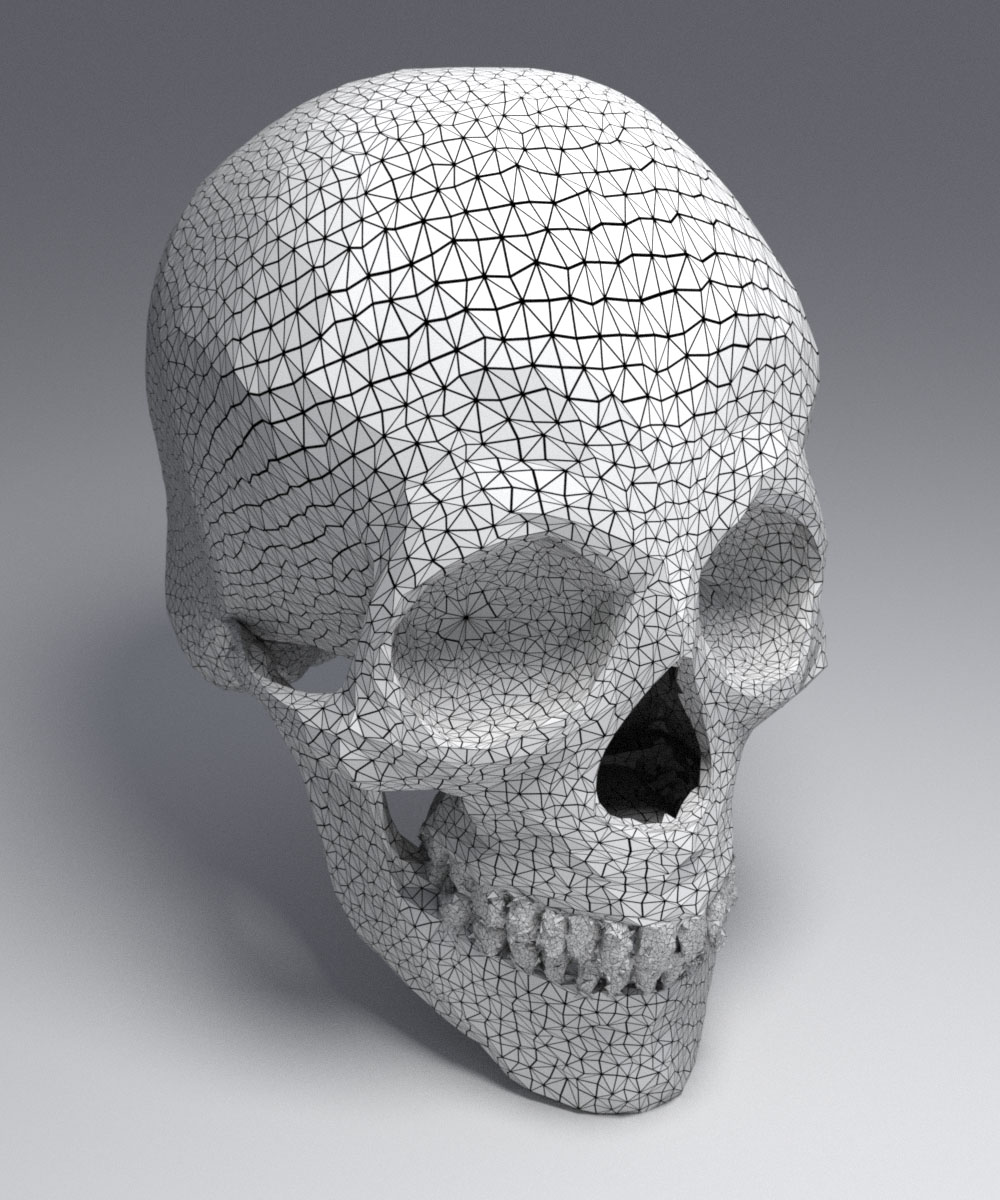}&
\includegraphics[width=0.122\linewidth]{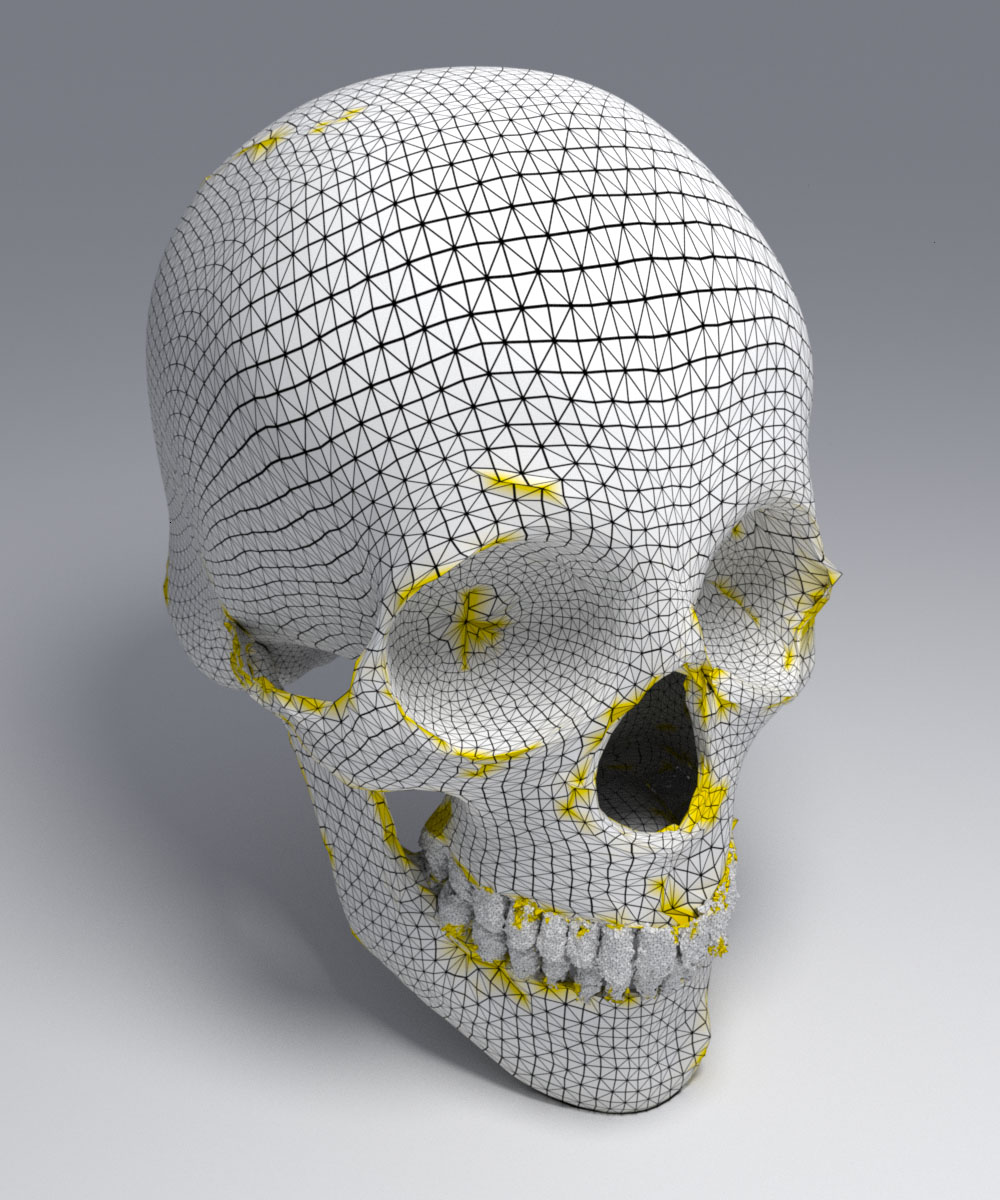} \\
\includegraphics[width=0.122\linewidth]{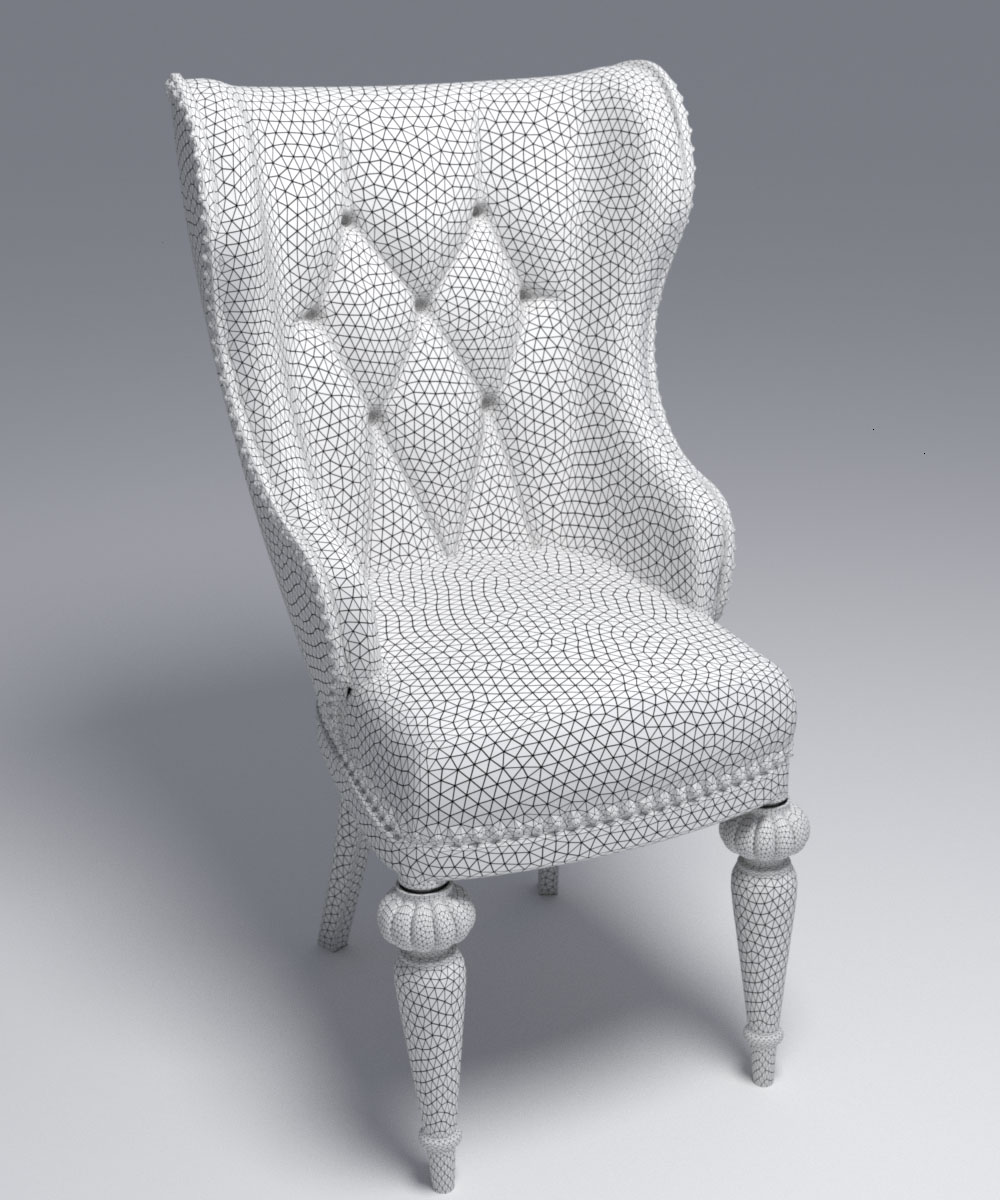}&
\includegraphics[width=0.122\linewidth]{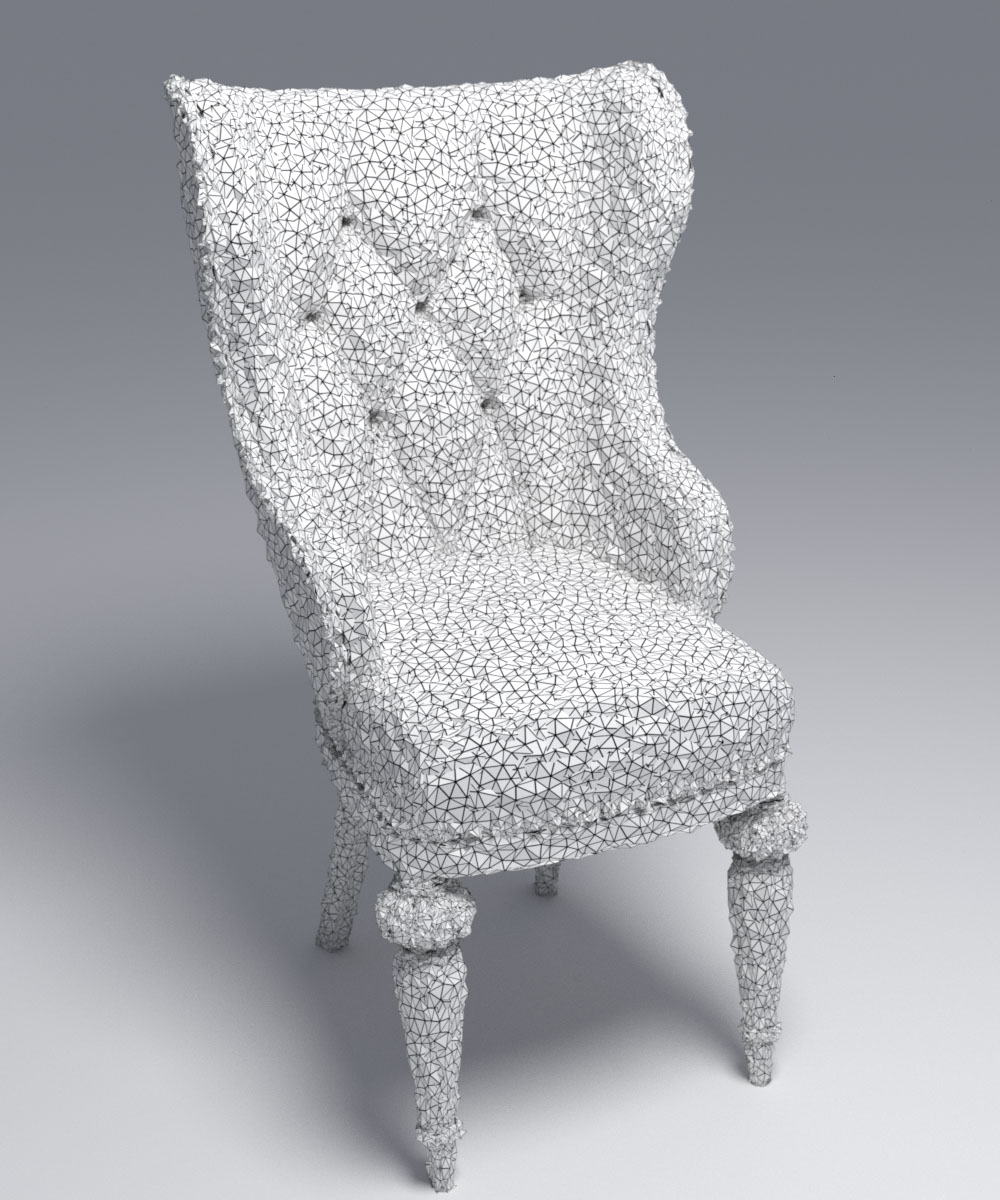}&
\includegraphics[width=0.122\linewidth]{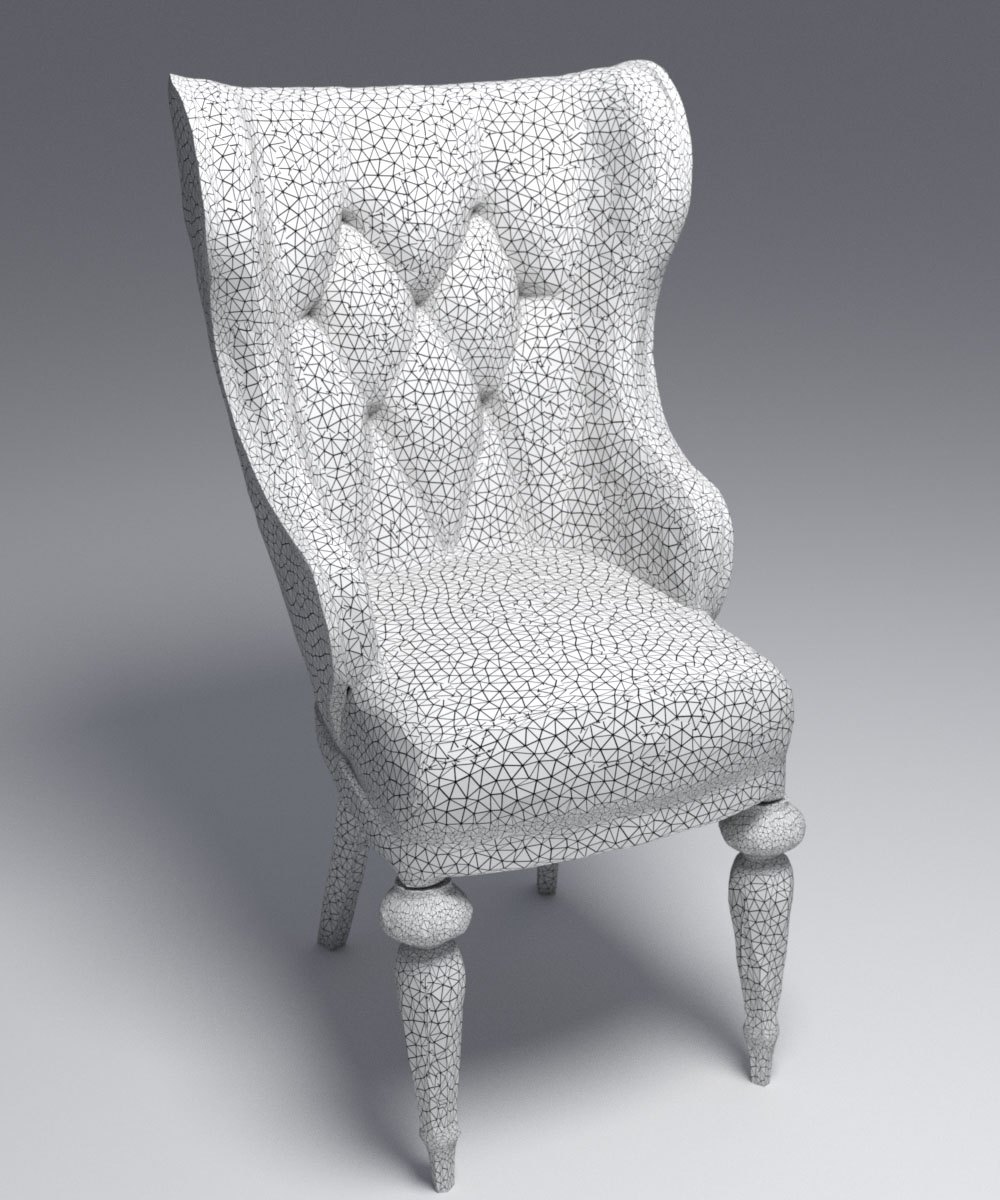}&
\includegraphics[width=0.122\linewidth]{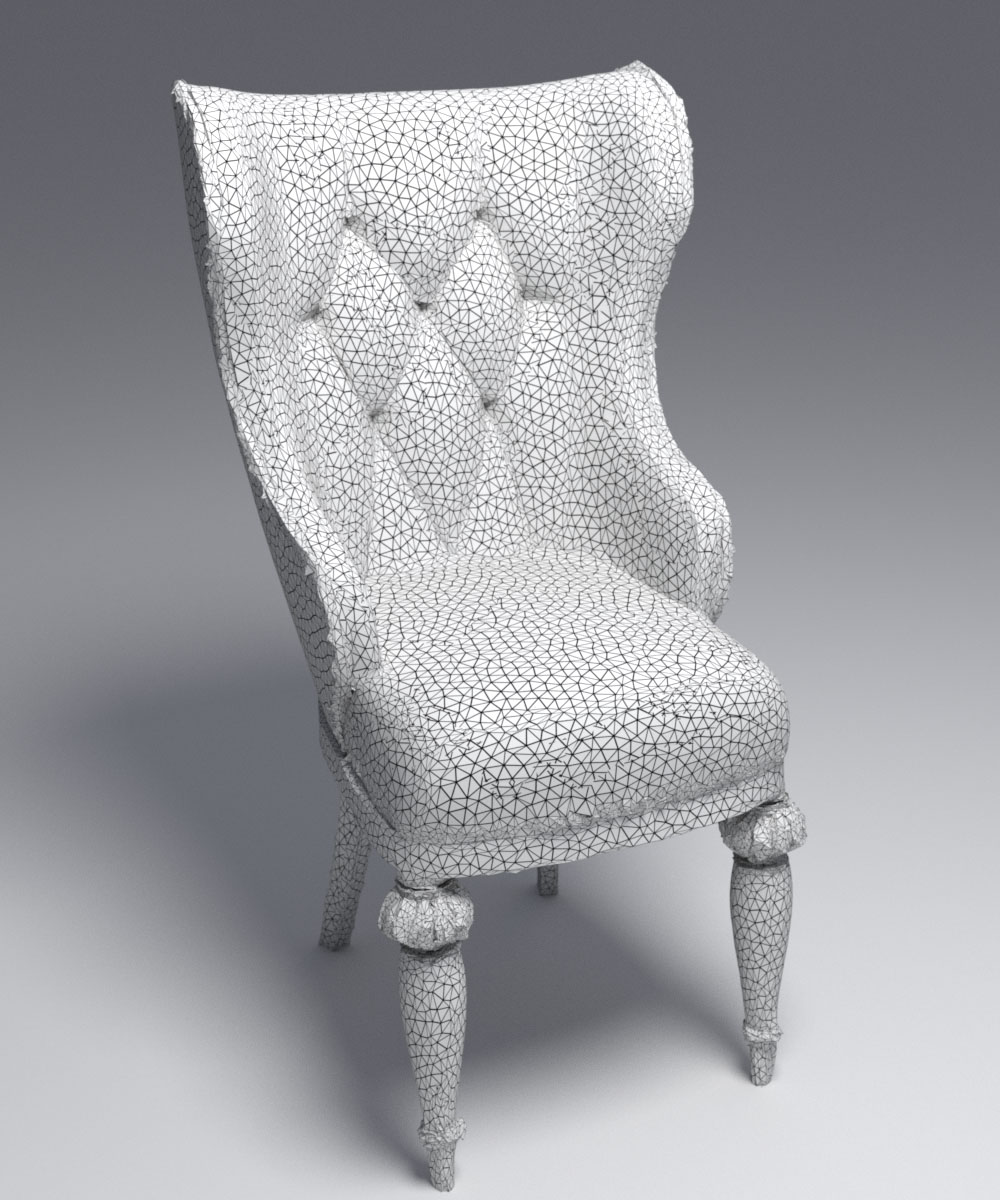}&
\includegraphics[width=0.122\linewidth]{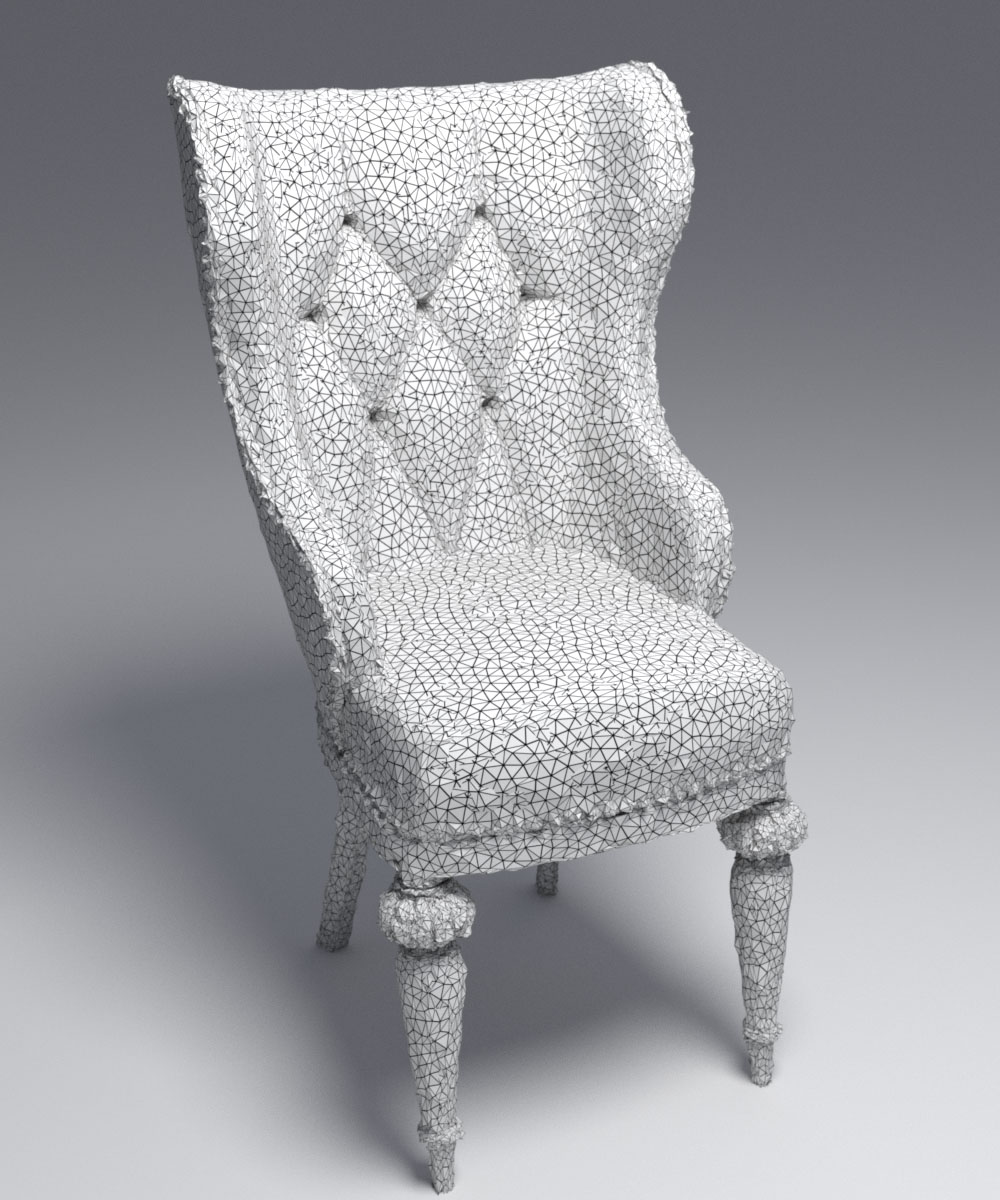}&
\includegraphics[width=0.122\linewidth]{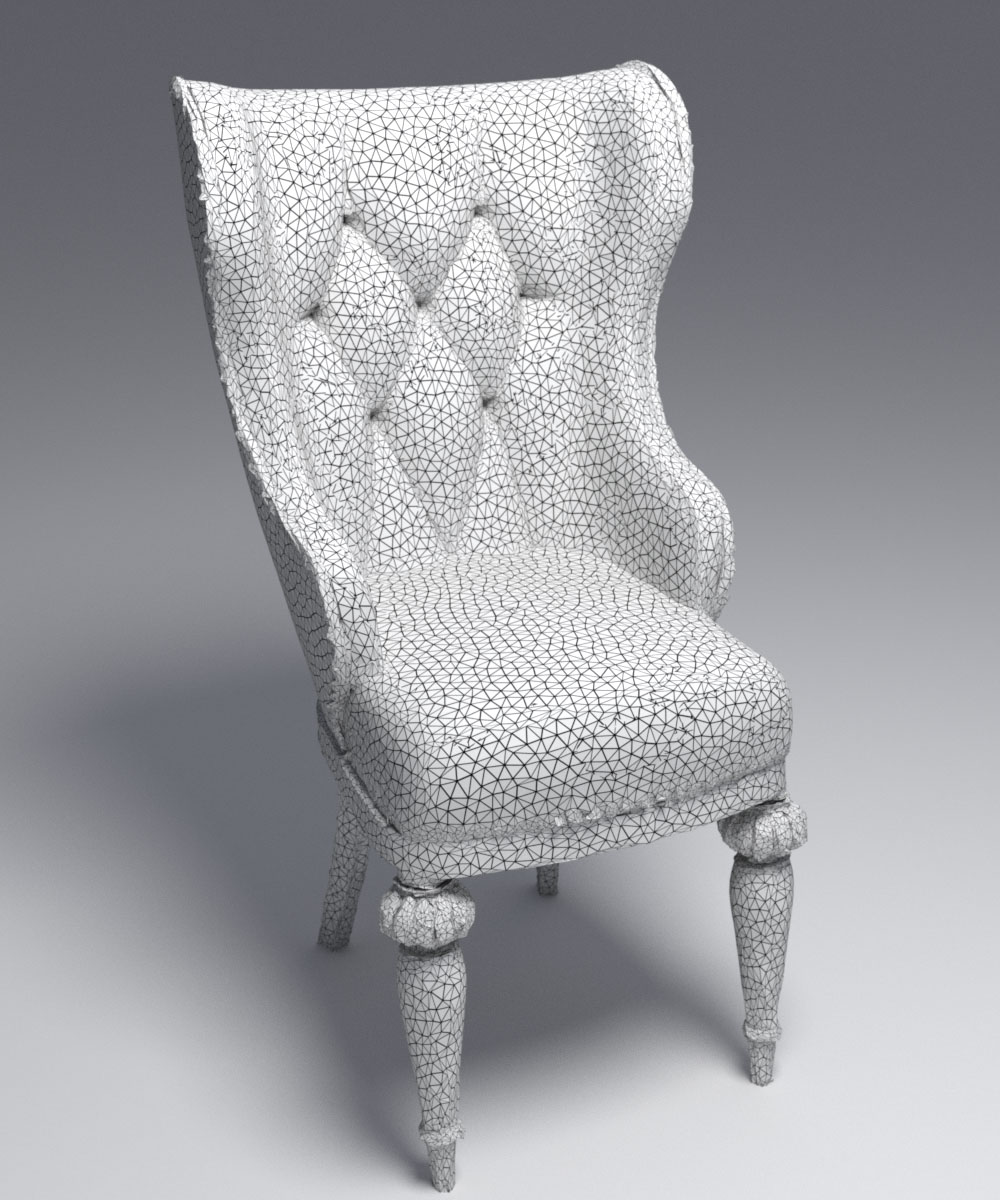}&
\includegraphics[width=0.122\linewidth]{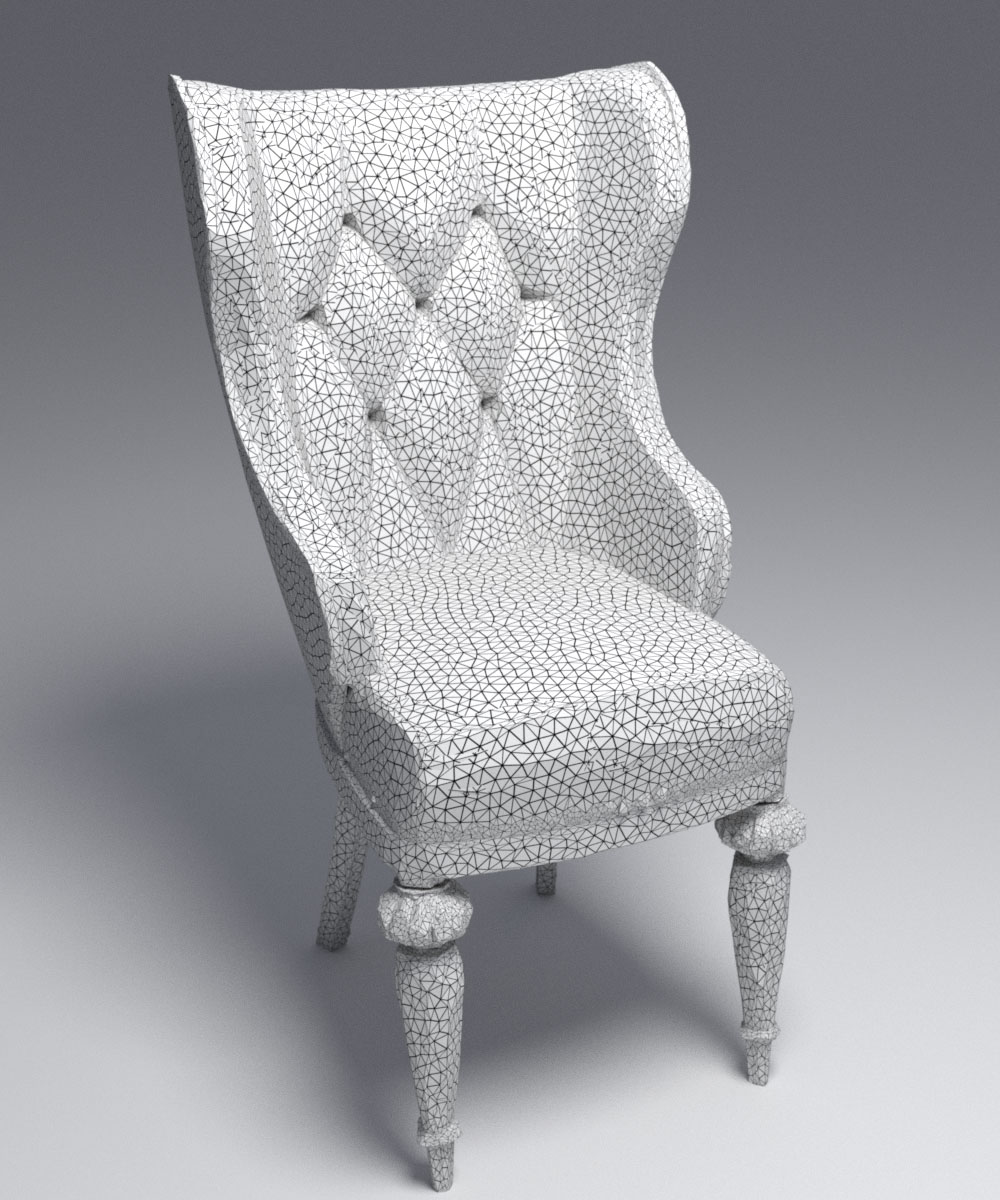}&
\includegraphics[width=0.122\linewidth]{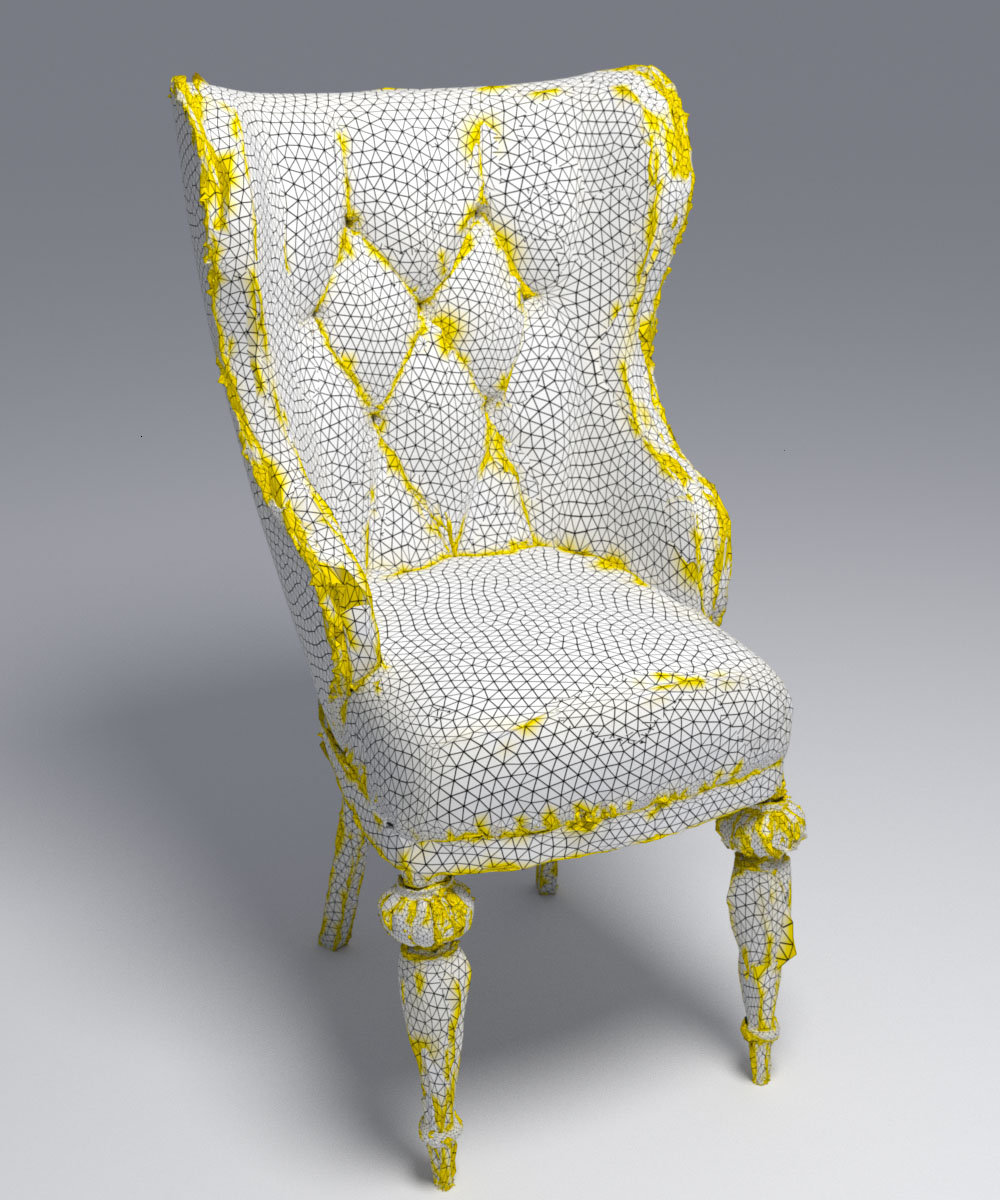} \\
\includegraphics[width=0.122\linewidth]{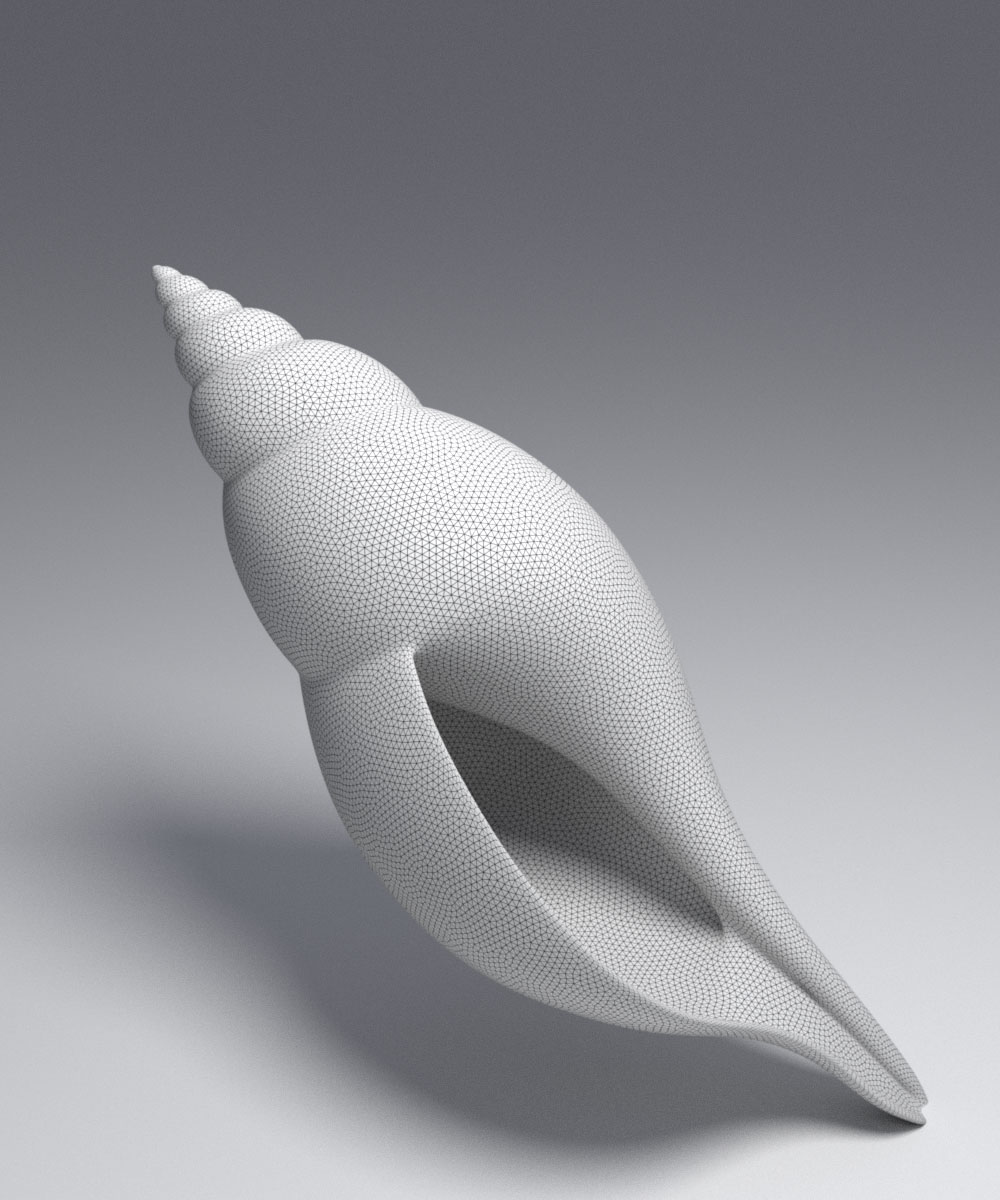}&
\includegraphics[width=0.122\linewidth]{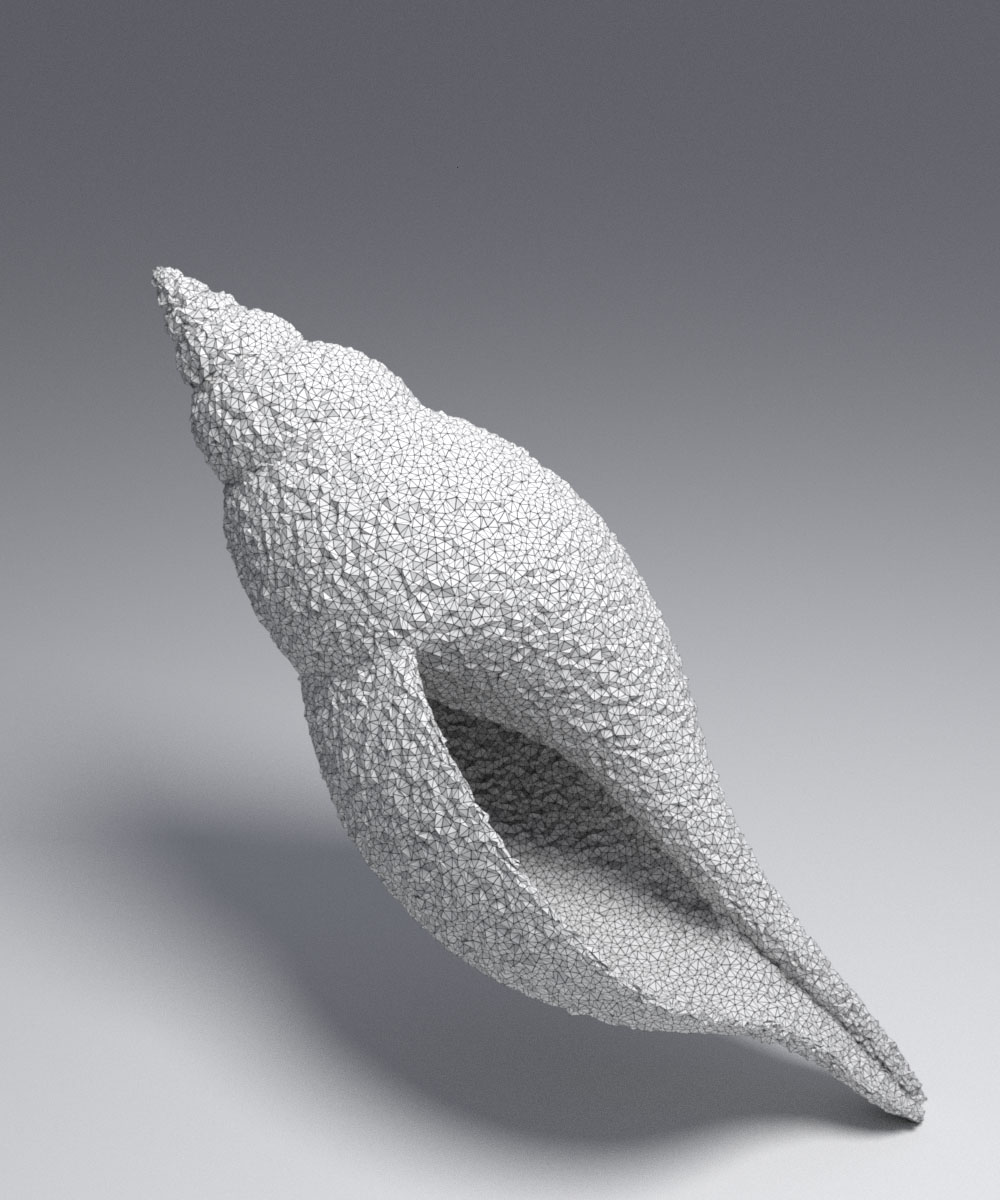}&
\includegraphics[width=0.122\linewidth]{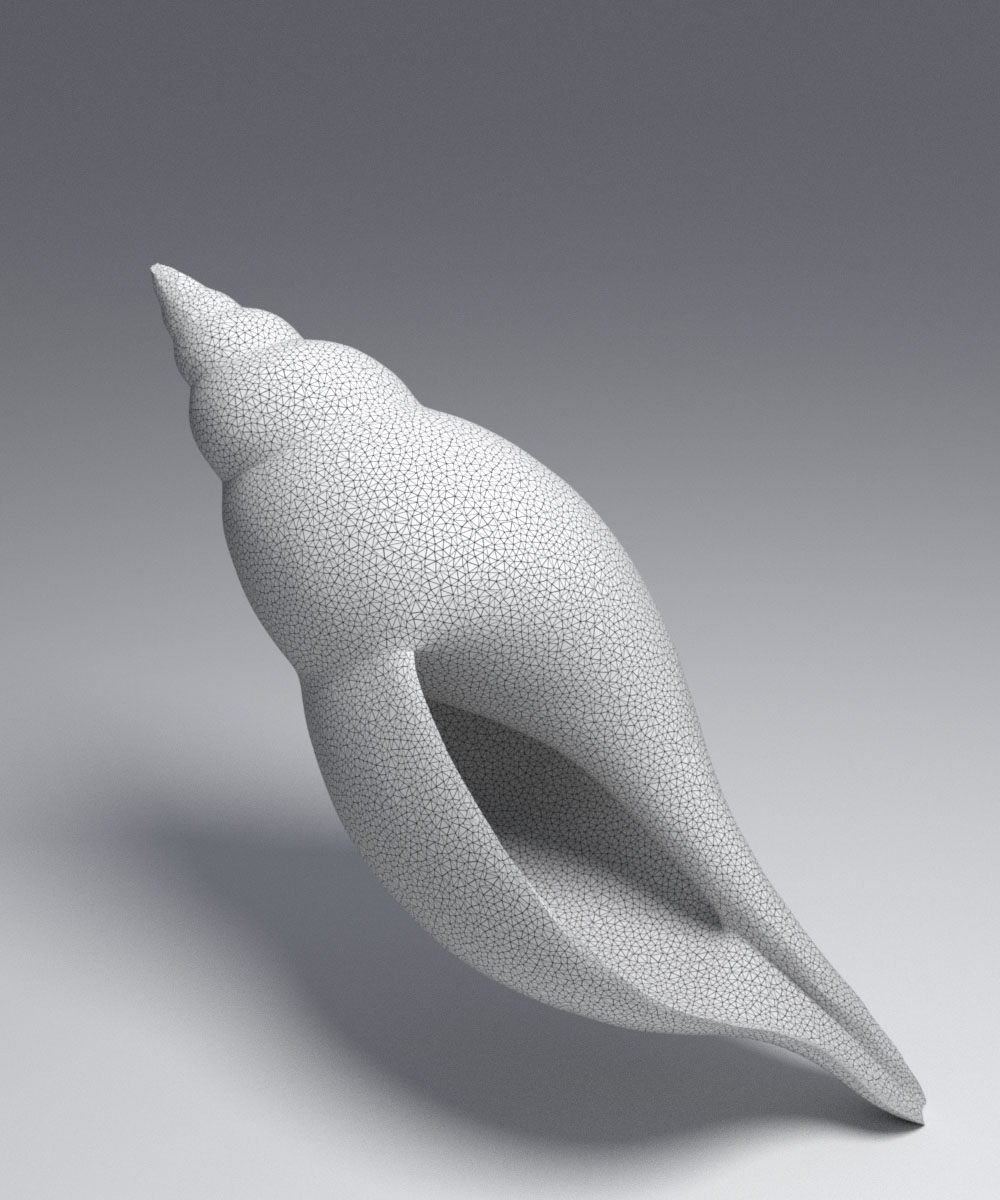}&
\includegraphics[width=0.122\linewidth]{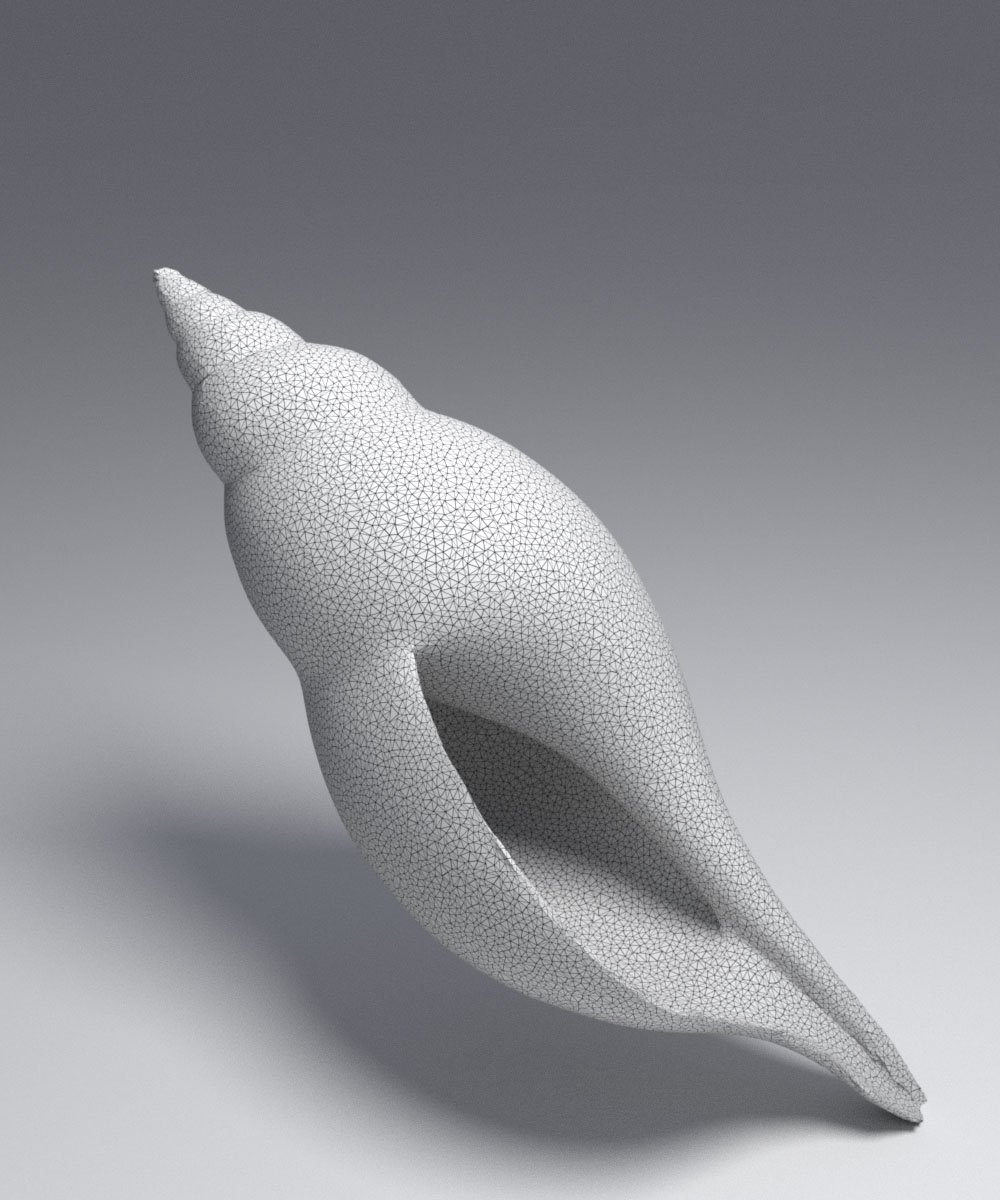}&
\includegraphics[width=0.122\linewidth]{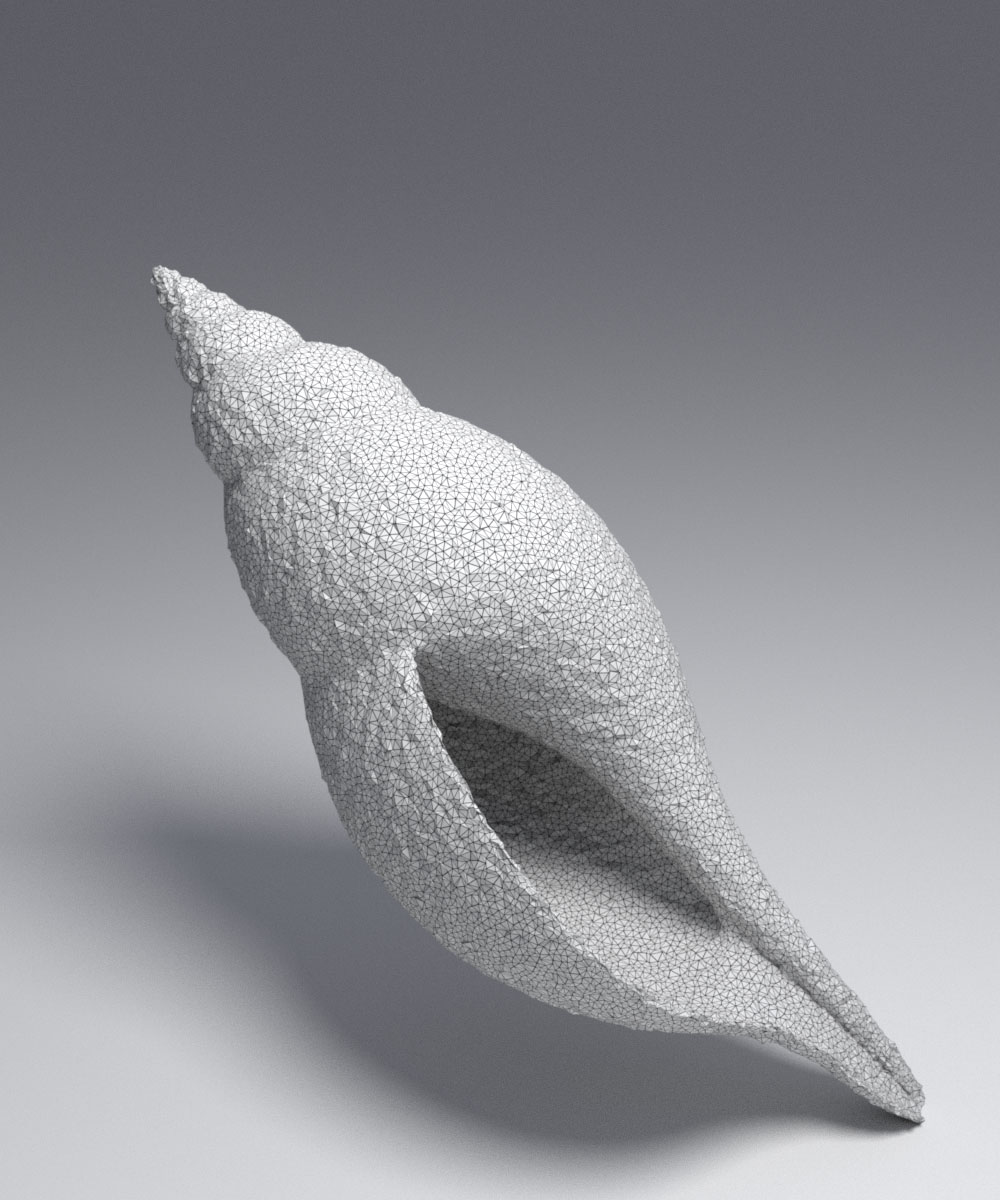}&
\includegraphics[width=0.122\linewidth]{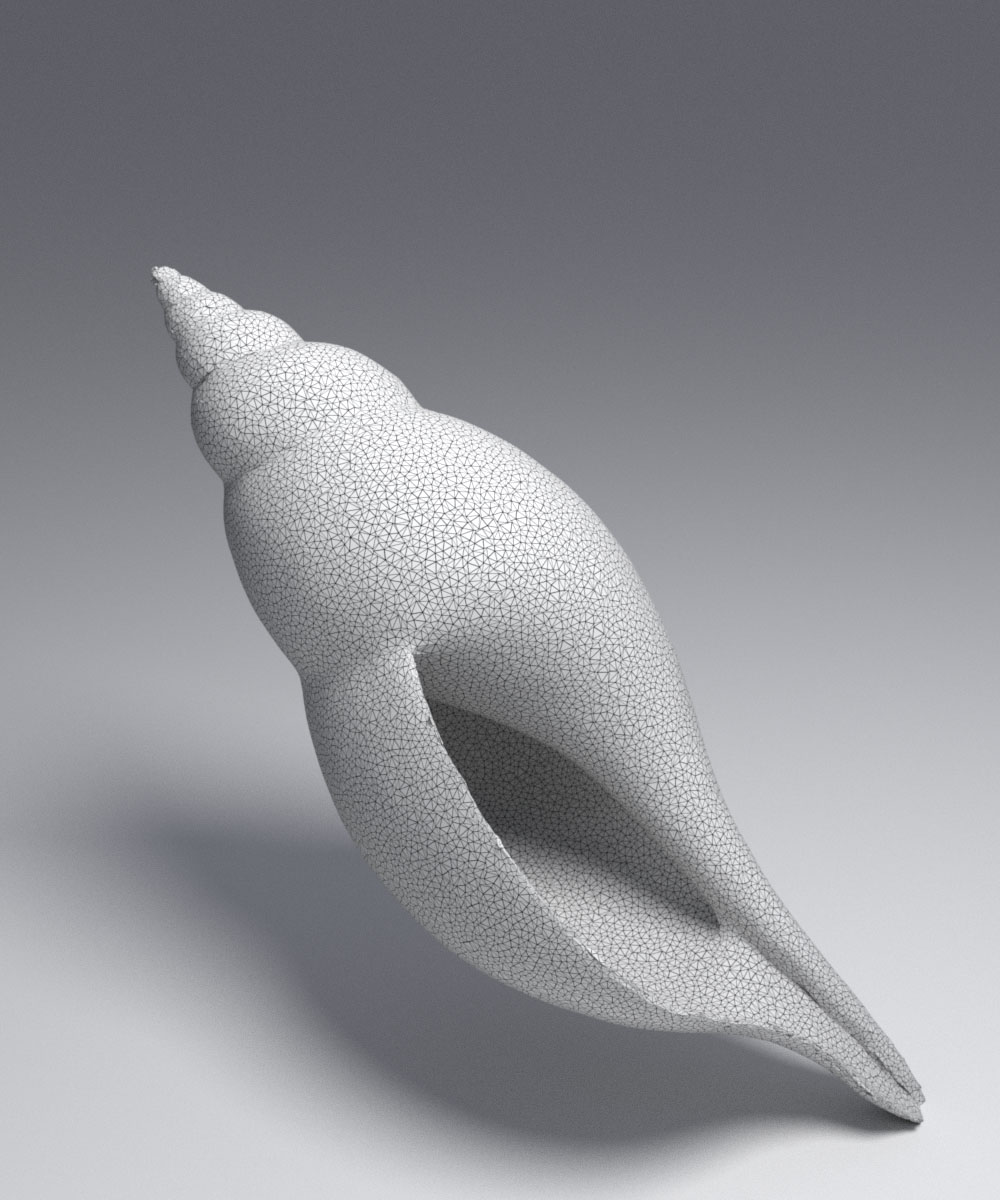}&
\includegraphics[width=0.122\linewidth]{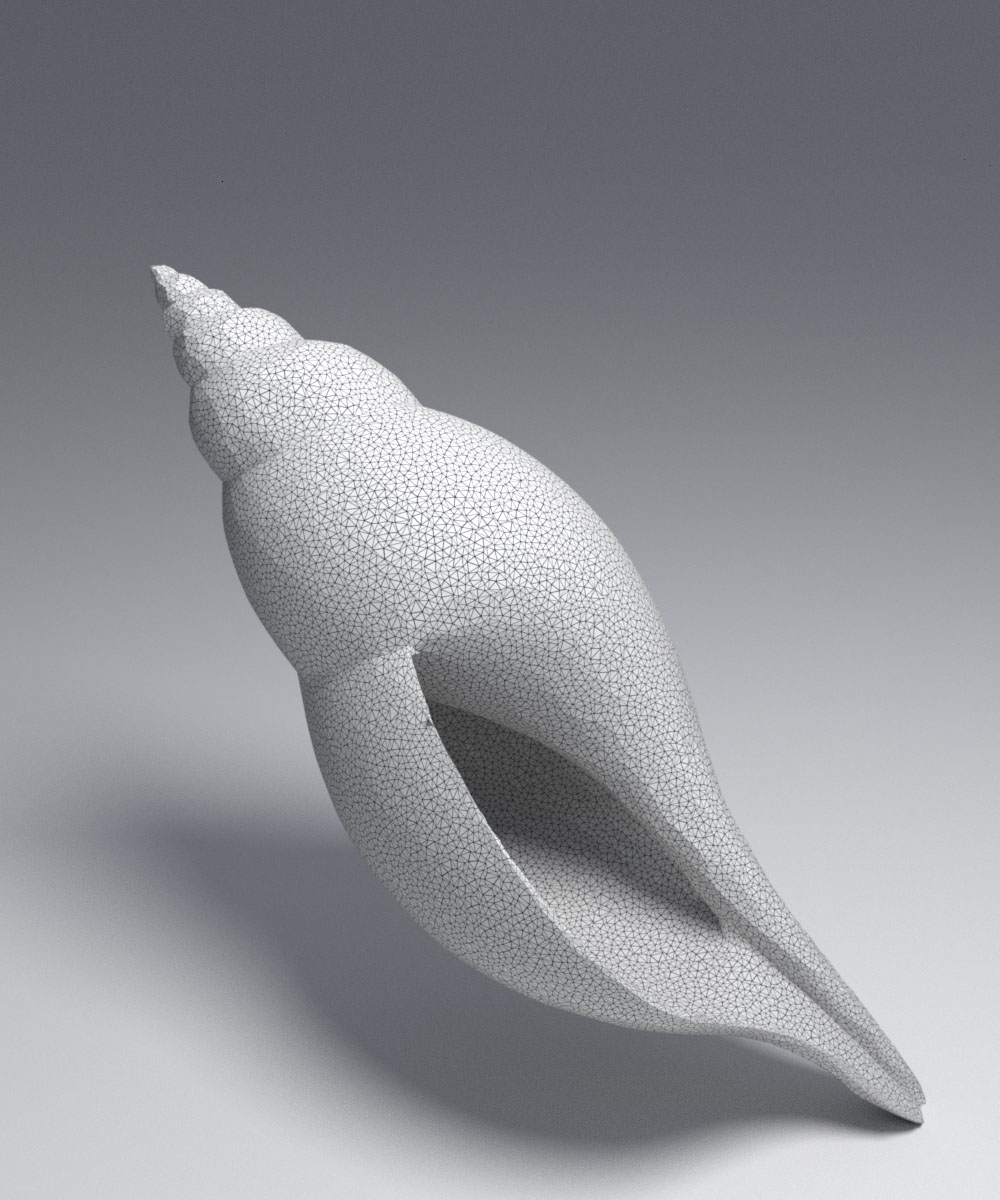}&
\includegraphics[width=0.122\linewidth]{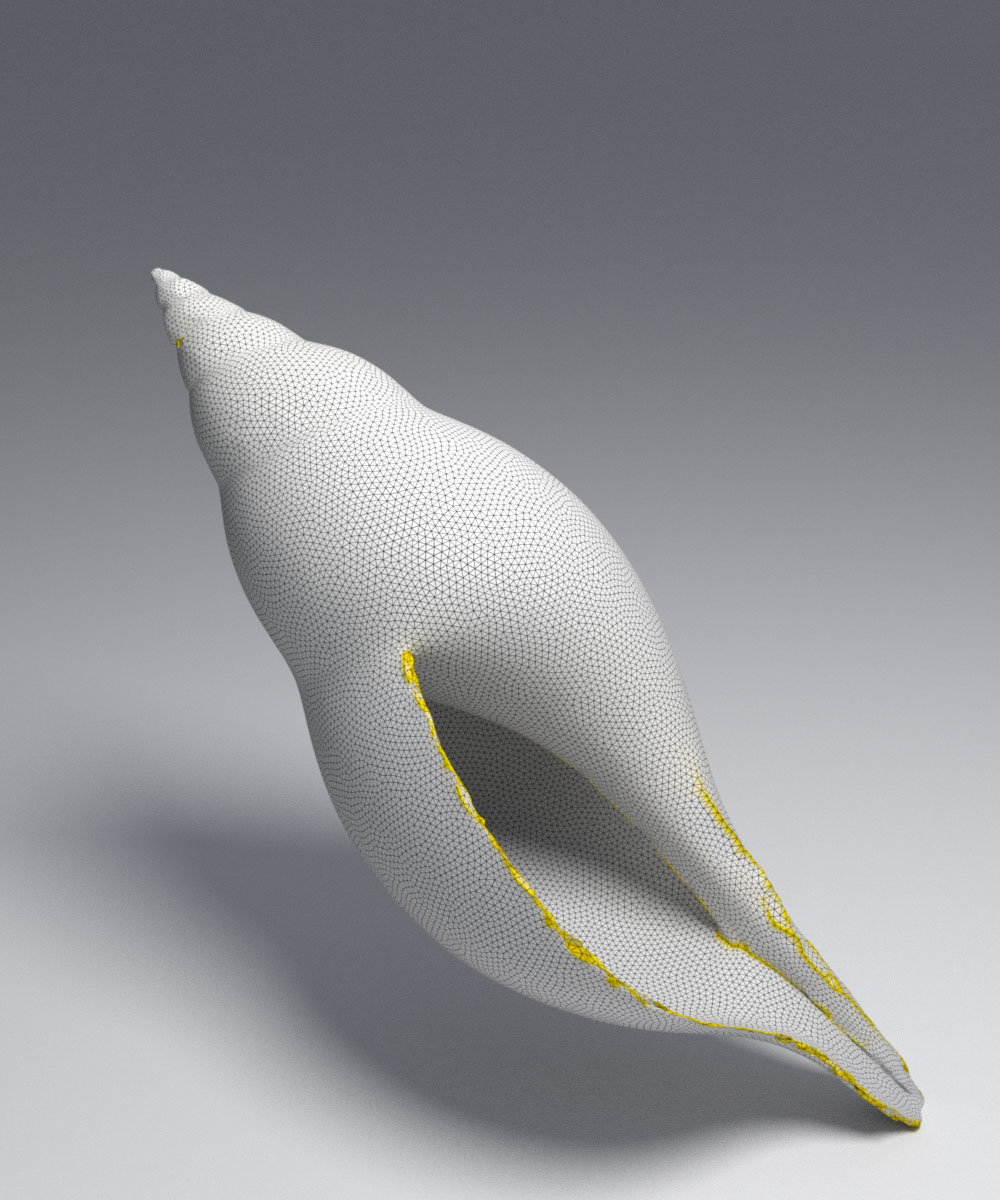} \\
&
\includegraphics[width=0.122\linewidth]{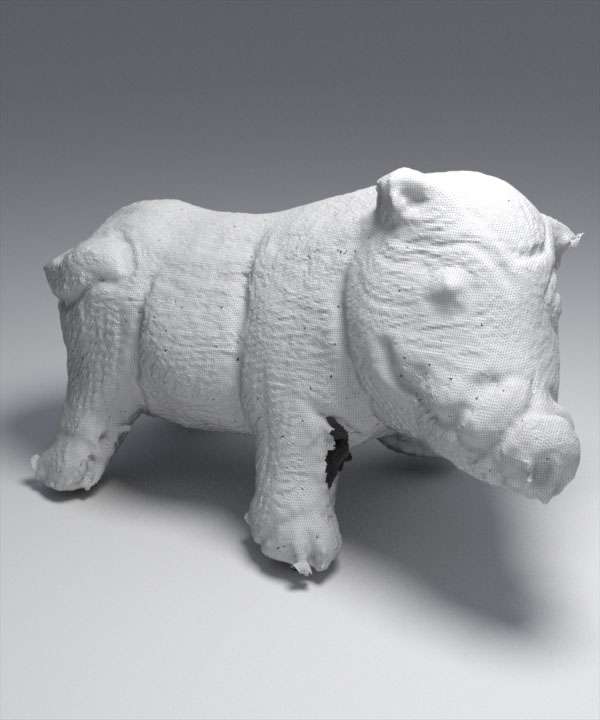}&
\includegraphics[width=0.122\linewidth]{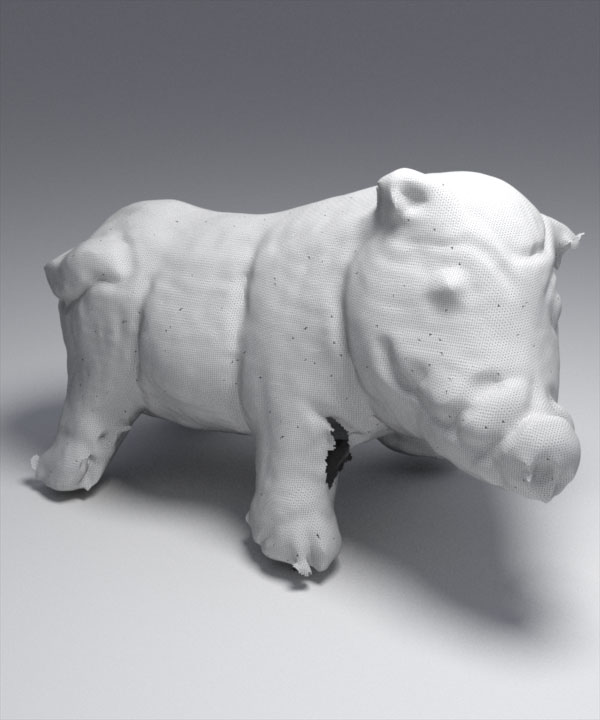}&
\includegraphics[width=0.122\linewidth]{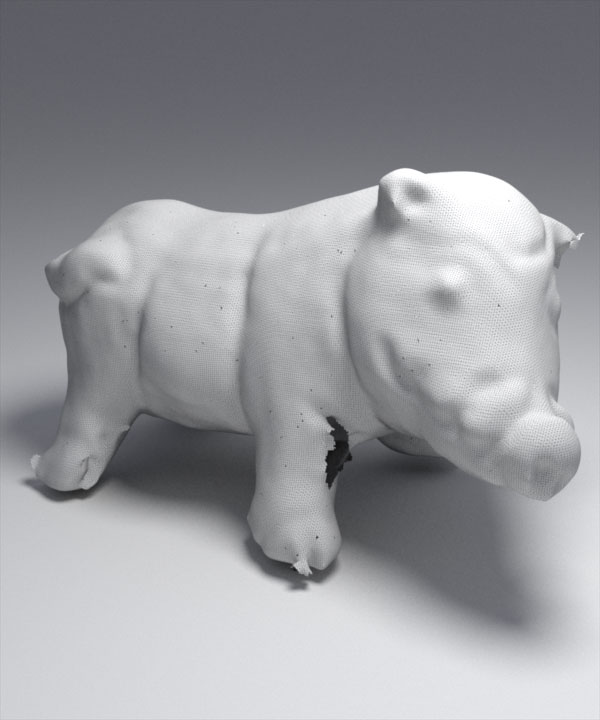}&
\includegraphics[width=0.122\linewidth]{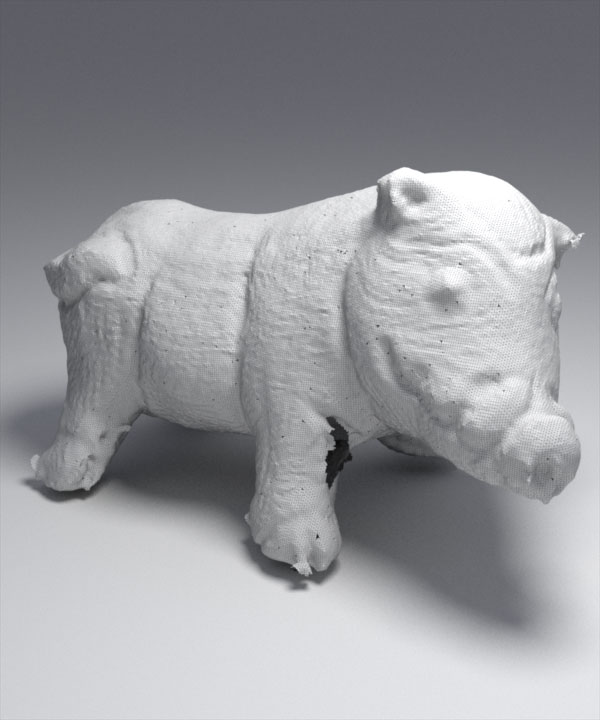}&
\includegraphics[width=0.122\linewidth]{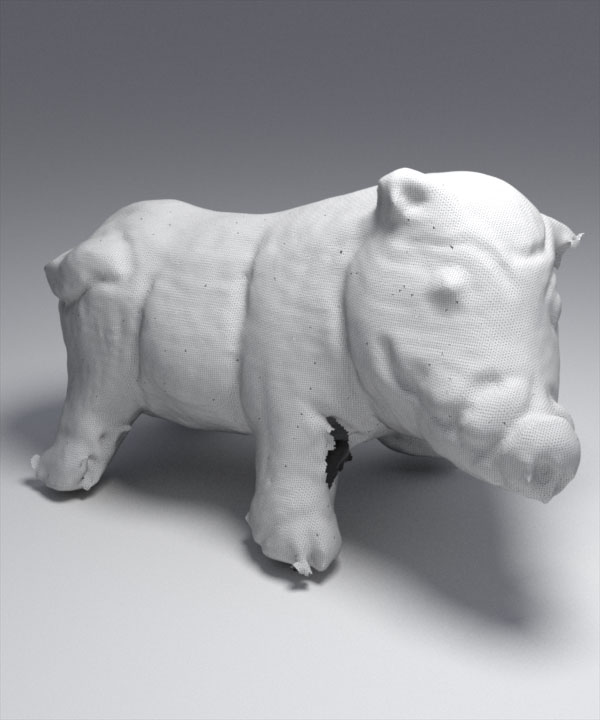}&
\includegraphics[width=0.122\linewidth]{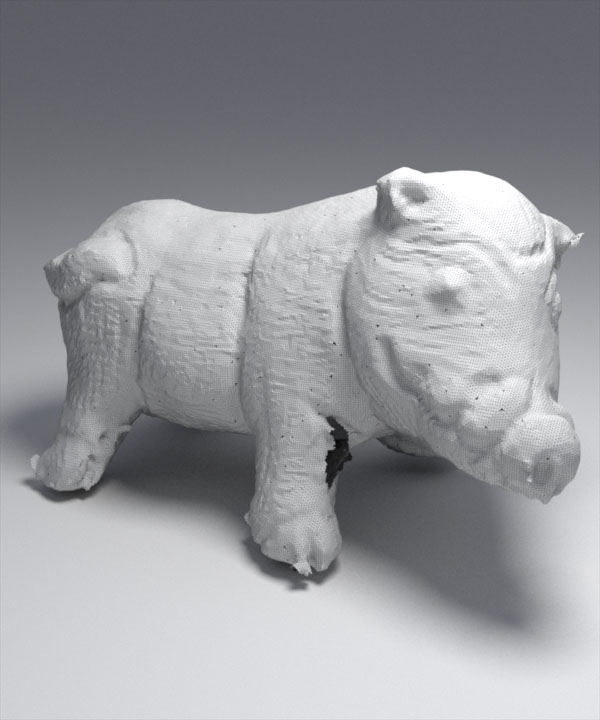}&
\includegraphics[width=0.122\linewidth]{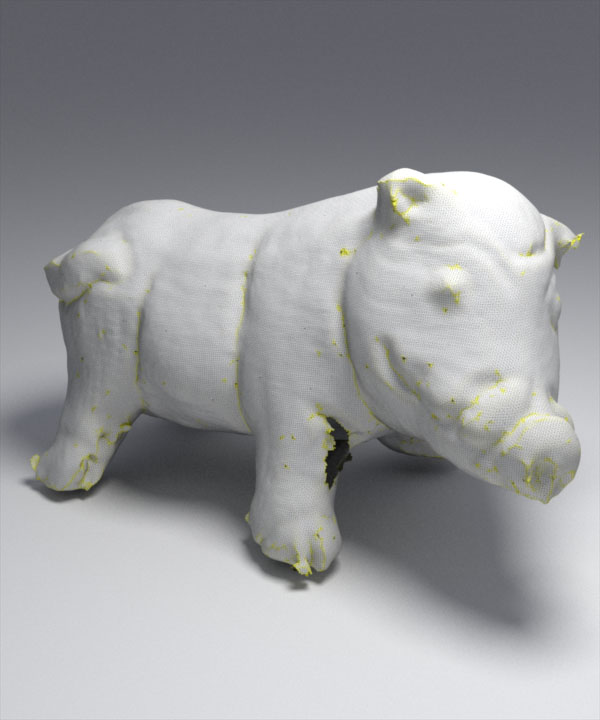} \\
 &
\includegraphics[width=0.122\linewidth]{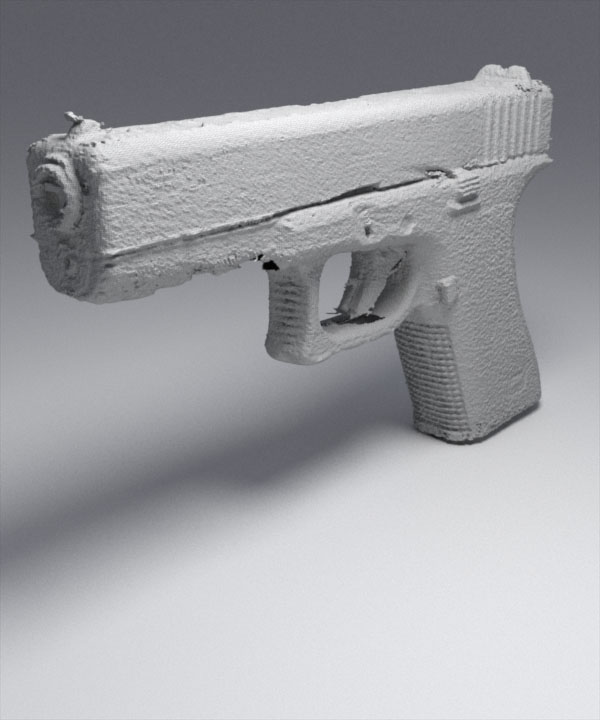}&
\includegraphics[width=0.122\linewidth]{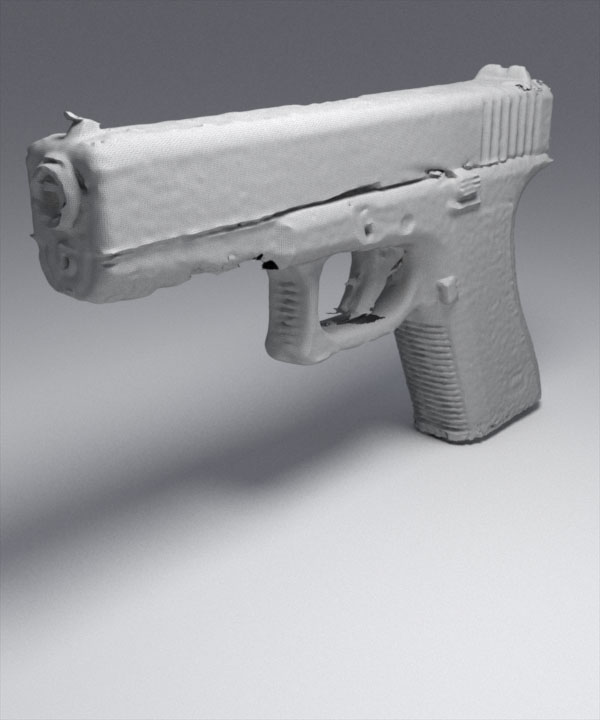}&
\includegraphics[width=0.122\linewidth]{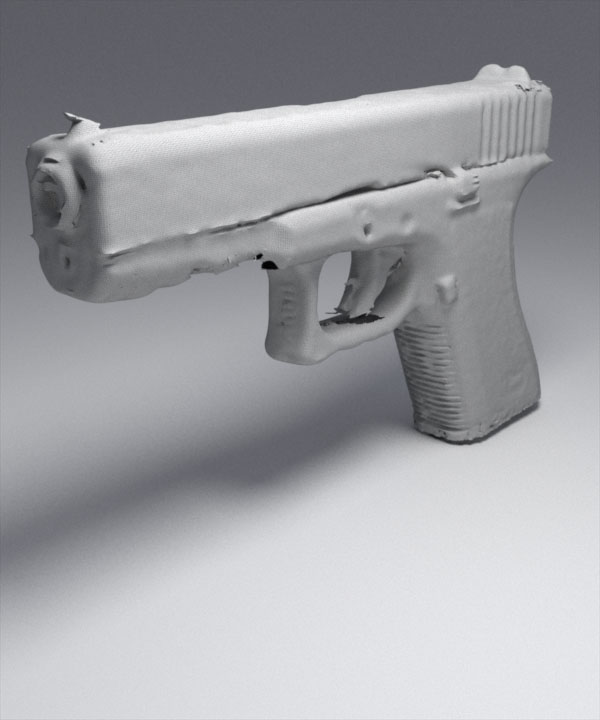}&
\includegraphics[width=0.122\linewidth]{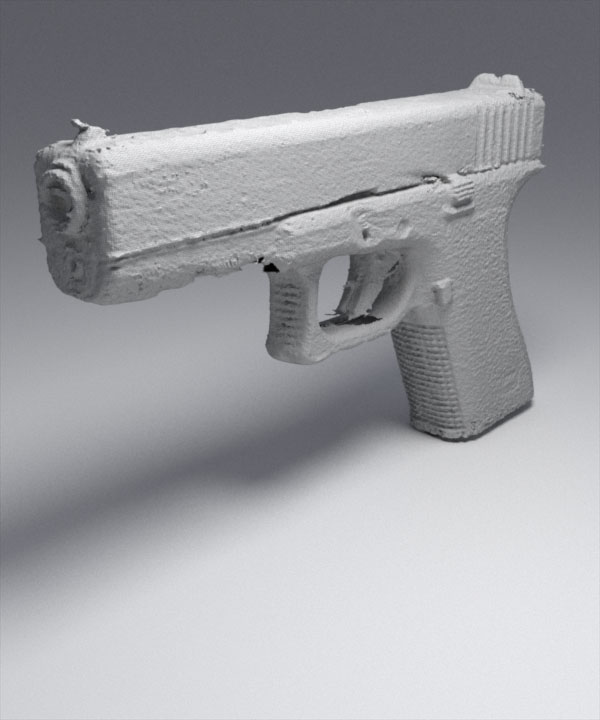}&
\includegraphics[width=0.122\linewidth]{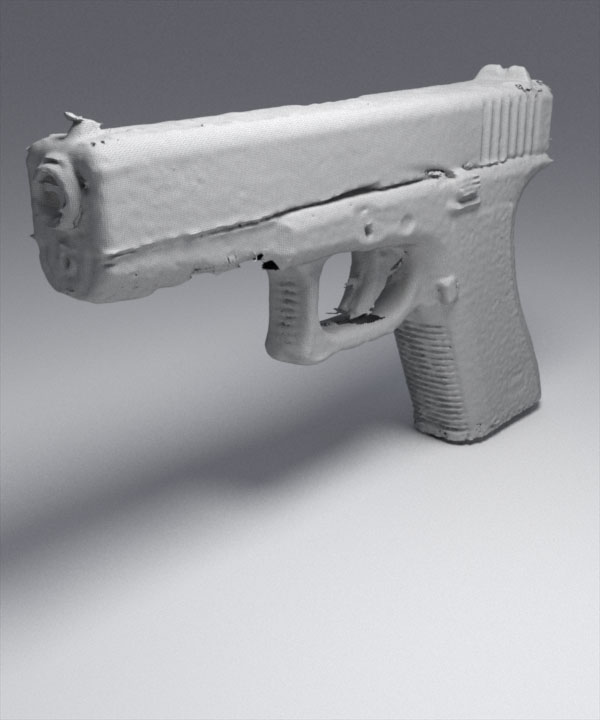}&
\includegraphics[width=0.122\linewidth]{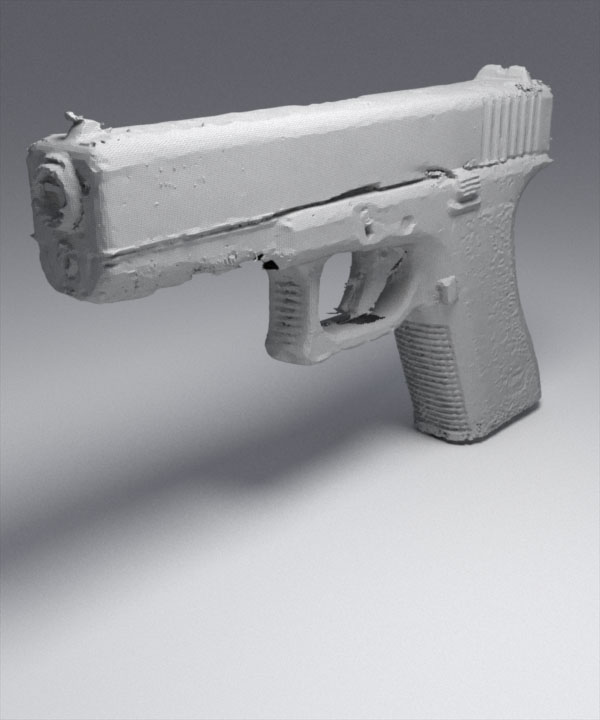}&
\includegraphics[width=0.122\linewidth]{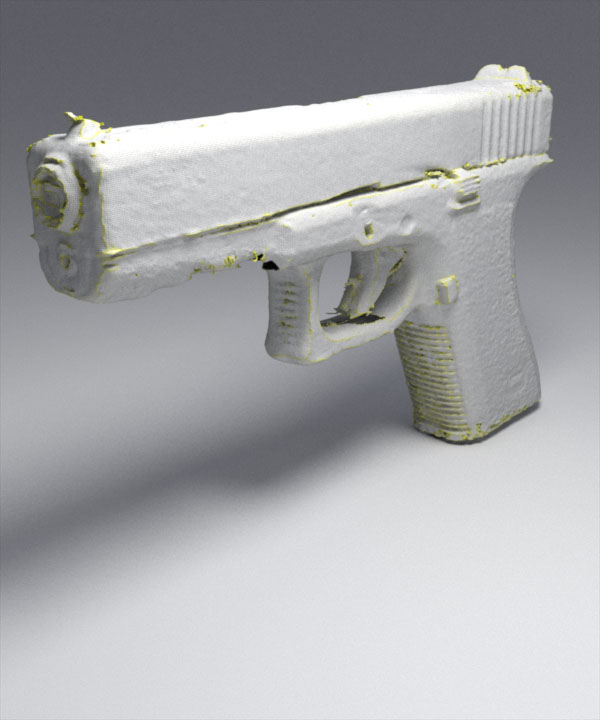} \\

Ground truth & Noisy & \footnotesize{\protect\cite{ZhangFilter2015}} & 
\footnotesize{\protect\cite{sun2007fast}}& \footnotesize{\protect\cite{fleishman2003bilateral}}& 
\footnotesize{\protect\cite{zheng2011bilateral}} &  \footnotesize{\protect\cite{he2013mesh}} & Our denoising
\end{tabular}
\caption{Denoising results and comparisons. For our method, features ($v$) are shown in yellow. The last two rows illustrate denoising on LiDAR acquired geometries where no ground-truth information is available.}
\label{fig:denoising}
\end{figure*}

\begin{figure*}[tbh]
\centering
\begin{tabular}{@{}c@{}c@{}c@{}c@{}c@{}c@{}}
\includegraphics[width=0.16\linewidth]{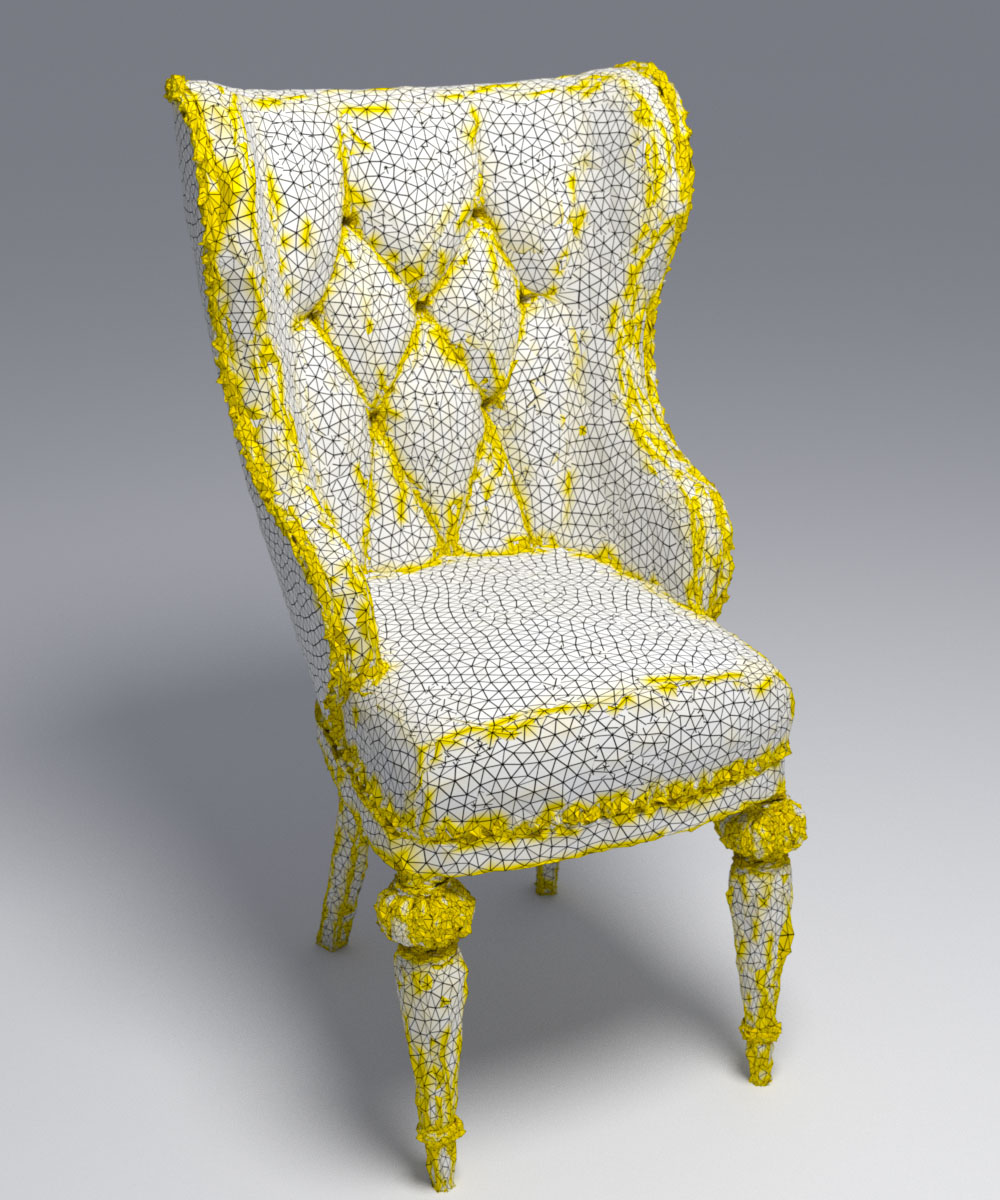}&
\includegraphics[width=0.16\linewidth]{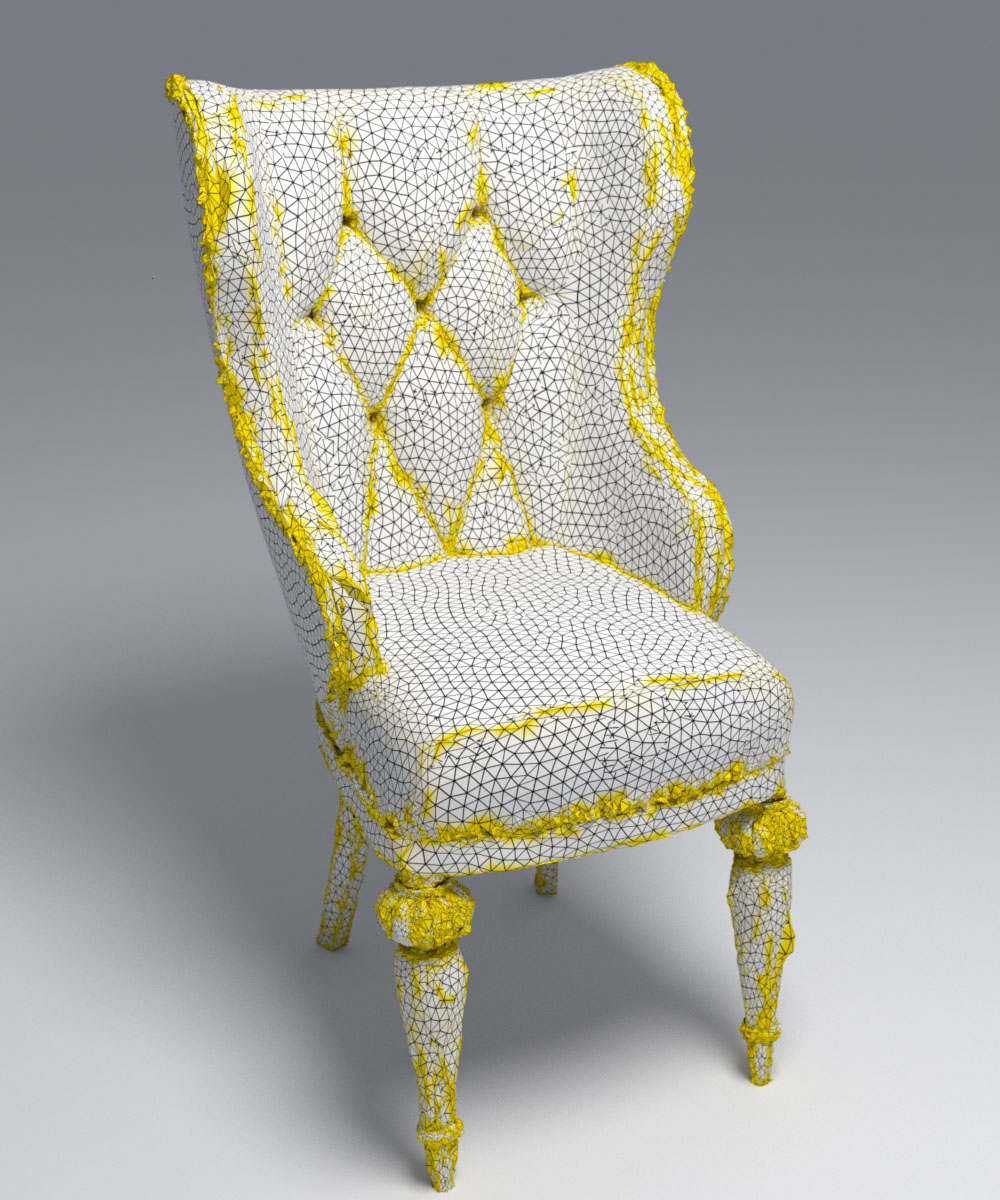} &
\includegraphics[width=0.16\linewidth]{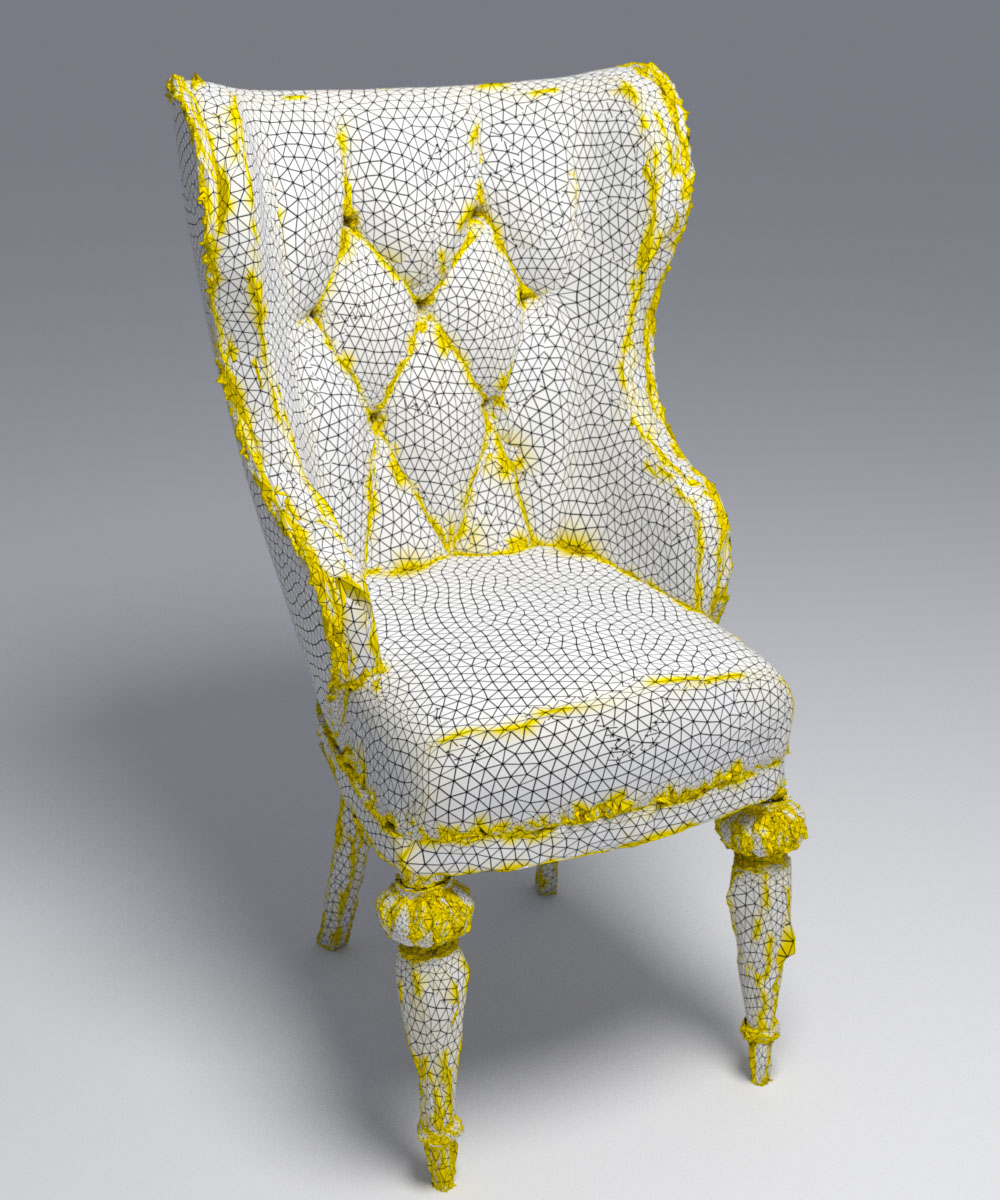}&
\includegraphics[width=0.16\linewidth]{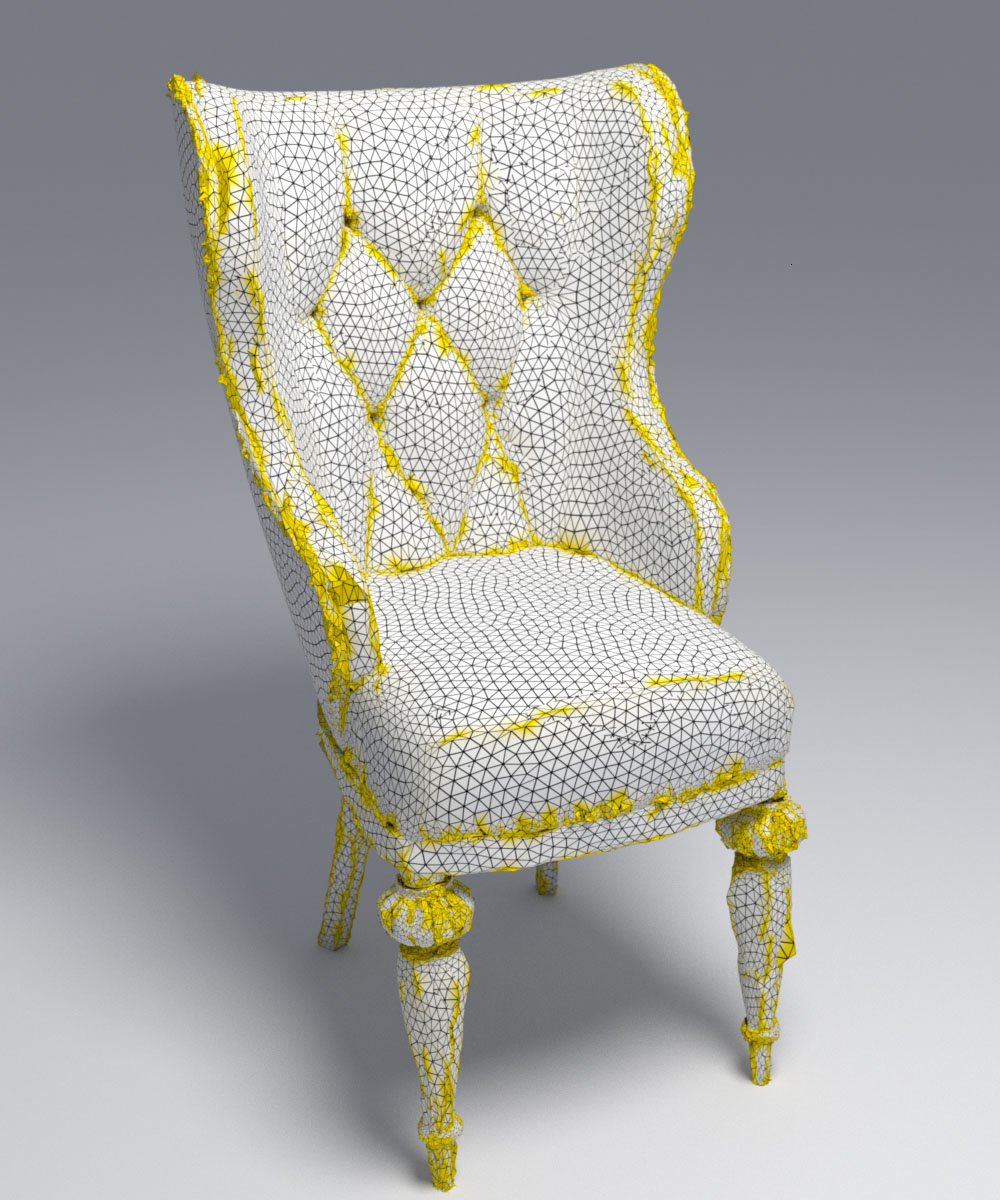} &
\includegraphics[width=0.16\linewidth]{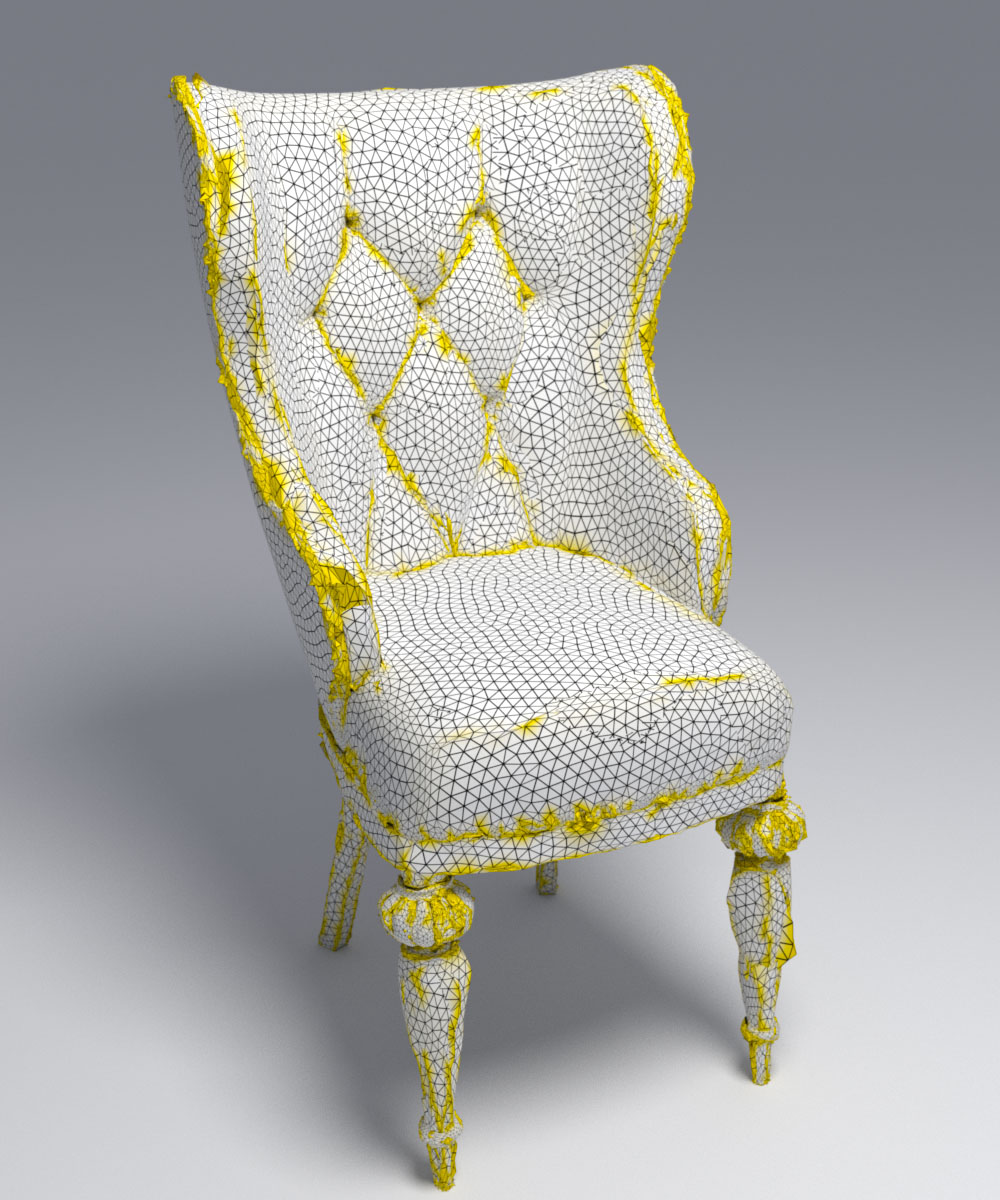}&
\includegraphics[width=0.16\linewidth]{images/chair_denoised_6iter.jpg} \\
(a) 1 iter. & (b) 2 iter. & (c) 3 iter. & (d) 4 iter.& (e) 5 iter.& (f) 6 iter.
\end{tabular}
\caption{Influence of the number of MS / vertex projection iterations.}
\label{fig:iterations}
\end{figure*}

\begin{figure*}[tbh]
\centering
\begin{tabular}{@{}c@{}c@{}c@{}c@{}c@{}}
\includegraphics[width=0.25\linewidth]{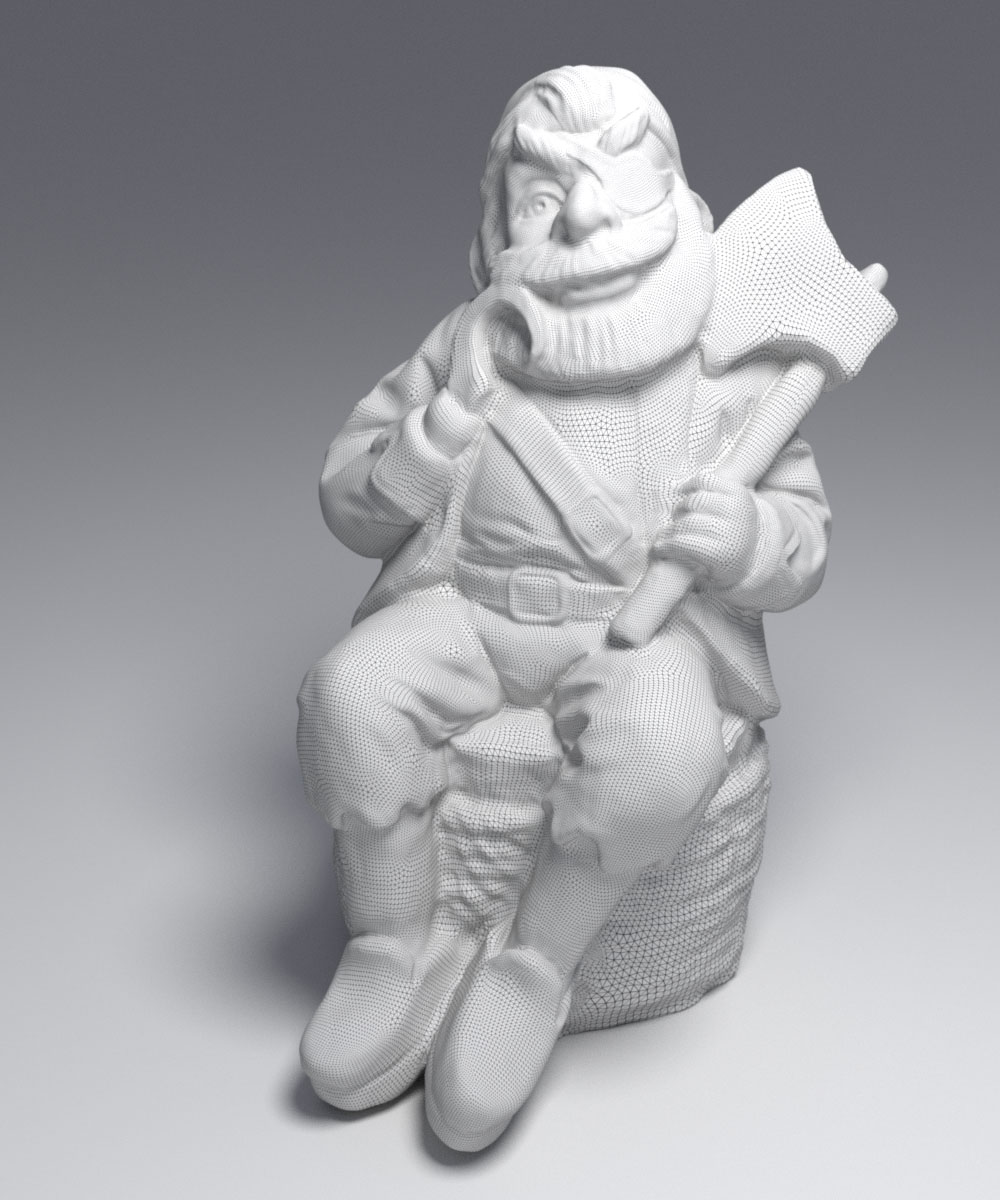} &
\includegraphics[width=0.25\linewidth]{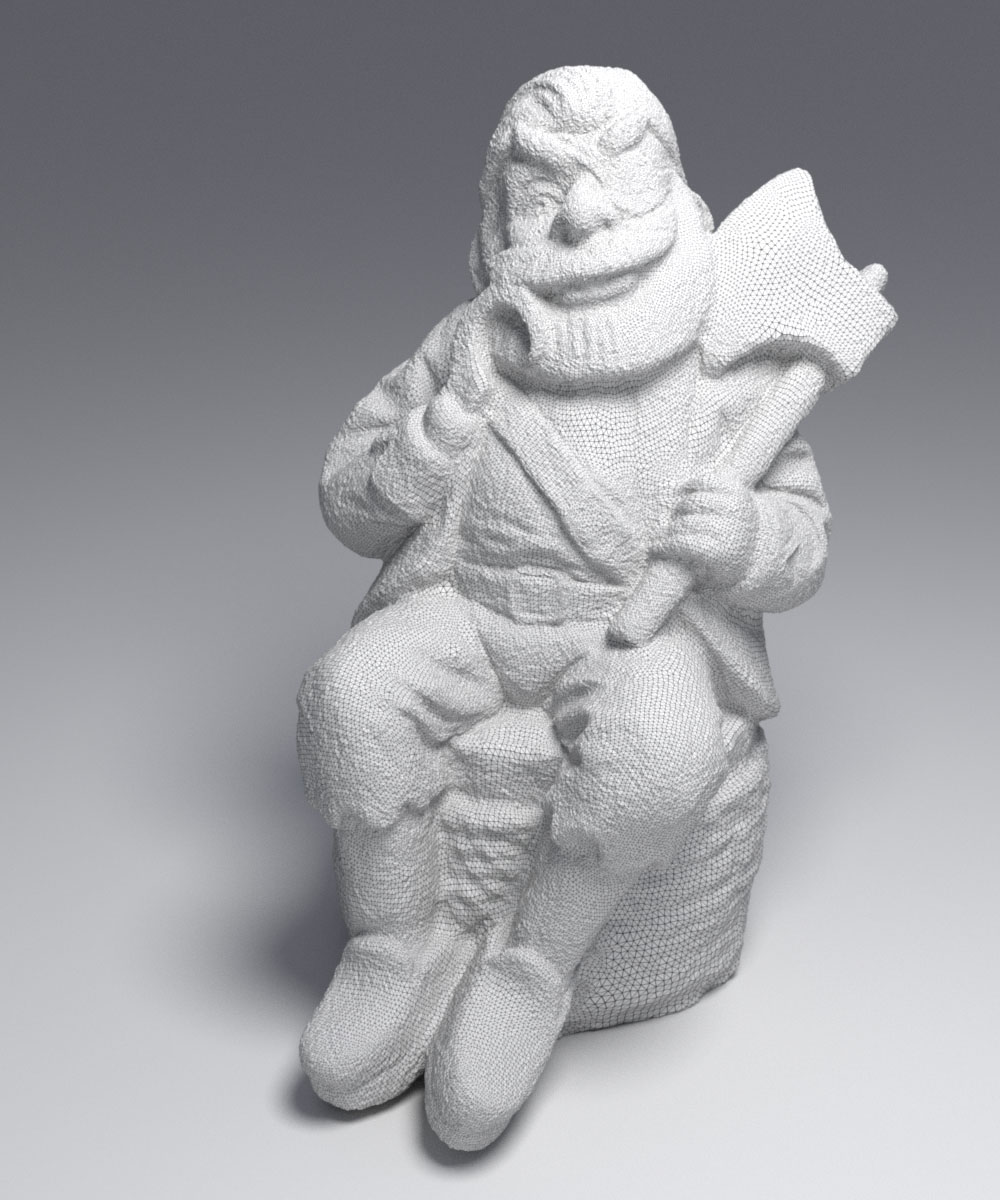}&
\includegraphics[width=0.25\linewidth]{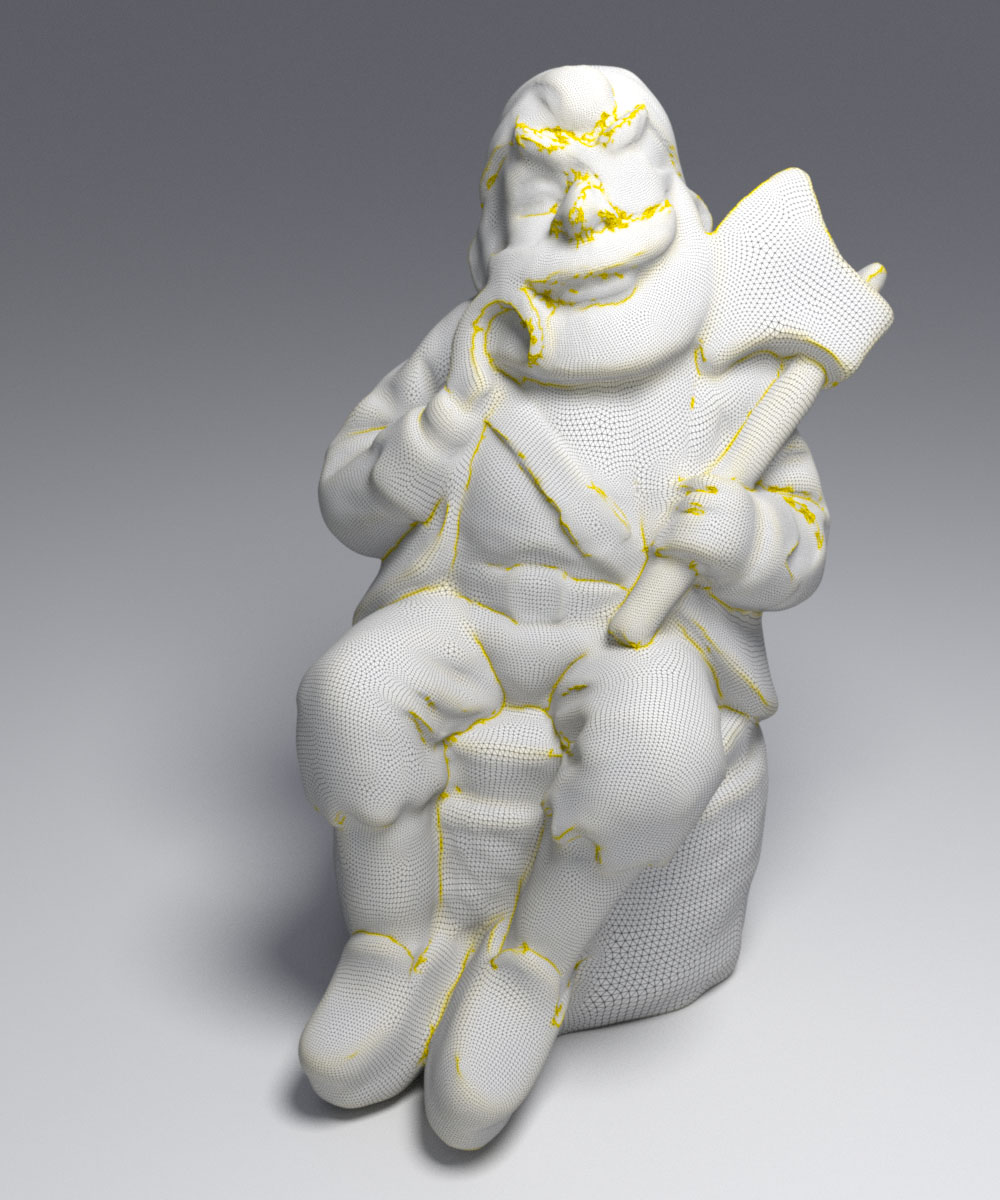} &
\includegraphics[width=0.25\linewidth]{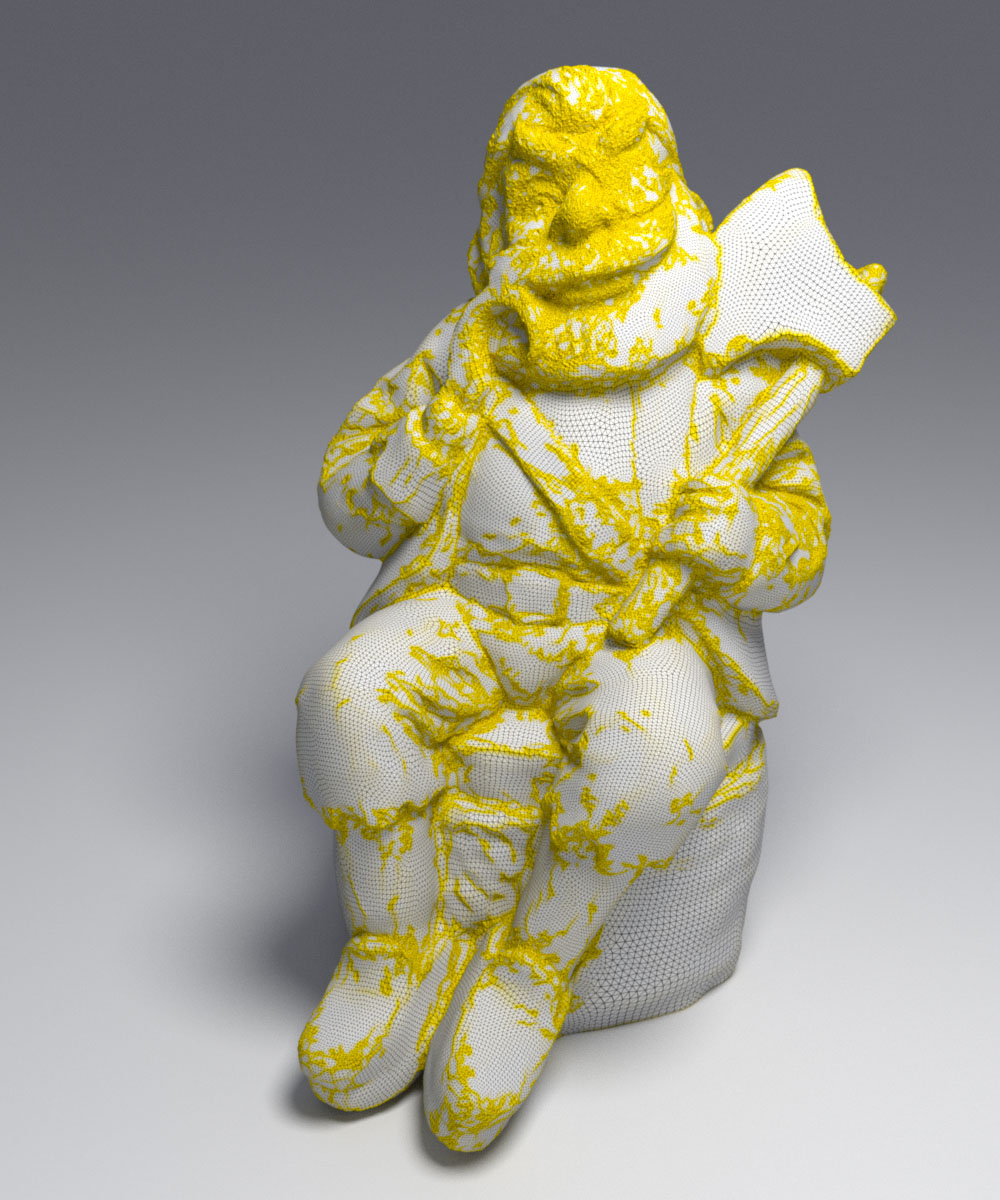}\\
\end{tabular}
\caption{Influence of $\lambda$ on denoising results. First row: Original and noisy models. Second row: denoising with $\lambda=0.05$ (left) and $\lambda=0.01$ (right). Decreasing $\lambda$ produces longer features, and more discontinuities.}
\label{fig:lambda}
\end{figure*}

Given an input noisy mesh, our method smoothes out noise while
preserving features. For this application, we alternate between
solving for the MS functional to regularize the normal vector field,
and conforming the mesh vertices to the estimated normals. In our
examples, our method achieves visually satisfactory results within 2
to 6 such iterations (see Fig.~\ref{fig:iterations}).  \dav{In this
  experiment, we have considered both synthetic shapes perturbed using
  a Gaussian noise in order to have a ground-truth, and more realistic
  noisy LiDAR shapes.}  We illustrate these results in
Fig.~\ref{fig:denoising} and show the effect of varying $\lambda$ in
Fig.~\ref{fig:lambda}. In practice, decreasing $\lambda$ produces
longer discontinuities, and thus less smooth denoising
results. Numerical and visual comparisons with state-of-the-art
denoising techniques can be found respectively in
\jaco{Table}~\ref{tab:metrics} and in \jaco{Fig.}~\ref{fig:denoising}.

\begin{figure}[ht]
\centering
\begin{tabular}{@{}c@{}c@{}c@{}c@{}}
\rotatebox{90}{\hspace{0.9cm}\cite{tong2016variational}} & 
\includegraphics[width=0.27\linewidth]{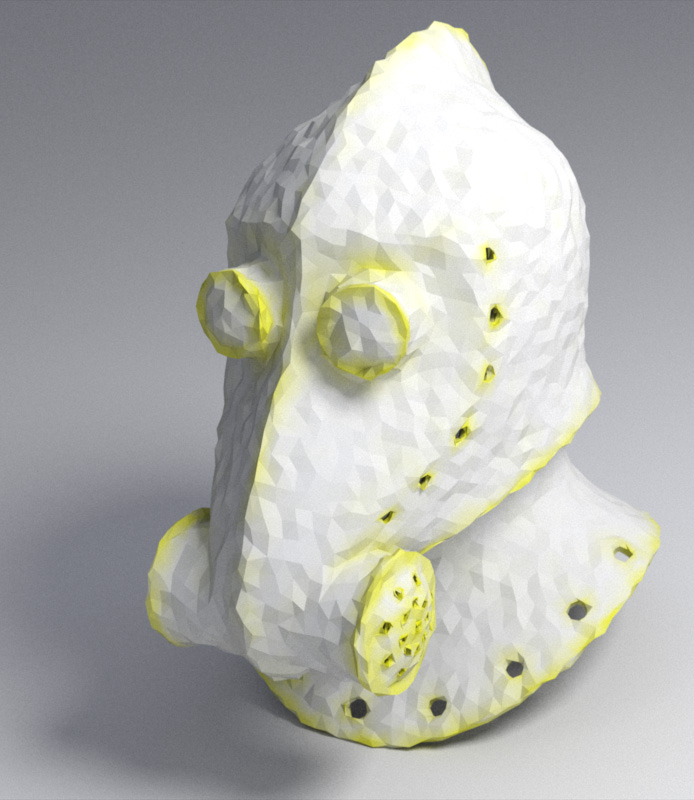}&
\includegraphics[width=0.27\linewidth]{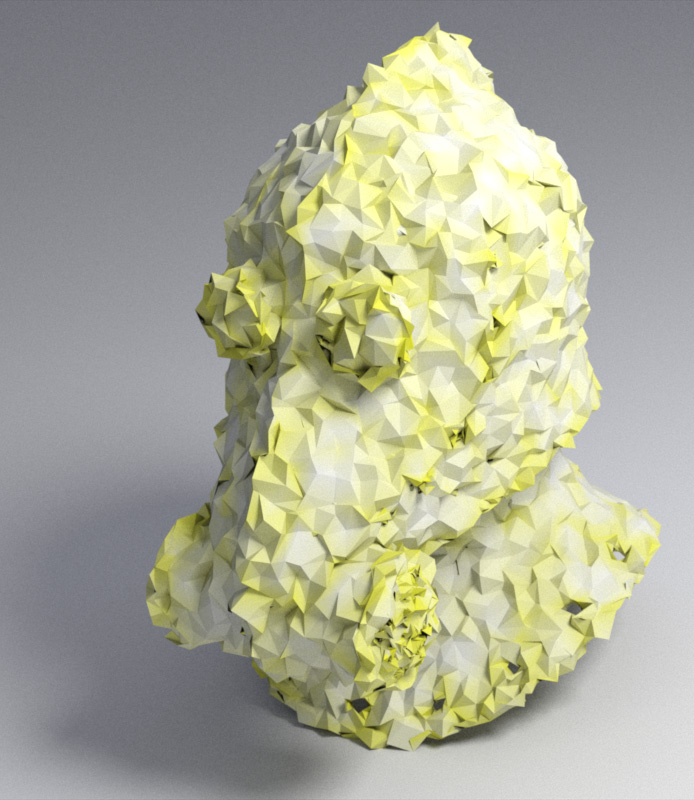}&
\includegraphics[width=0.27\linewidth]{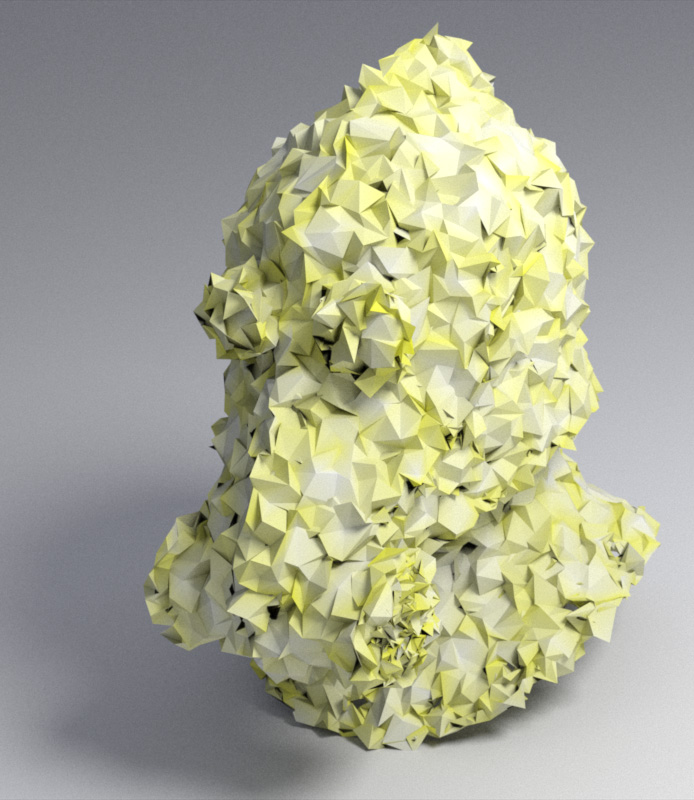}\\
\rotatebox{90}{\hspace{0.5cm}Our method} &
\includegraphics[width=0.27\linewidth]{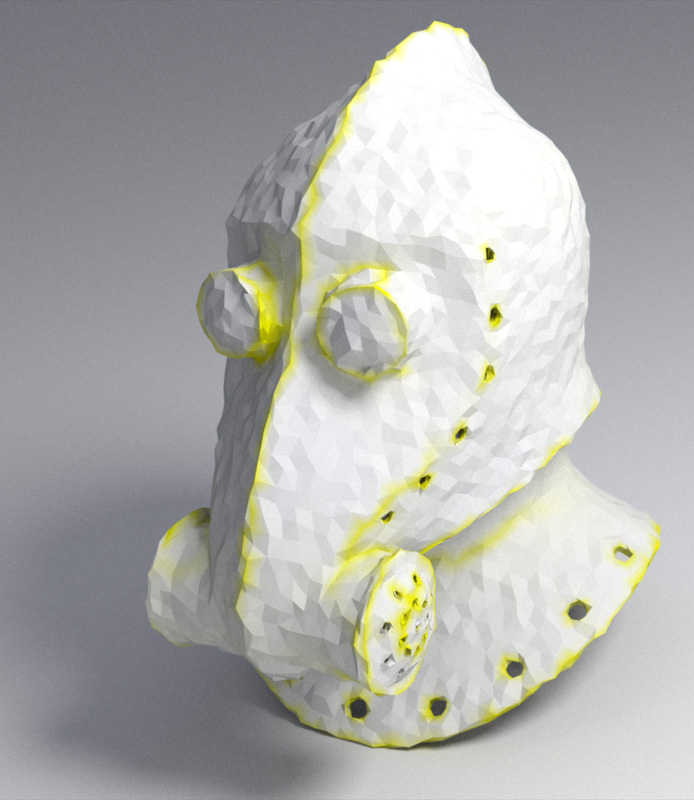} &
\includegraphics[width=0.27\linewidth]{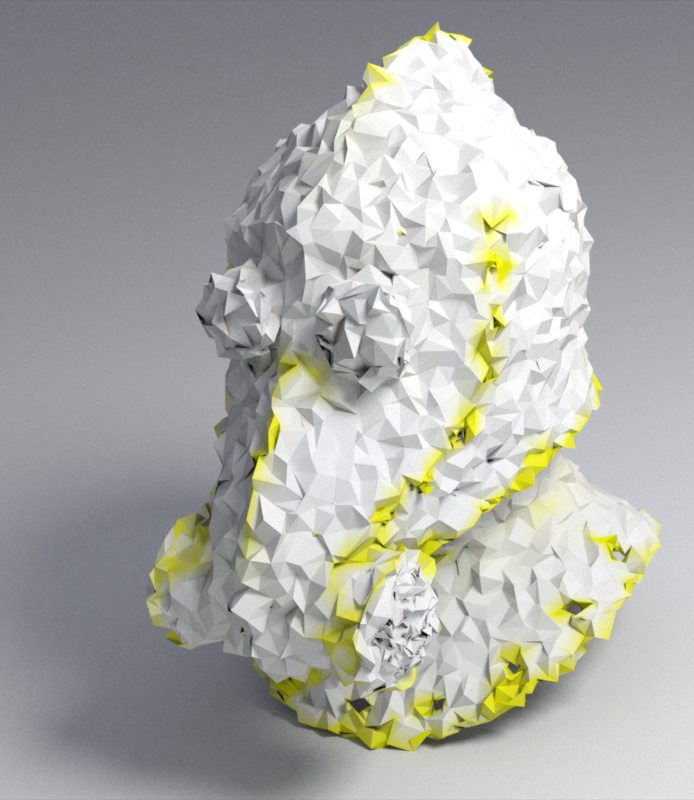} &
\includegraphics[width=0.27\linewidth]{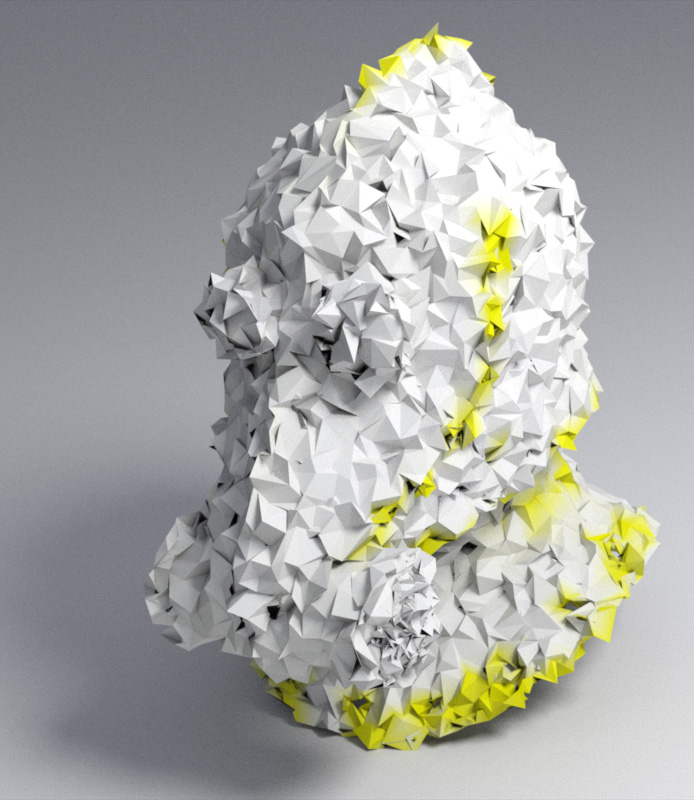} 
\end{tabular}
\caption{We compare our method (bottom) to that of Tong and Tai~\cite{tong2016variational} on meshes with weak, moderate and extreme noise levels. Our approach remains robust even to extreme noise levels. }
\label{fig:noisecompare}
\end{figure}

We compare our features to that of Tong and Tai~\shortcite{tong2016variational} who solve 
a similar AT functional via a Finite Elements discretization, using a noisy non-uniform mesh (Fig.~\ref{fig:femcomparison}) and meshes with varying amount of noise (Fig.~\ref{fig:noisecompare}). 
We show that our features are more robust to noise, making our method suitable for denoising applications. 
{The MS functional reconstructs a piecewise smooth vector field
  in terms of direction and orientation.  Starting from a smooth
  surface with some few flipped normal vectors, the MS functional will
  reorient the vectors as incorrect orientations introduce spurious
  discontinuities in the normal field. Combined with the projection
  step (Sect. \ref{sec:projection}) where the fairness term unfolds
  the surface, we do not observe flipped triangle in our experiments.}

\newcommand{\cone}{\cellcolor{blue!60}}  
\newcommand{\ctwo}{\cellcolor{blue!50}}  
\newcommand{\cthree}{\cellcolor{blue!40}}  
\newcommand{\cfour}{\cellcolor{blue!30}}  
\newcommand{\cfive}{\cellcolor{blue!20}}  
\newcommand{\csix}{\cellcolor{blue!10}}

\begin{table*}\centering
\footnotesize{
\begin{tabular}{|p{0.7cm}|c|c|c|c|c|c||c|c|c|c|c|c|}
\hline
   & \multicolumn{6}{c||}{Hausdorff Distance ($\times 100$)} & \multicolumn{6}{c|}{Perceptual Metric} \\
  \hline
 & \footnotesize{Z15} 
  & \footnotesize{S07} 
  & \footnotesize{F03} 
  & \footnotesize{Z11} 
  & \footnotesize{H13} 
  & {\footnotesize{Our}}
  & \footnotesize{Z15} 
  & \footnotesize{S07} 
  & \footnotesize{F03} 
  & \footnotesize{Z11} 
  & \footnotesize{H13} 
  & {\footnotesize{Our}}\\
  Chair & \ctwo\textbf{0.0677} & \cfive0.1221  &\cthree 0.0762 &\cone 0.0636
  & \cfour
                                                                  0.0786
  &\csix 0.1695

  & \csix 0.587  & \ctwo 0.539  & \cfive 0.577 & \cone\textbf{0.531}  &\cfour 0.557 &
                                                                   \cthree 0.546 \\
  Skull &  \cthree 0.1082& \csix 0.1773 &\cfour 0.1180 & \cone\textbf{0.1017} &
                                                                    \ctwo 0.1078
  & \cfive 0.1689
  & \cfive 0.635 &\cfour 0.626 &\csix 0.650 &\cthree 0.623 &\ctwo 0.603 &\cone\textbf{0.592}\\
  
  Helmet & \cthree 0.1748& \csix 0.2595 & \ctwo 0.1691 &
                                                   \cone\textbf{0.1408}  & \cfour 0.1874
  & \cfive 0.1963
  & \cfive 0.577 & \cthree 0.524 & \csix 0.596 & \ctwo 0.510  & \cfour 0.558 &\cone\textbf{0.507}\\
  Pirate & \ctwo 0.0853 & \csix 0.1170   &  \cfive 0.0898   & \cfour 0.0892  &
                                                                \cone\textbf{0.0838}
  &\cthree 0.0860
  & \cone\textbf{0.277}  & \csix 0.404  &\ctwo 0.304 & \cfour 0.326  &
                                                                  \cfive 0.387
  &\cthree 0.318 \\
  Shell & \ctwo 0.0386& \cfour 0.0500  &  \cfive 0.0641   &
                                                            \cone\textbf{0.0331}
  &\cthree 0.0464 &\csix 0.1324
  &\cone \textbf{0.346}& \cfive 0.499  &\csix 0.698 & \cthree 0.410  & \cthree 0.410 & \ctwo 0.349 \\
\hline
\end{tabular}
}
\caption{We provide numerical evaluation with state-of-the-art
  methods, with respect to the Hausdorff distance (RMS, normalized by
  the bounding box diameter), and the perceptual metric of Lavou\'{e}
  et al.~\protect\cite{lavoue2006perceptually} \dav{(ranks are given
    as cell colors from dark blue --rank 1-- to light-blue --rank 6--)}. Our method 
  performs better according to perceptual distances. This is due to a
  slight shrinkage of our denoising results which drastically impacts
  purely geometric metrics. Parameters were not adjusted to perform
  well under these metrics. Columns are named following: Z15~\protect\cite{ZhangFilter2015}, S07~\protect\cite{sun2007fast}, F03~\protect\cite{fleishman2003bilateral}, Z11~\protect\cite{zheng2011bilateral}, and H13~\protect\cite{he2013mesh}, and AT is our method.}
\label{tab:metrics}
\end{table*}

\subsection{Segmentation}

We segment meshes into piecewise smooth parts. Note that, contrary to semantic mesh segmentation, our segmentation is purely geometric and bears no semantic meaning.
This can find applications in mesh parameterization where seams should remain as little visible as possible~\cite{ray2010invisible}, or as a preprocessing for further feature extraction (see the mesh segmentation survey of Attene et al.~\cite{attene2006mesh}). However, this makes numerical evaluation difficult since no ground truth segmentation is available, in contrast to semantic segmentation benchmarks~\cite{chen2009benchmark}.

\begin{figure*}[ht]
\centering
\begin{tabular}{@{}c@{}c@{}c@{}c@{}}
\includegraphics[width=0.24\linewidth]{images/helmet_segments.jpg}&
\includegraphics[width=0.24\linewidth]{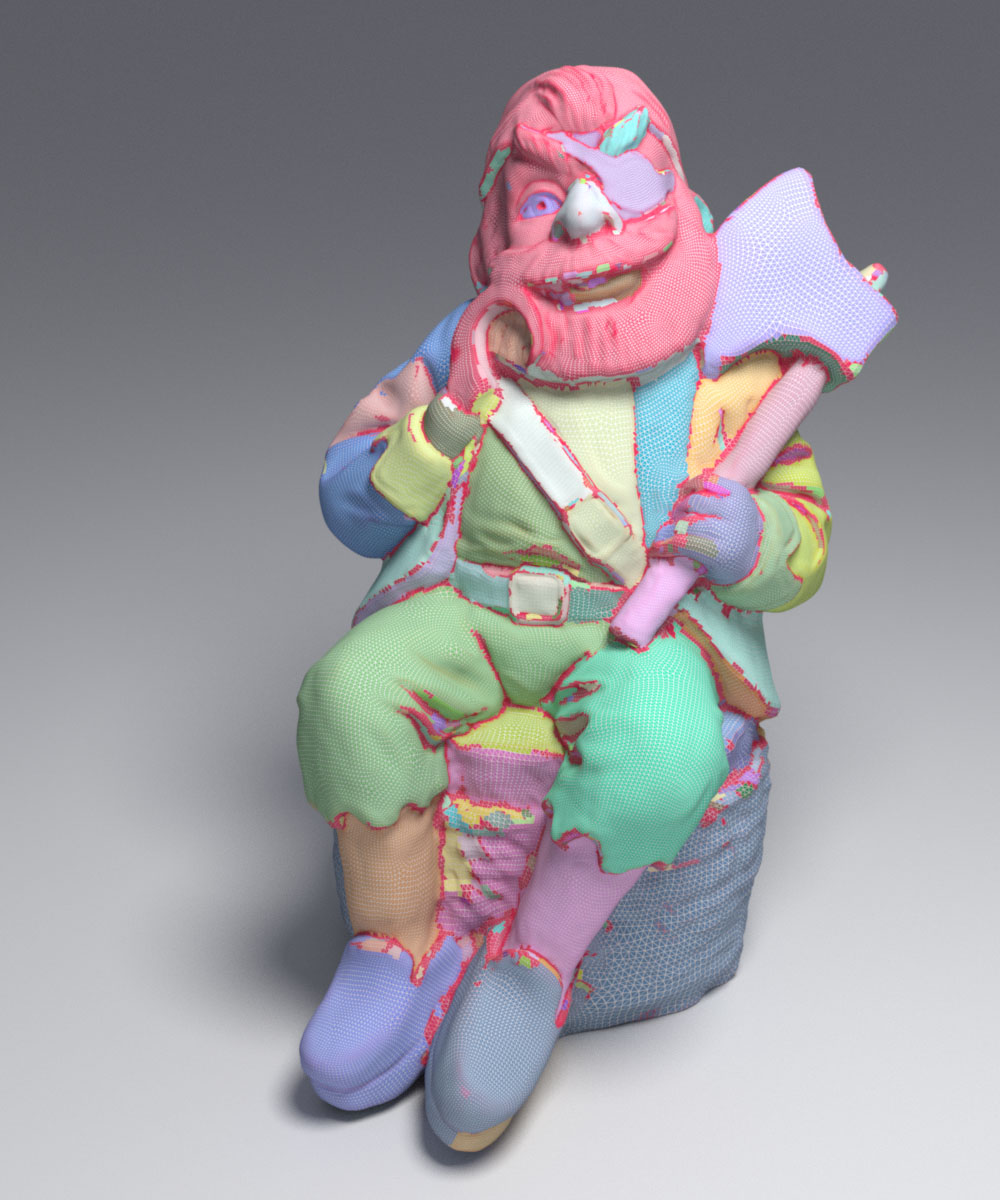}&
\includegraphics[width=0.24\linewidth]{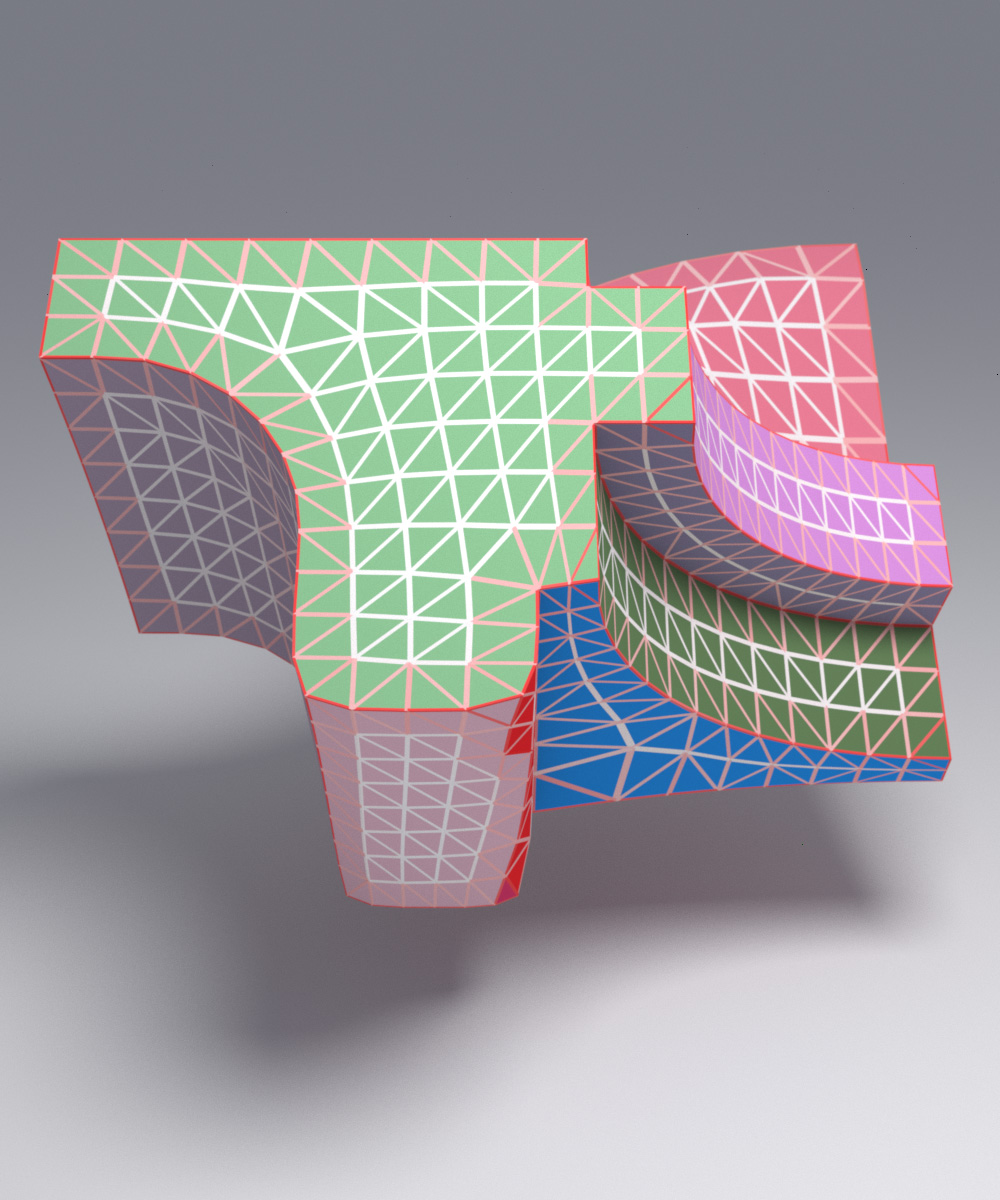}&
\includegraphics[width=0.24\linewidth]{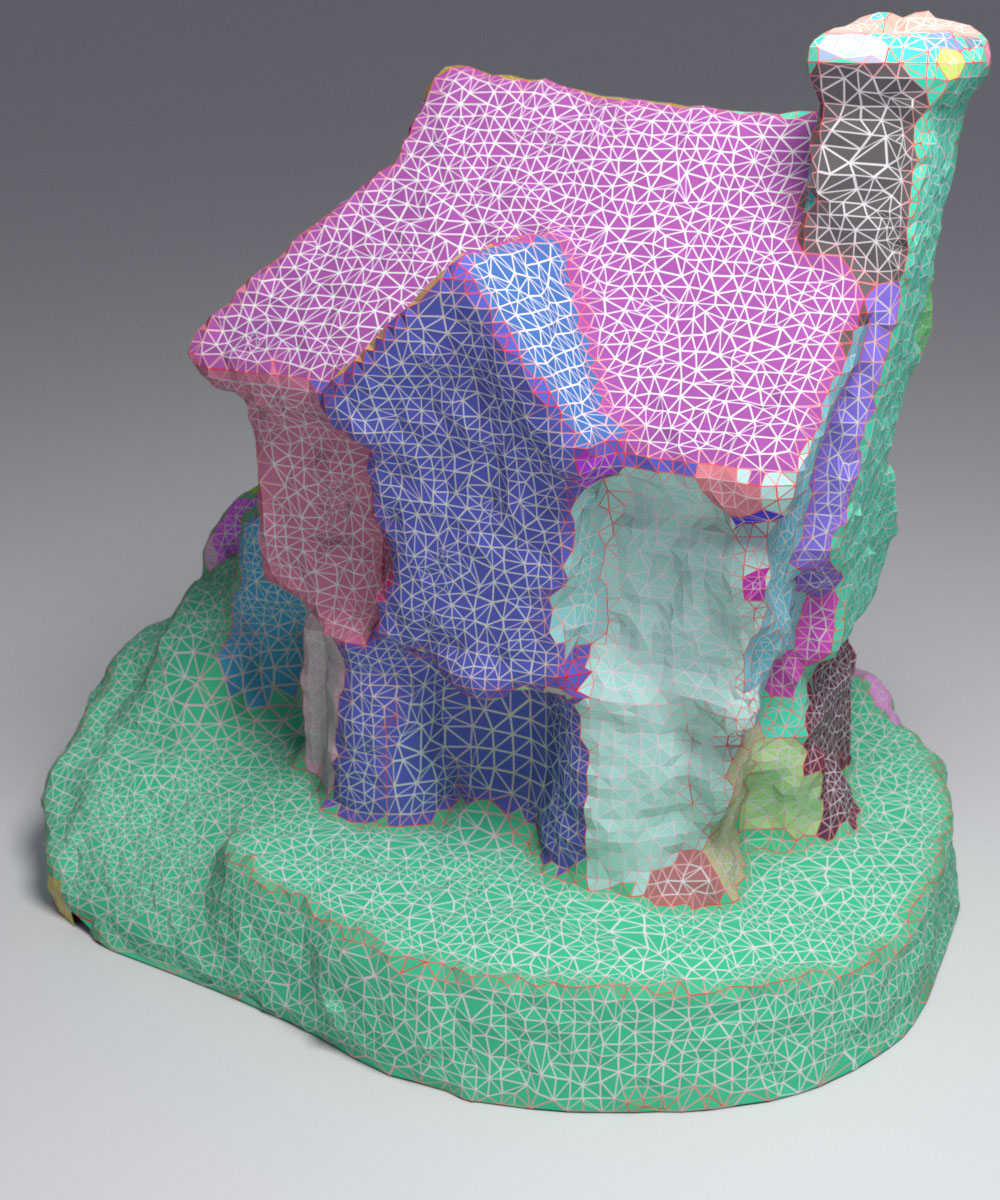} \\
\includegraphics[width=0.24\linewidth]{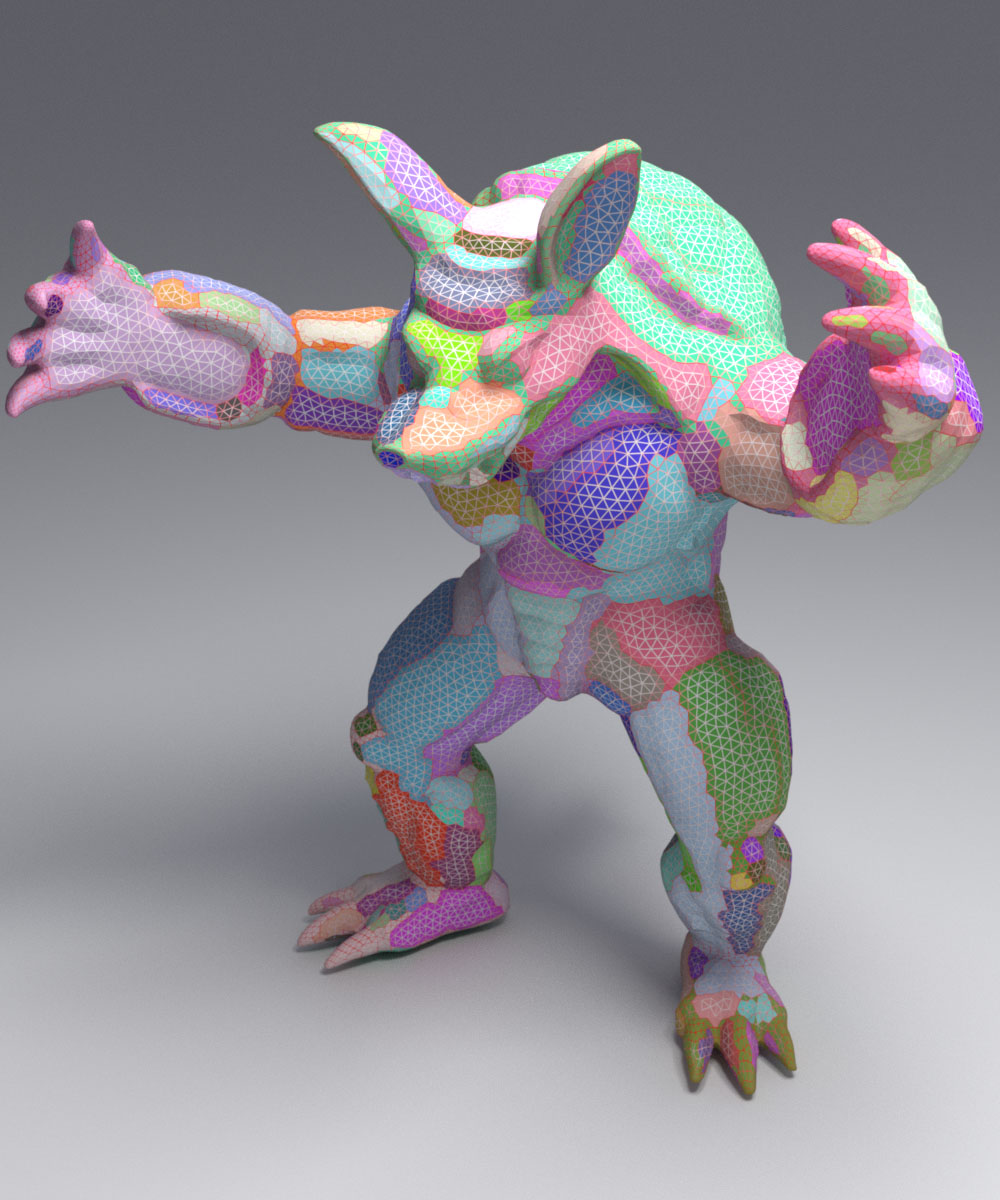} &
\includegraphics[width=0.24\linewidth]{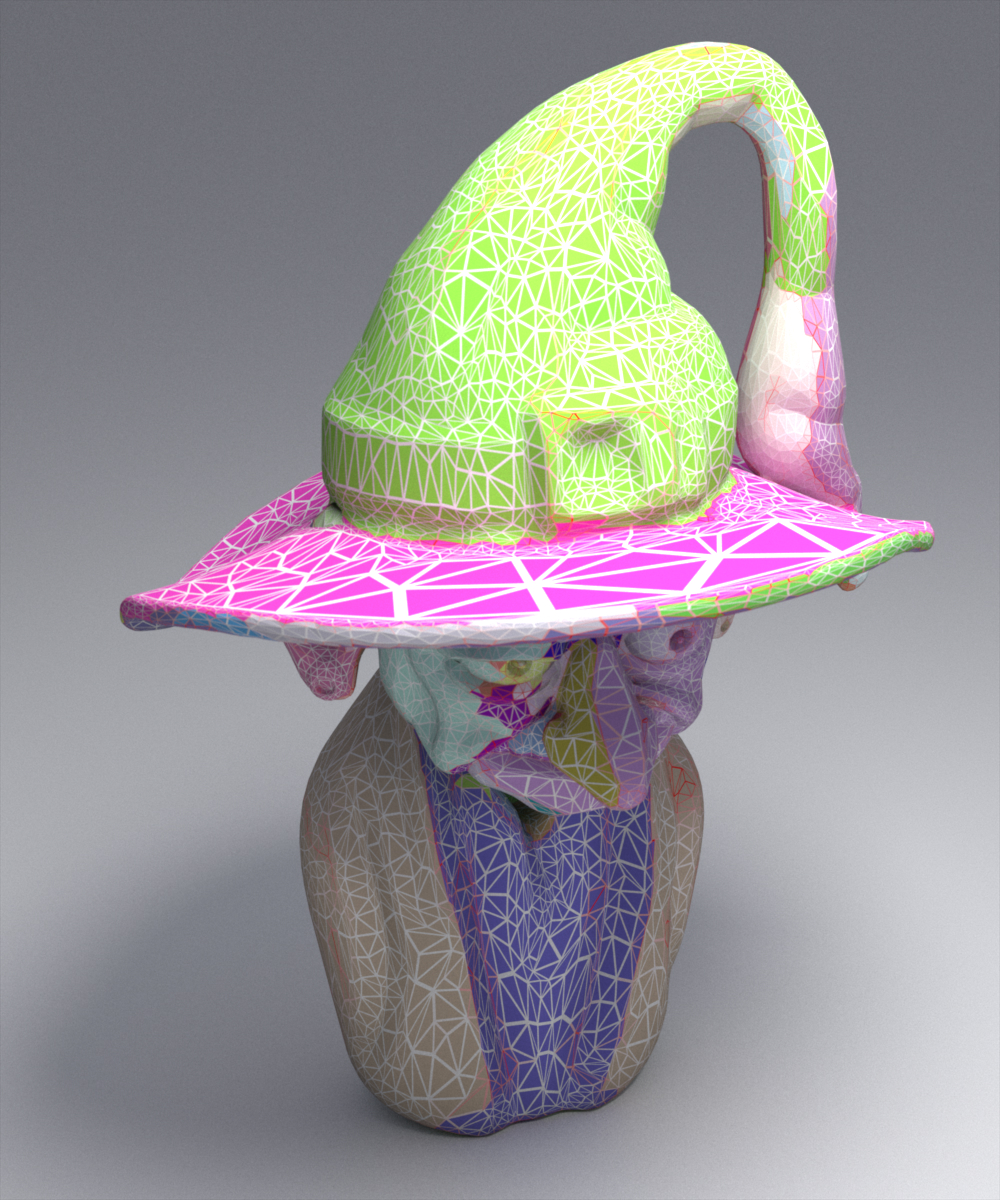} &
\includegraphics[width=0.24\linewidth]{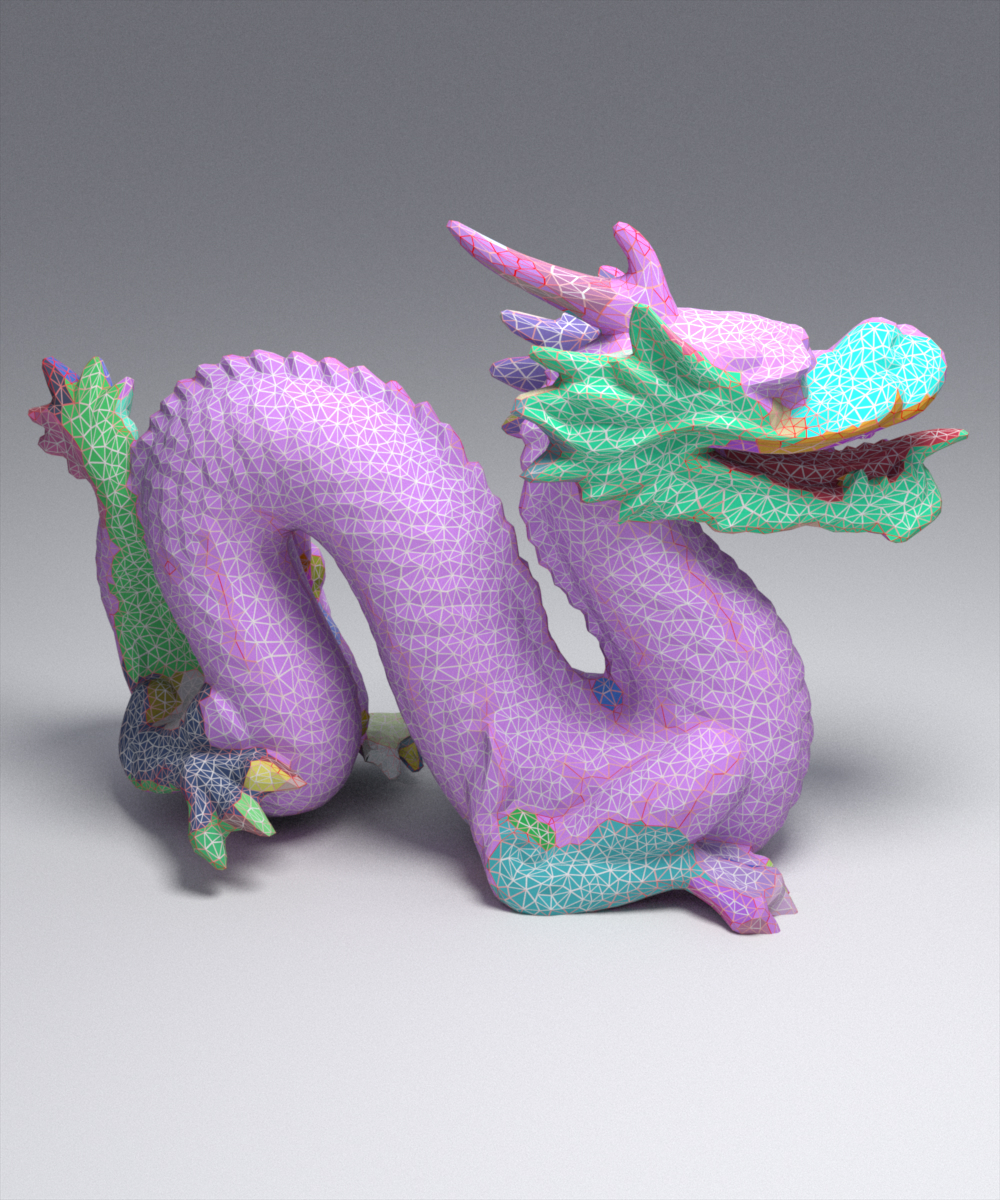} &
\includegraphics[width=0.24\linewidth]{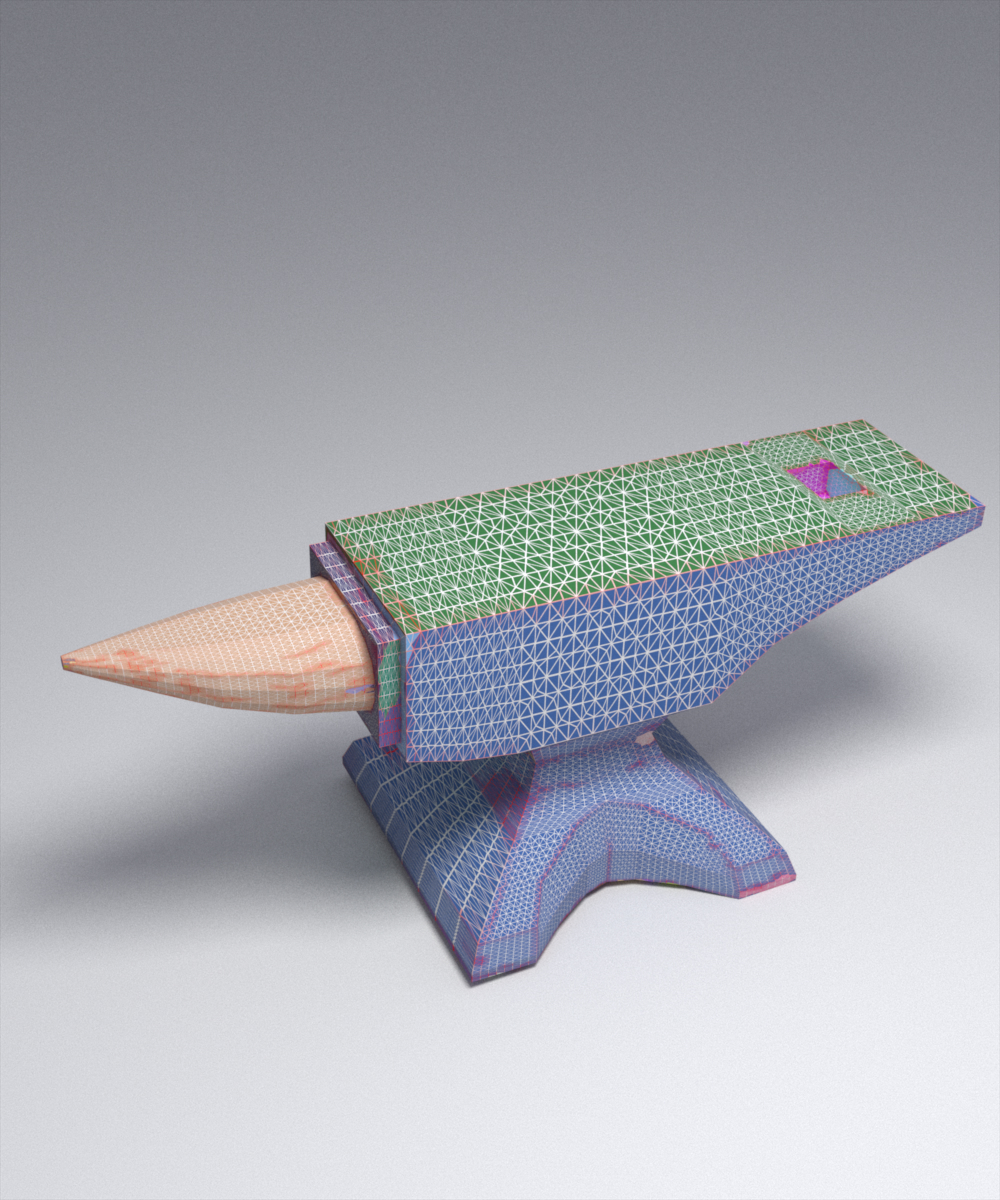} 
\end{tabular}
\caption{Segmentation results. Edge splitting probabilities are shown in red.}
\label{fig:segmentation}
\end{figure*}

To perform this segmentation, we make use of our MS solver to obtain the feature field $v$ for each vertex of the mesh. We further clamp $v$ since it can occasionally take values outside of the range $[0, 1]$ due to approximations. We then make use of the method of Keuper et al.~\cite{keuper2015efficient} which solves a minimum multicut problem in the triangle adjacency graph using the Kernighan Lin method. This method takes as input the probability of two adjacent triangles to belong to different segment, which amounts to bringing the feature $v$ from vertices to edges. For each edge, we obtain a splitting probability by averaging the value $1 - v$ of both vertices of the edge, while forcing this probability to $0.1\%$ when adjacent triangles have their (regularized) normals further than $5\degree$ apart. Segmentation results can be seen in Fig.~\ref{fig:segmentation}.

\begin{figure*}[tbh]
\centering
\begin{tabular}{@{}c@{}c@{}c@{}c@{}c@{}c@{}c@{}}
\includegraphics[width=0.14\linewidth]{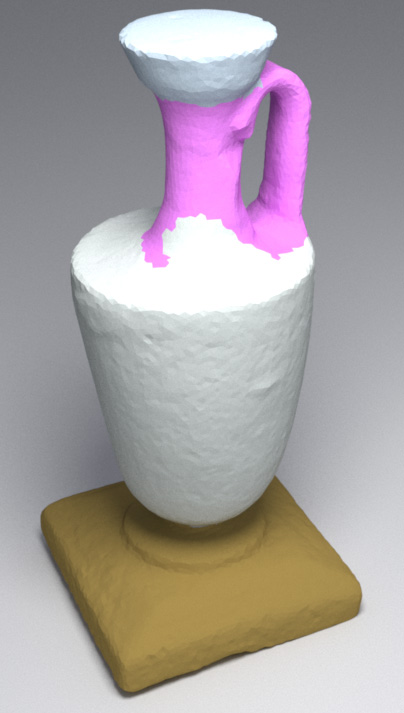}&
\includegraphics[width=0.14\linewidth]{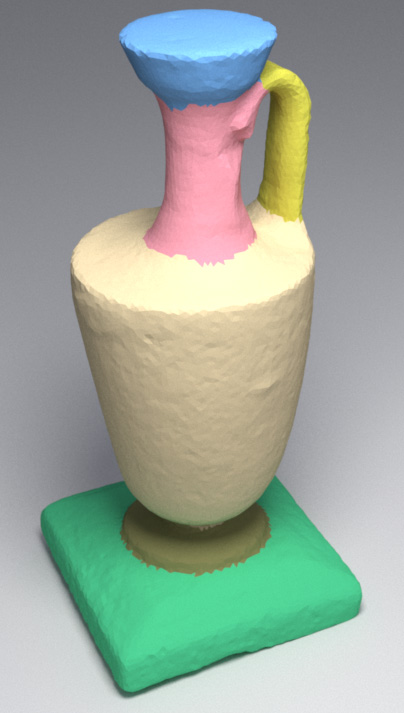}&
\includegraphics[width=0.14\linewidth]{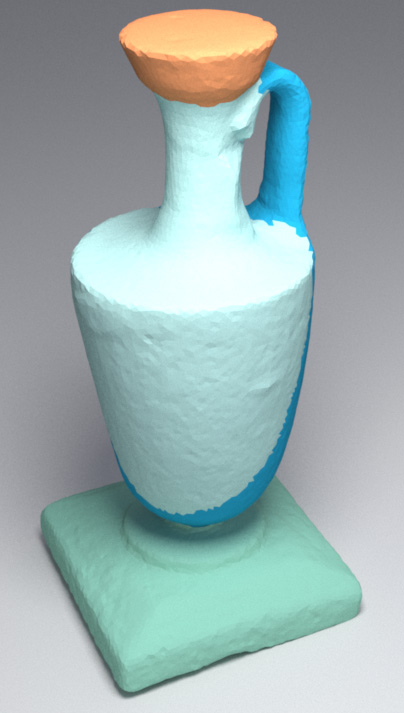}&
\includegraphics[width=0.14\linewidth]{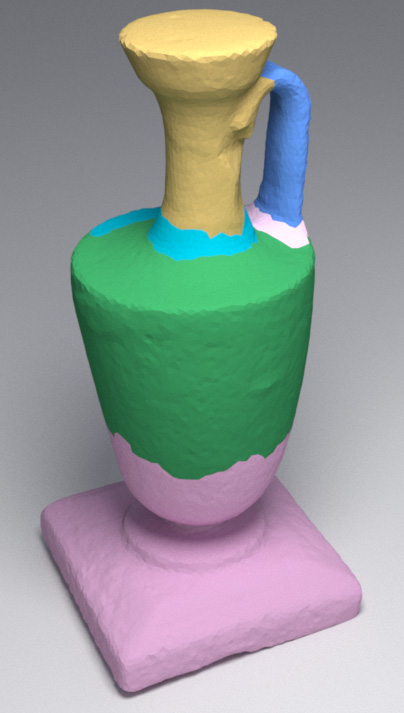} &
\includegraphics[width=0.14\linewidth]{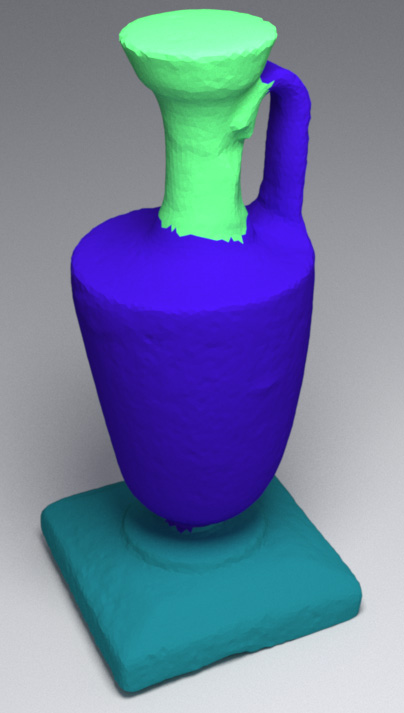} &
\includegraphics[width=0.14\linewidth]{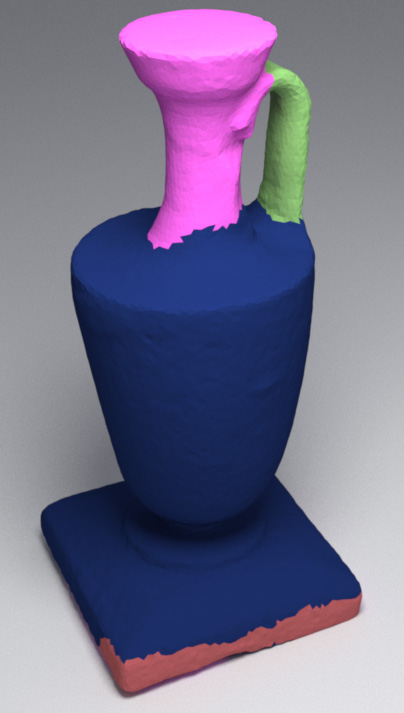} &
\includegraphics[width=0.14\linewidth]{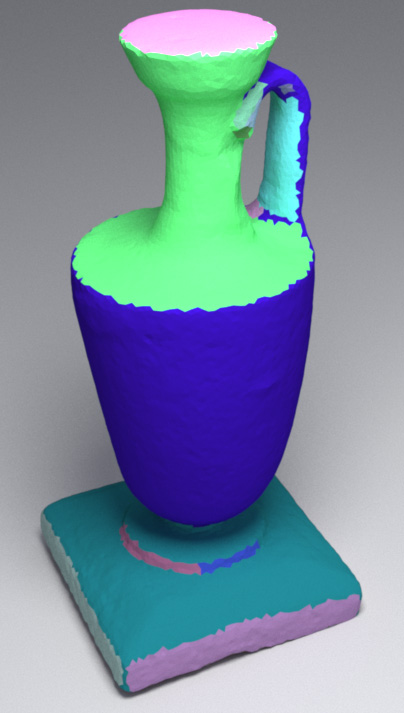}\\
\scriptsize{(a) Normalized}  & \scriptsize{(b) Randomized}  & \scriptsize{(c) Random Walks}  & \scriptsize{(d) Shape Diameter}  & \scriptsize{(e) Lifted Multicut}  & \scriptsize{(f) Projective}  & \scriptsize{(g) Our result}\\
\scriptsize{Cuts \protect\cite{golovinskiy2008randomized}} & \scriptsize{Cuts \protect\cite{golovinskiy2008randomized}} & \scriptsize{\protect\cite{lai2008fast}} & \scriptsize{Function \protect\cite{shapira2008consistent}} & \scriptsize{\protect\cite{keuper2015efficient}} & \scriptsize{ConvNet  \protect\cite{kalogerakis20173d}} &
\end{tabular} 
\caption{Comparison of semantic segmentations (a--f) with our geometric segmentation (g). While semantic segmentation produces parts of meaningful semantics such as the base, the neck, or the handle of this jar, our geometric segmentation detects sharp edges and results in piecewise smooth segments regardless of the underlying semantic.}
\label{fig:segmentationCompare}
\end{figure*}

\NB{In Fig.~\ref{fig:segmentationCompare}, we illustrate the differences between our geometric segmentation and state-of-the-art semantic segmentation methods. Our method is agnostic to the underlying semantics and produces geometrically consistent segmentations with piecewise smooth parts minimizing a well-defined AT energy, while semantic segmentation methods often produce semantically meaningful parts from heuristics or learning from large manually segmented databases~\cite{chen2009benchmark}.}

\subsection{Inpainting}

\begin{figure}[h]
\centering
\begin{tabular}{@{}c@{}c@{}c@{}c@{}}
\includegraphics[width=0.24\linewidth]{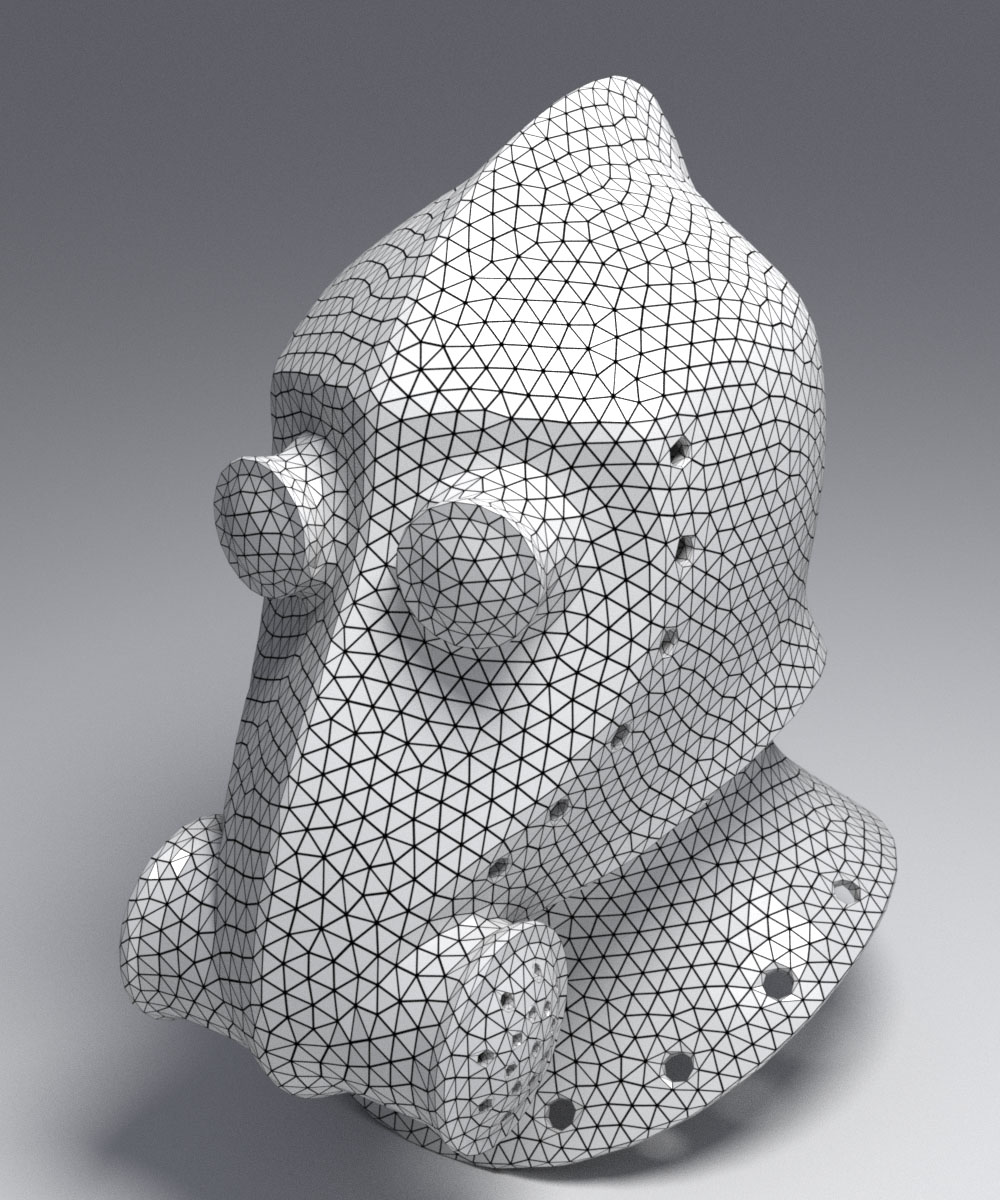}&
\includegraphics[width=0.24\linewidth]{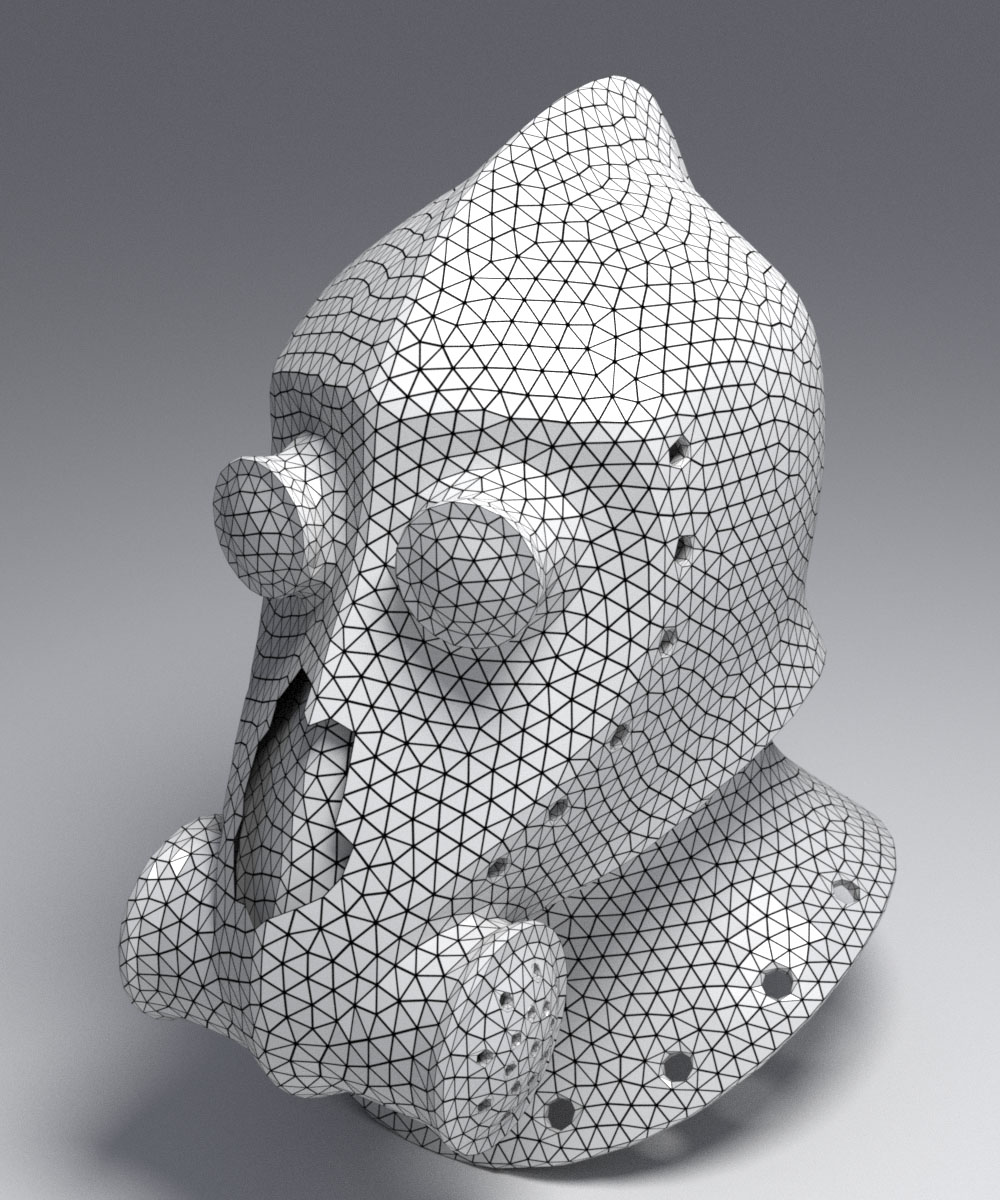}&
\includegraphics[width=0.24\linewidth]{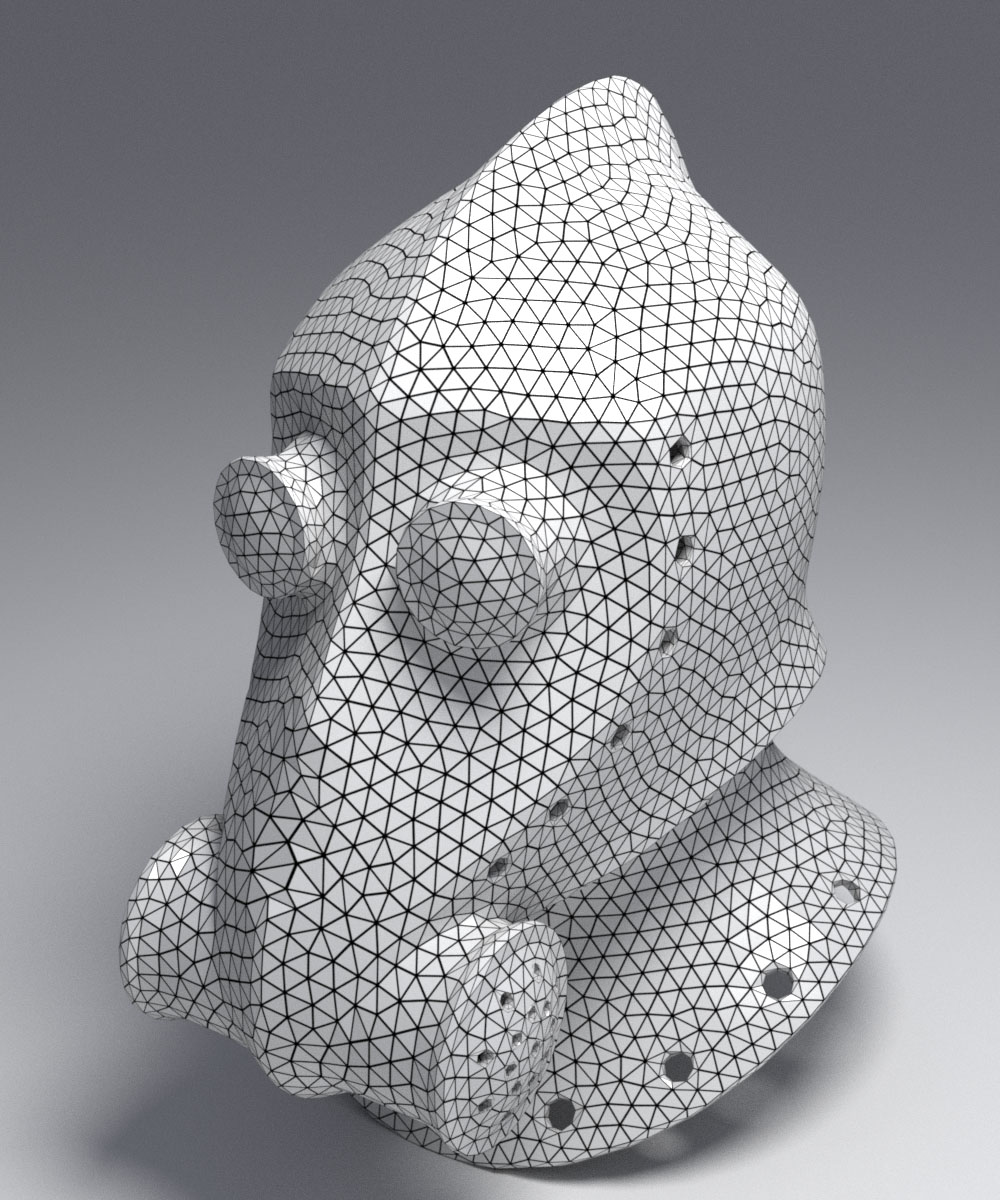}&
\includegraphics[width=0.24\linewidth]{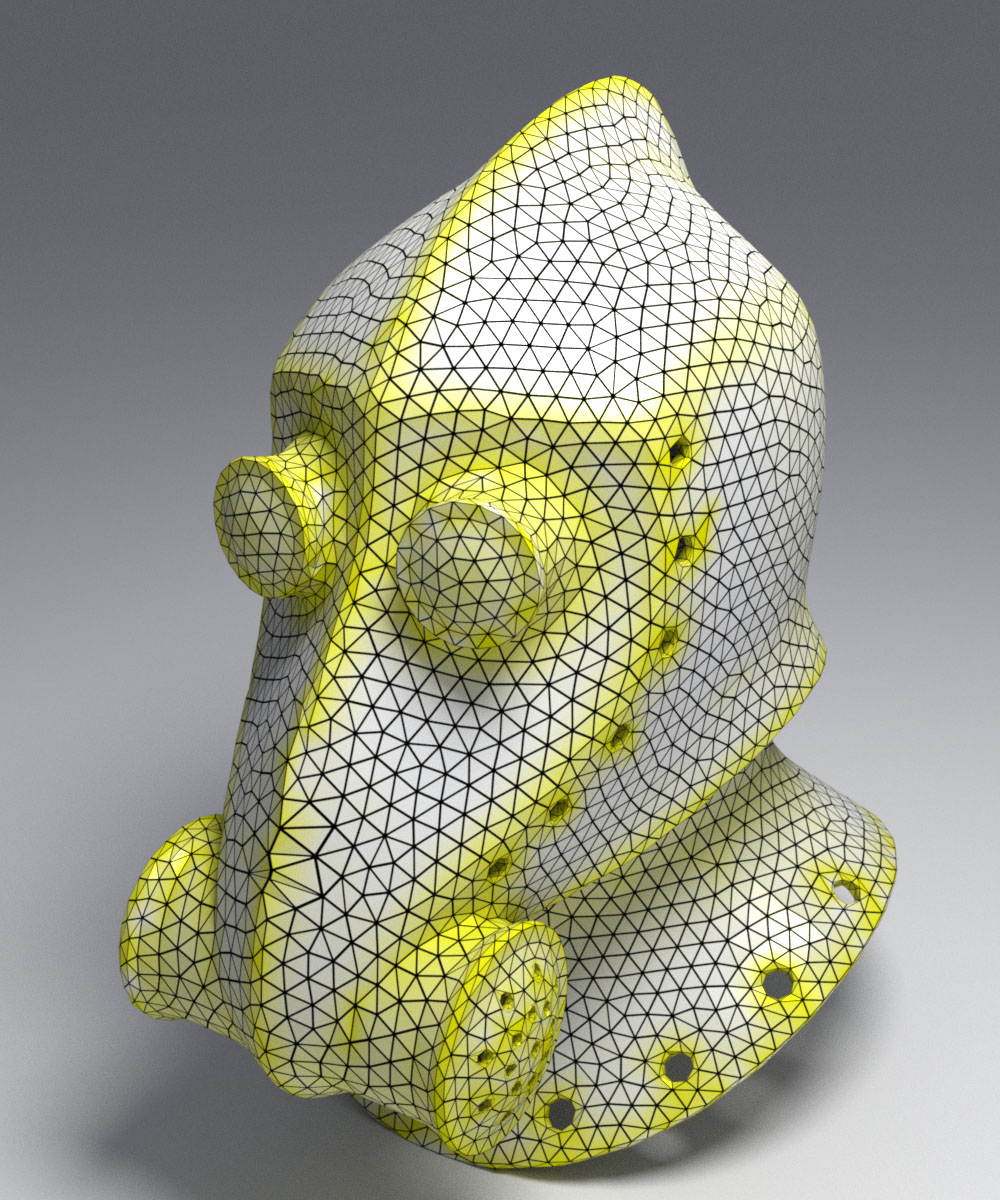}\\
\includegraphics[width=0.24\linewidth]{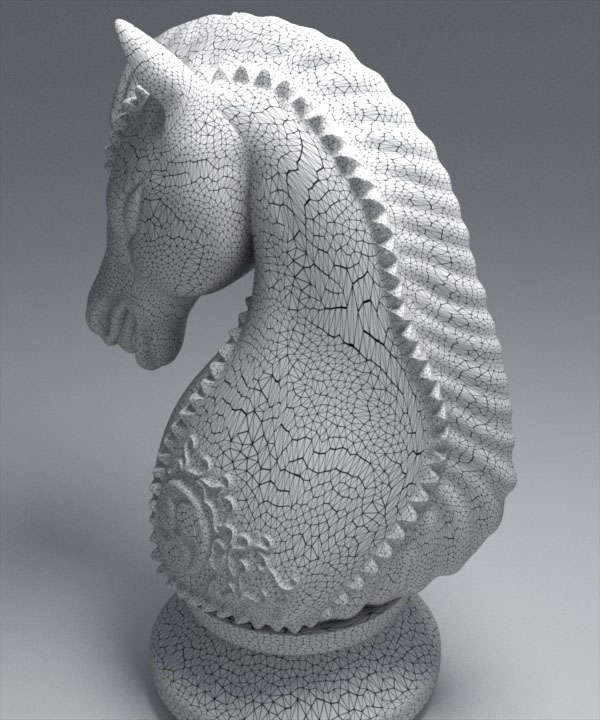}&
\includegraphics[width=0.24\linewidth]{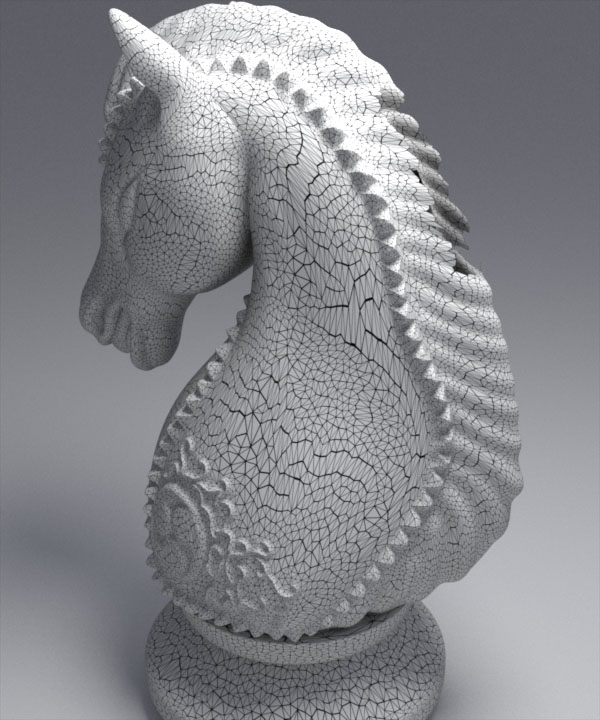}&
\includegraphics[width=0.24\linewidth]{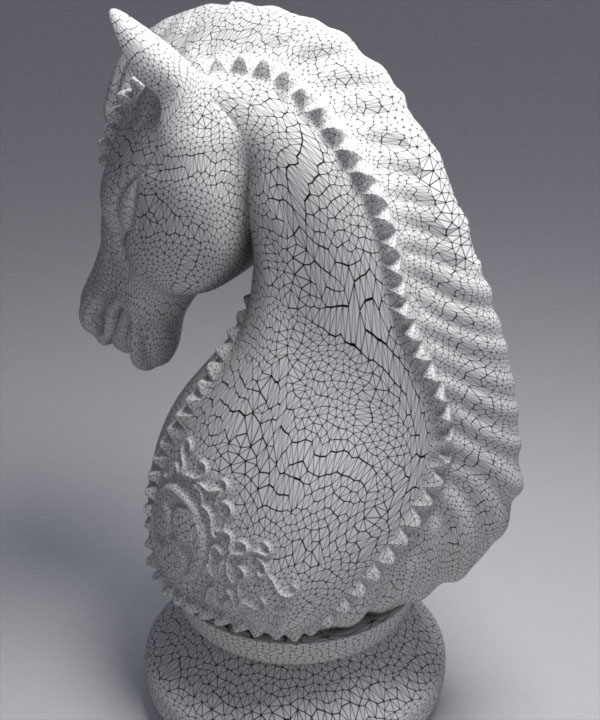}&
\includegraphics[width=0.24\linewidth]{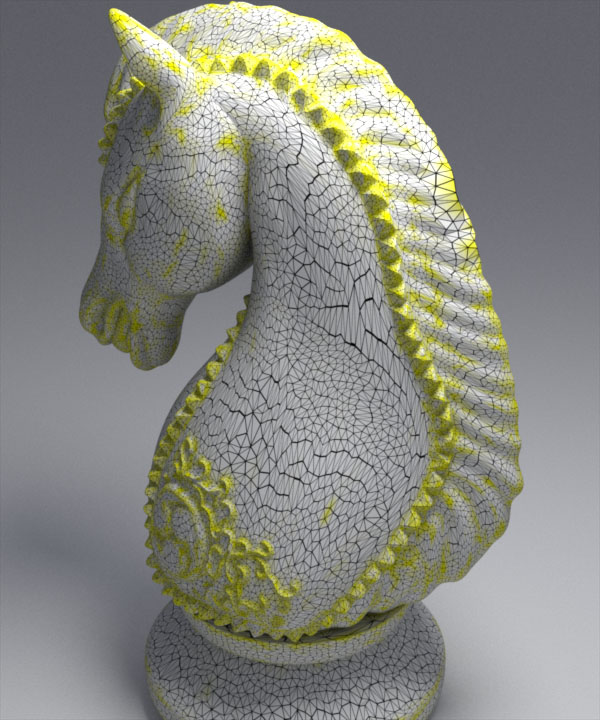}\\
\includegraphics[width=0.24\linewidth]{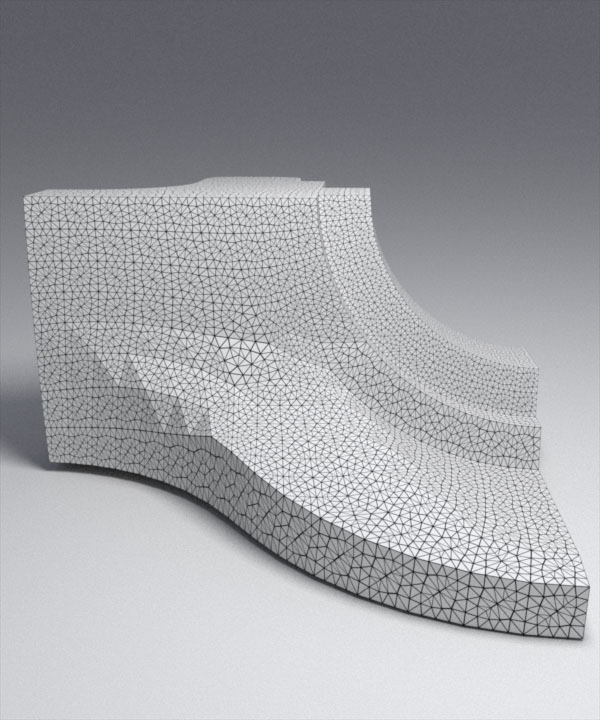}&
\includegraphics[width=0.24\linewidth]{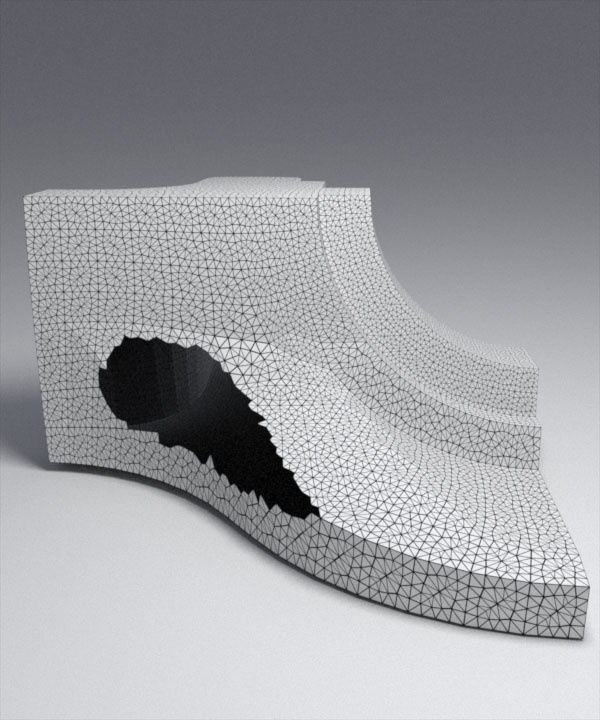}&
\includegraphics[width=0.24\linewidth]{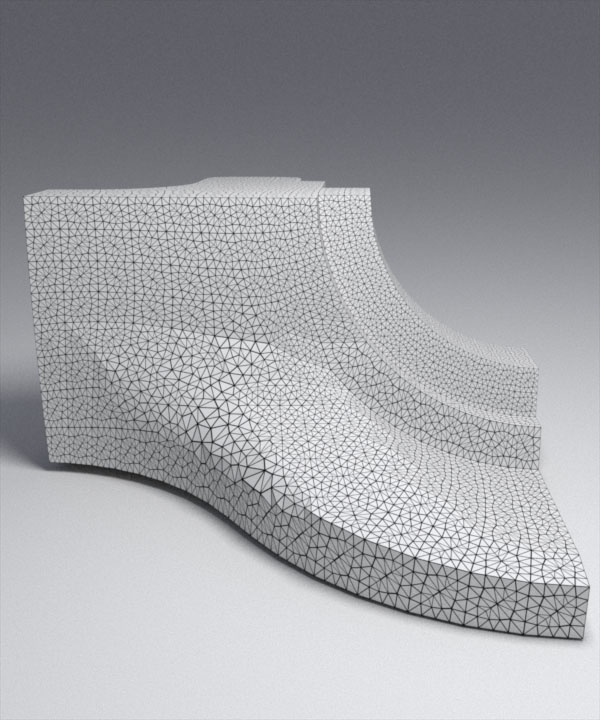}&
\includegraphics[width=0.24\linewidth]{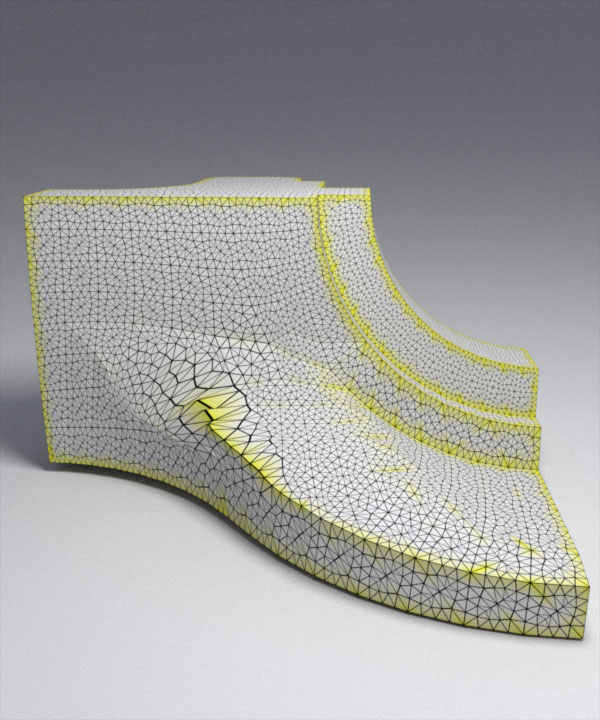}\\
\includegraphics[width=0.24\linewidth]{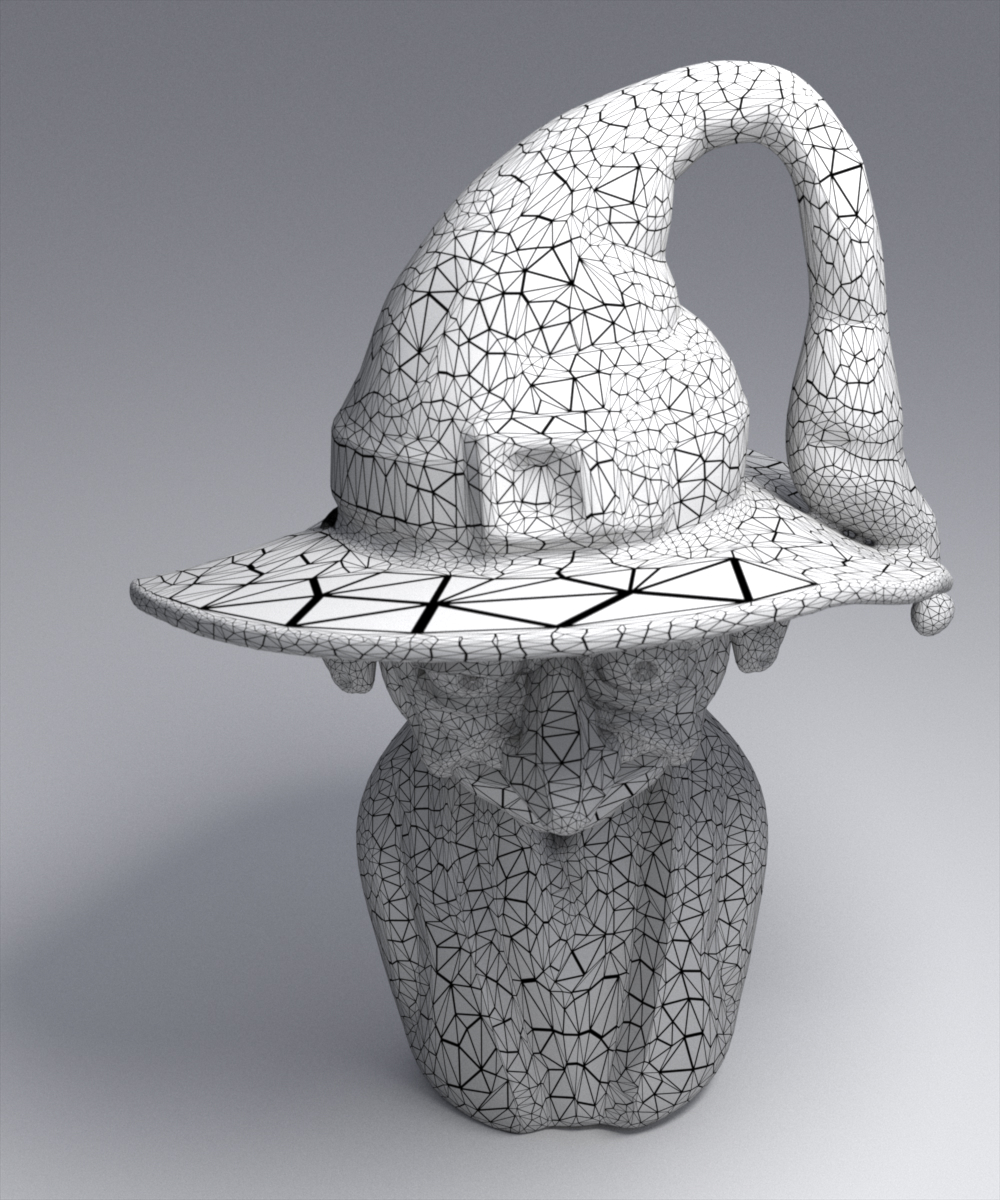}&
\includegraphics[width=0.24\linewidth]{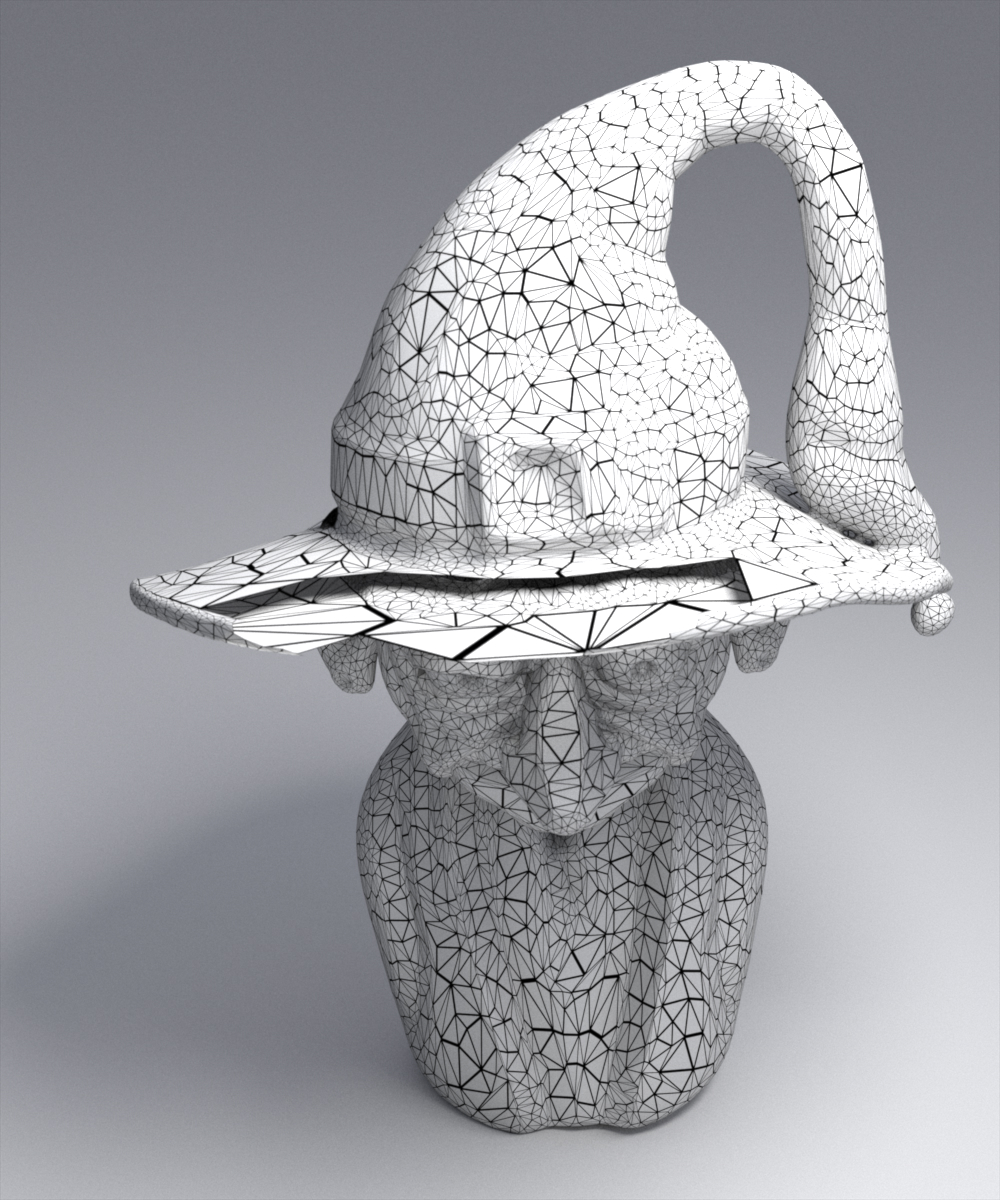}&
\includegraphics[width=0.24\linewidth]{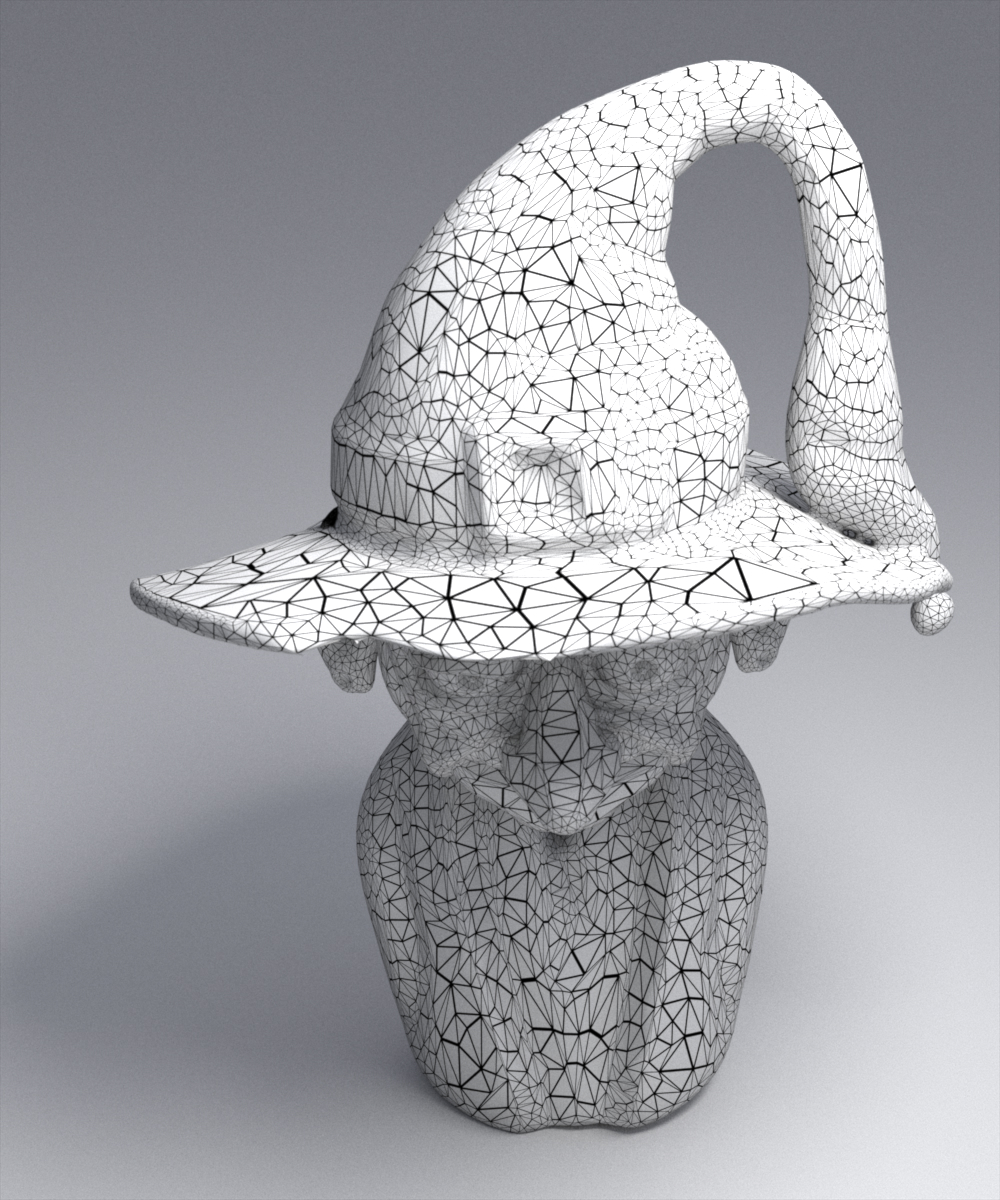}&
\includegraphics[width=0.24\linewidth]{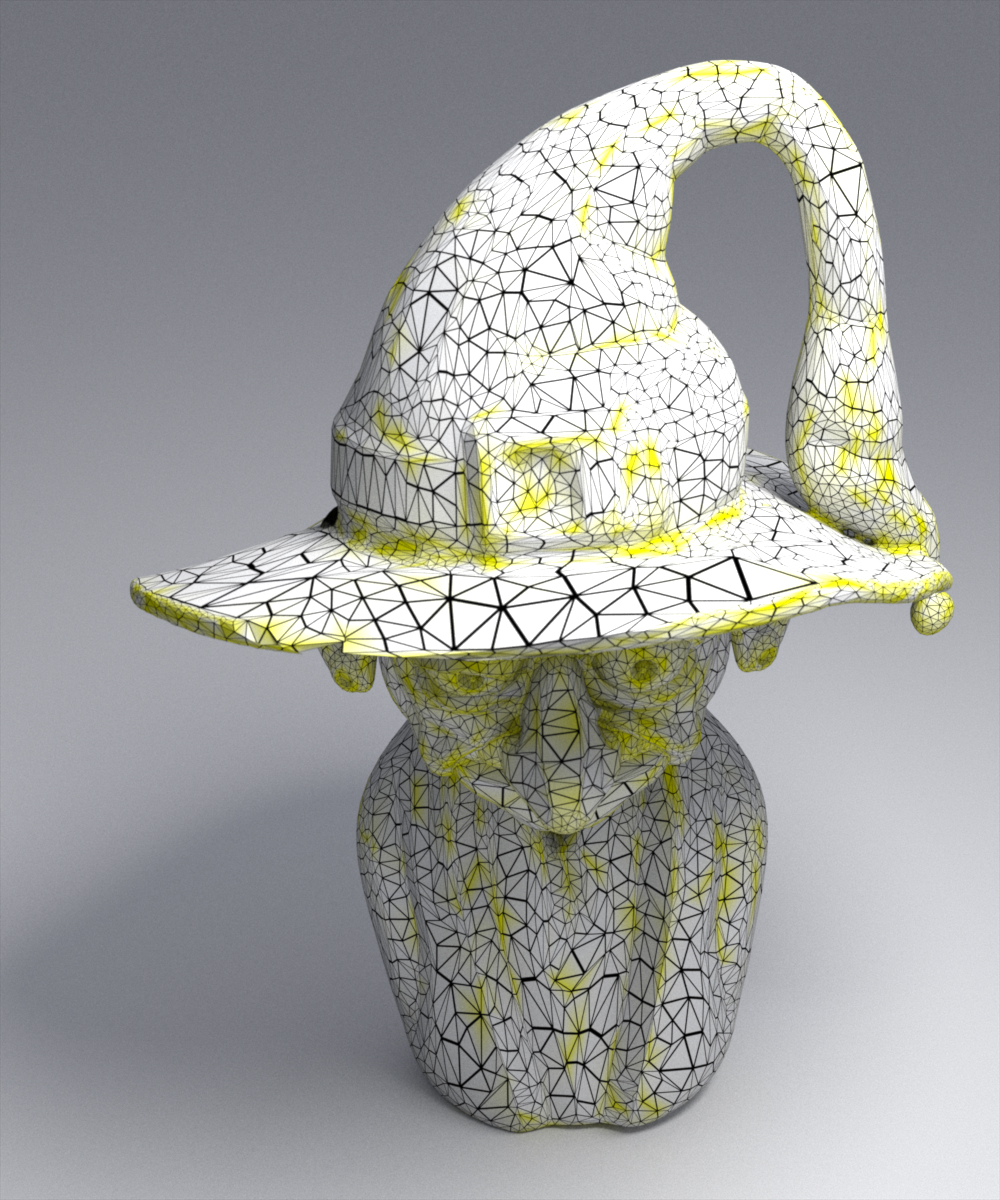}\\
\includegraphics[width=0.24\linewidth]{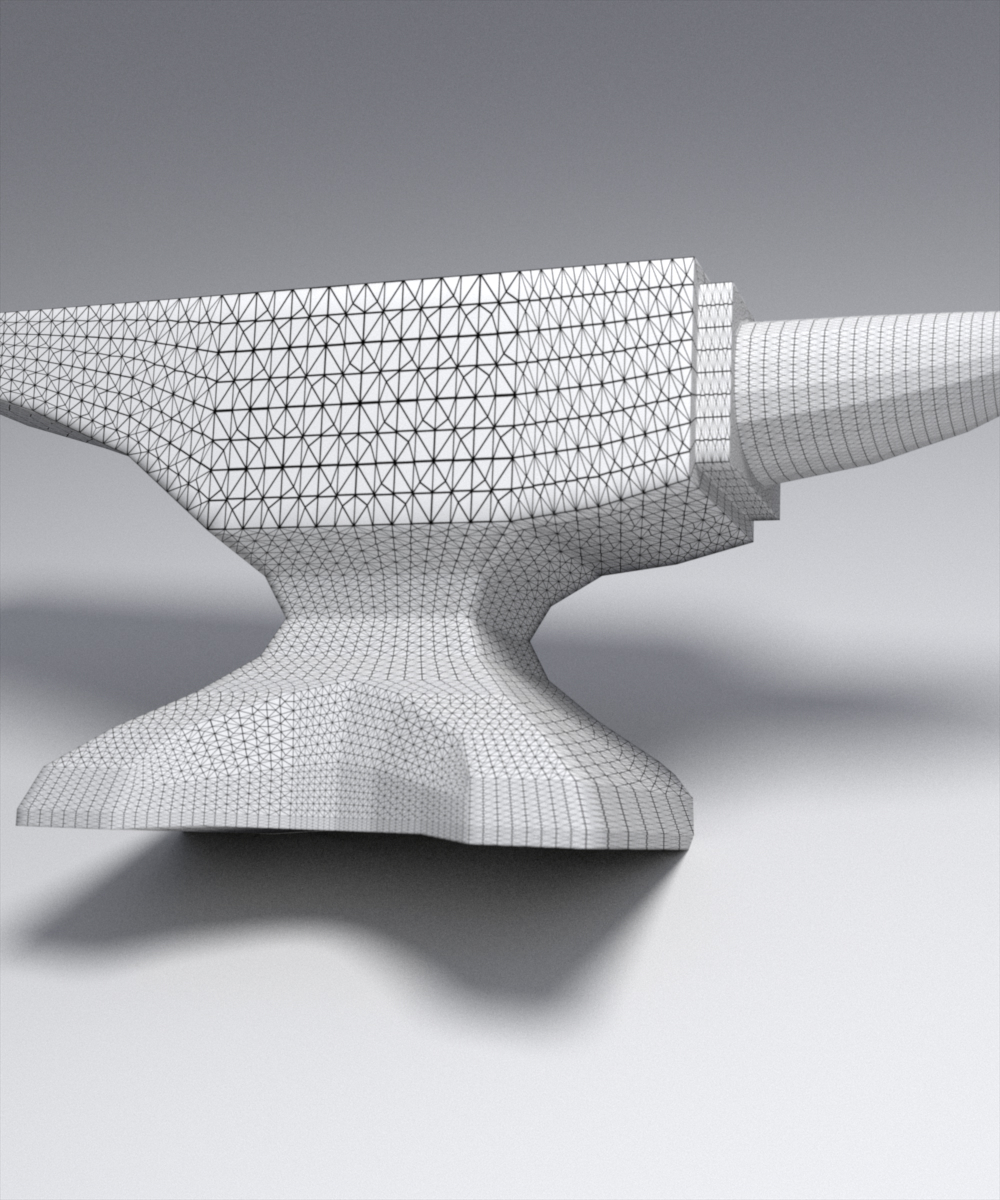}&
\includegraphics[width=0.24\linewidth]{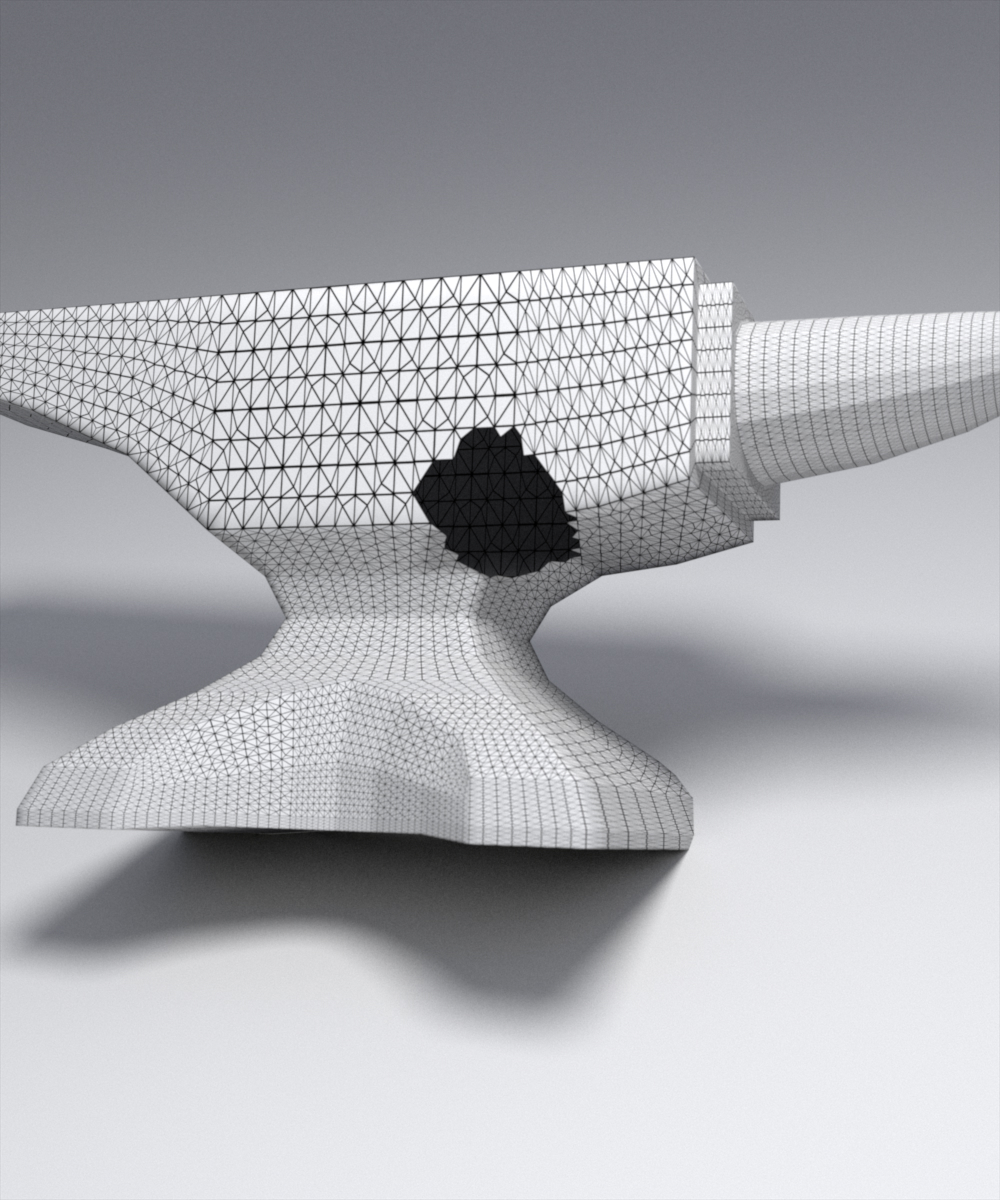}&
\includegraphics[width=0.24\linewidth]{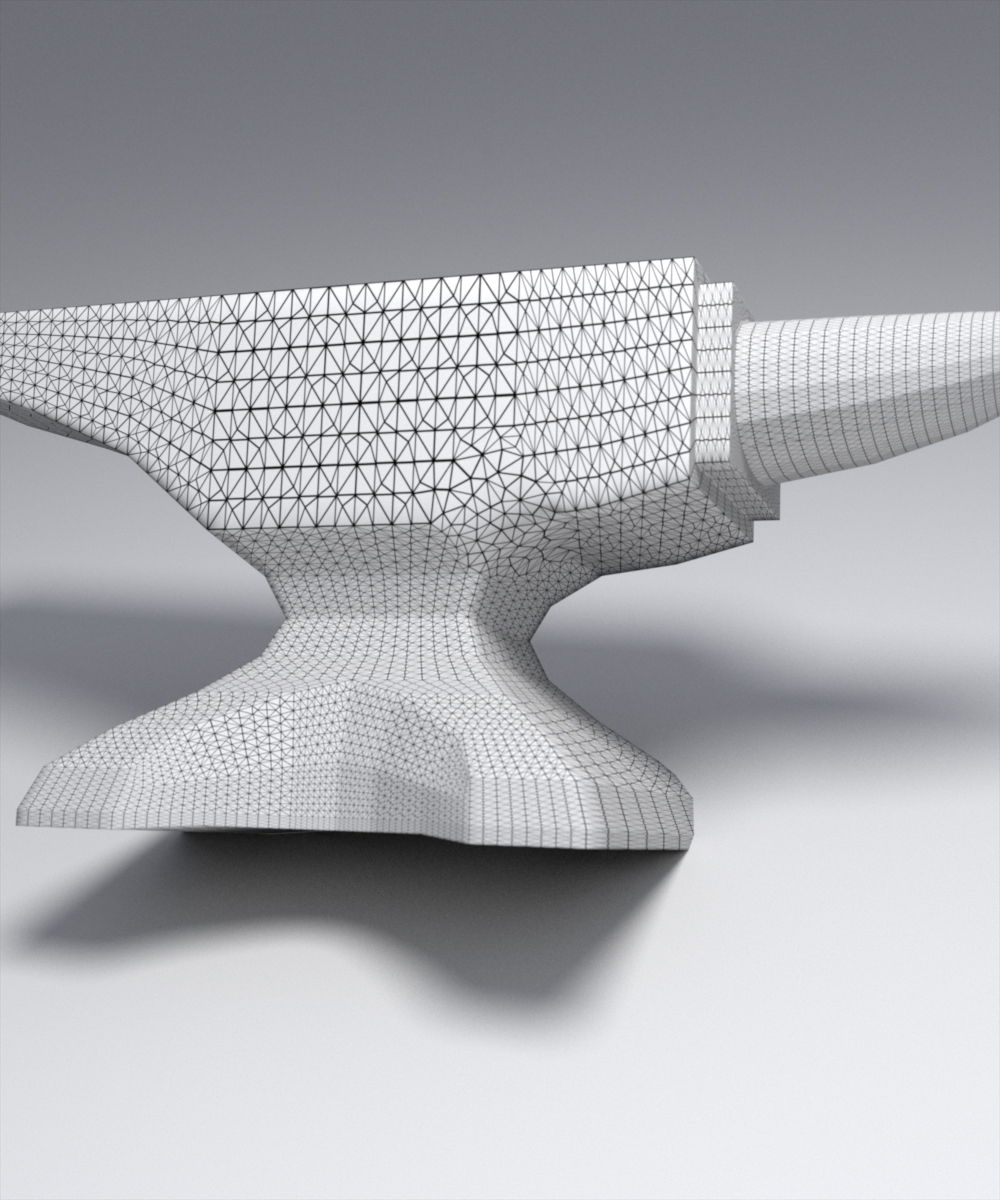}&
\includegraphics[width=0.24\linewidth]{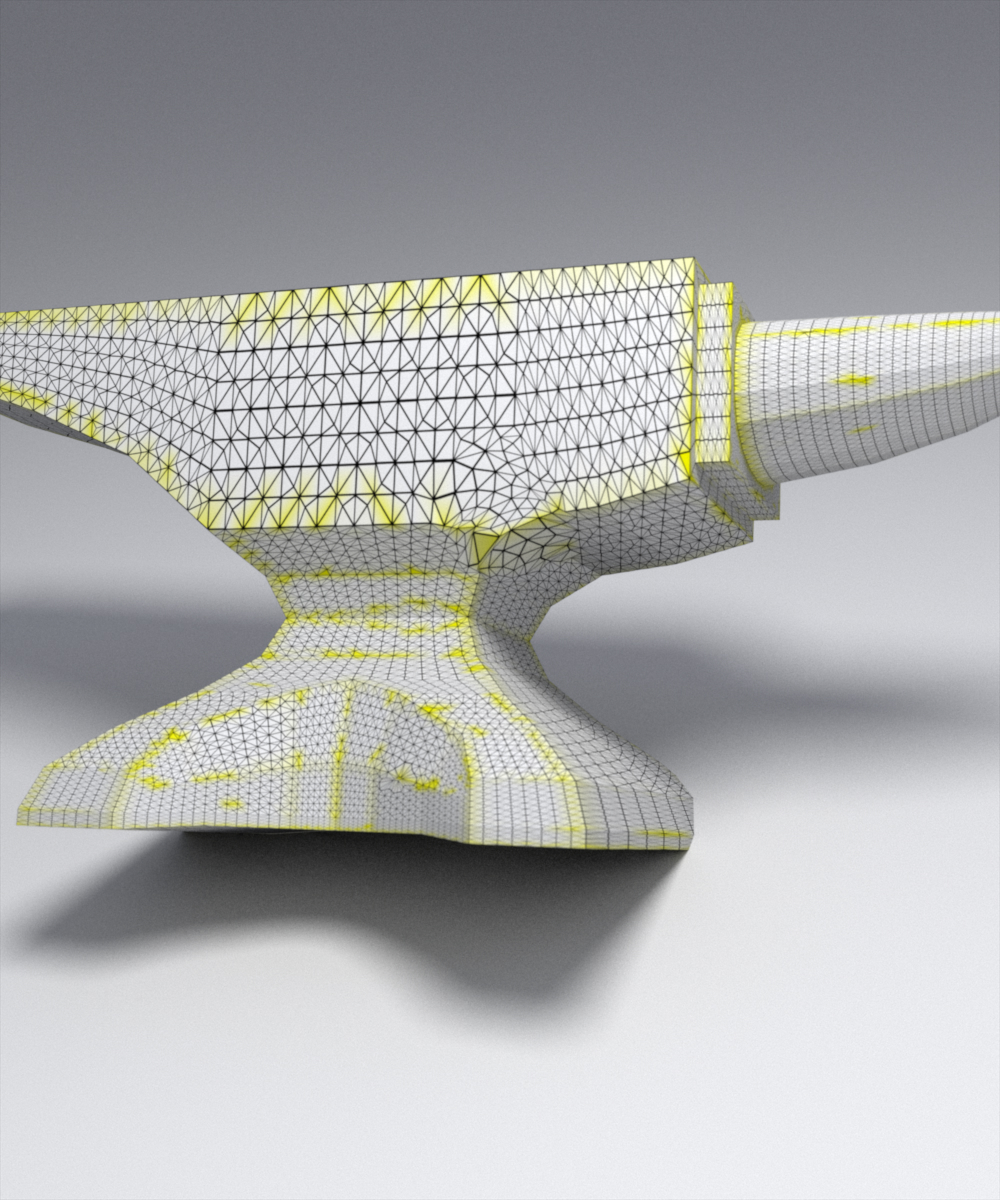}\\
Original & Missing area & CGAL filling & Our inpainting
\end{tabular}
\caption{Inpainting results. The recovered feature field $v$ is shown in yellow.}
\label{fig:inpainting}
\end{figure}

Inpainting seeks to fill in missing areas of a mesh. To perform mesh
inpainting, one should first triangulate the missing area. The first step is 
to recover the topology of the missing area, which we perform using the hole 
filling algorithm~\cite{liepa2003filling} implemented in CGAL~\cite{fabri2009cgal}.  
Then, we set the data attachment terms $\alpha$ and $w_2$ to a large value (typically, $\alpha = w_2 = 1000$) outside 
of the inpainted area to prevent known vertices from moving, and $w_2$ to zero inside. Since the hole filling tends to oversmooth the inpainted area, 
we set $\lambda$ to one tenth of its original value inside the inpainted area, hence favoring longer features.
We finally solve the MS problem and conform the mesh to the obtained
normals. Inpainting results are shown in Fig.~\ref{fig:inpainting} {and
Hausdorff distance evaluation is given in Table~\ref{tab:inpainting}}.

\begin{table}
\centering\begin{tabular}{|c|c|c|}
\hline
 &  CGAL  &  Our\\
\hline
Helmet & 0.160 & 0.147  \\
Knight &  1.179 & 1.167\\
Fandisk &   1.004 &  0.953\\
Witch & 7.503 & 7.098 \\
Anvil & 7.910 & 5.126 \\
\hline
\end{tabular}
\caption{{Hausdorff distance ($\times 1000$) between the original shape
  and CGAL filling and our inpainting}.}
\label{tab:inpainting}
\end{table}

\subsection{Embossing Normal Maps}

\begin{figure}[h]
\centering
\begin{tabular}{@{}c@{}c@{}c@{}c@{}}
\includegraphics[width=0.25\linewidth]{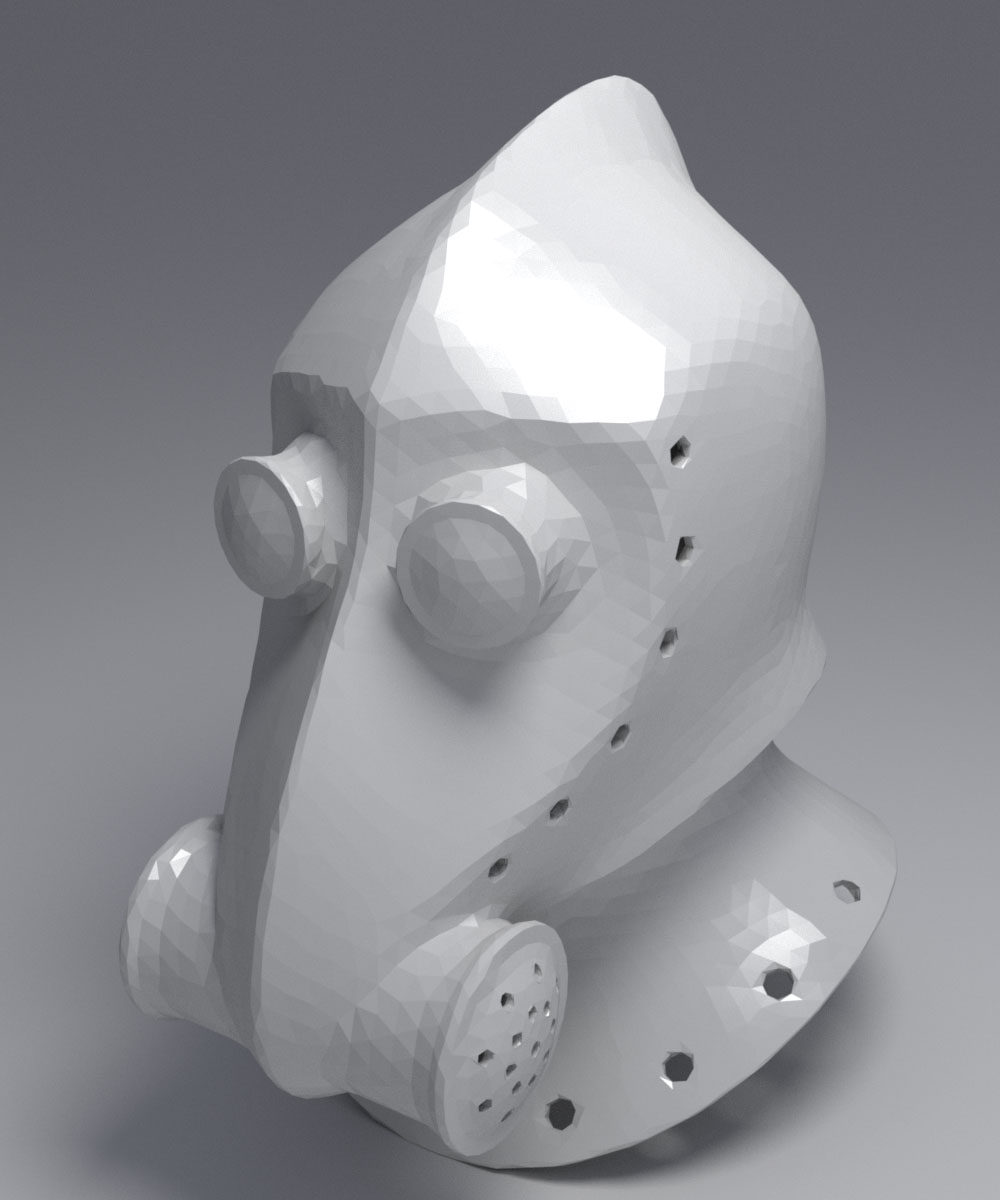}&
\includegraphics[width=0.25\linewidth]{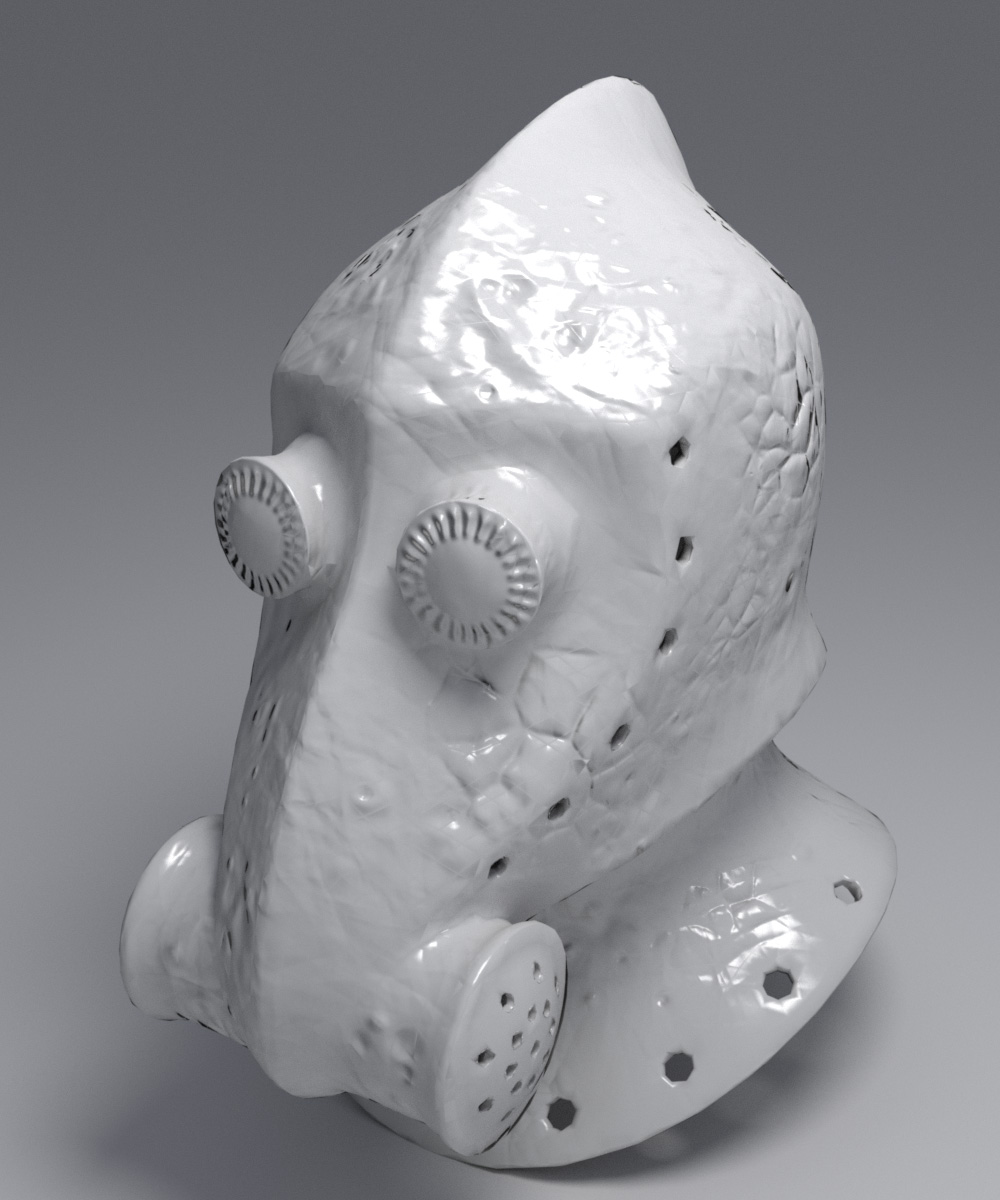}&
\includegraphics[width=0.25\linewidth]{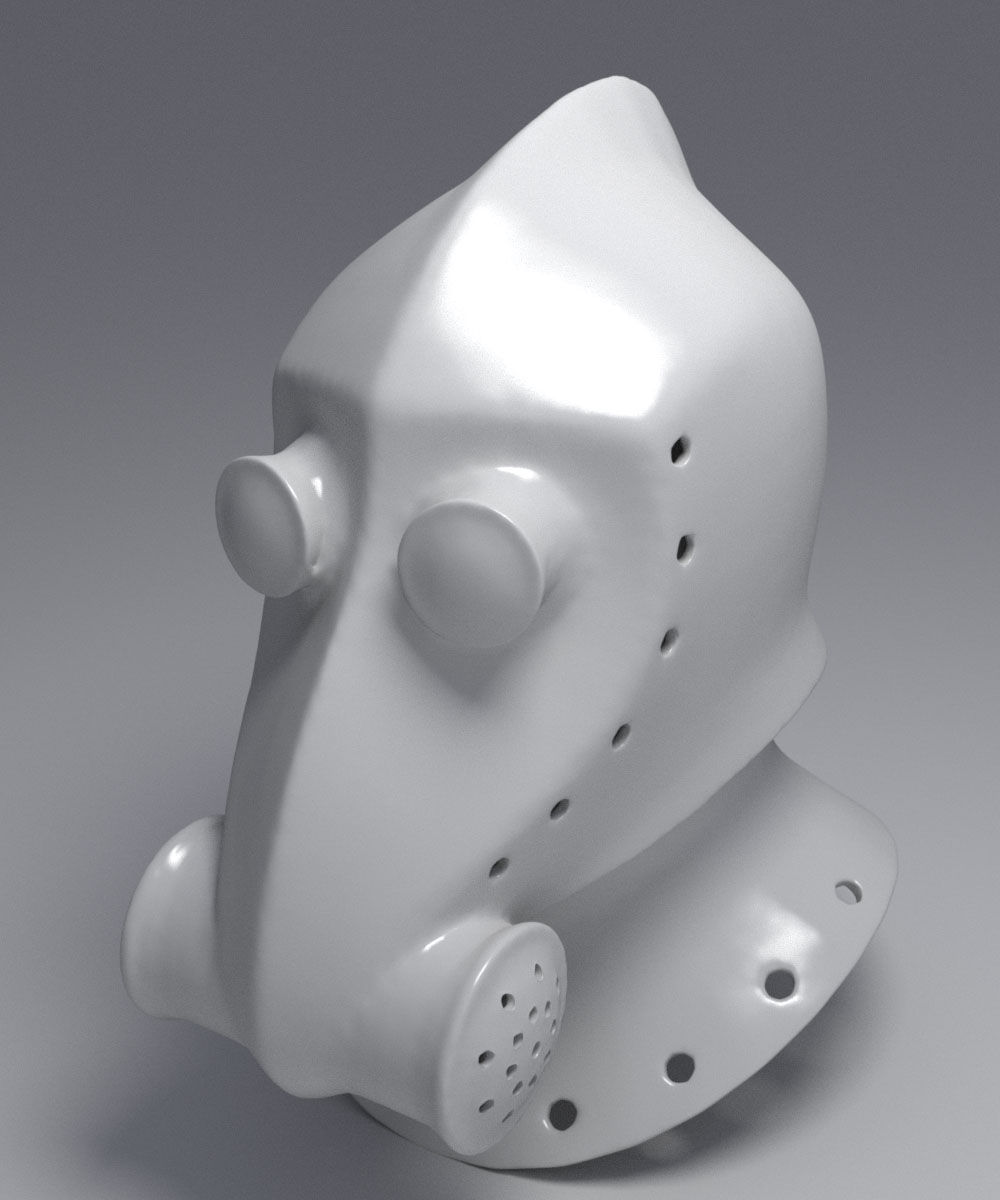}&
\includegraphics[width=0.25\linewidth]{images/helmet_embossed.jpg}\\
\includegraphics[width=0.25\linewidth]{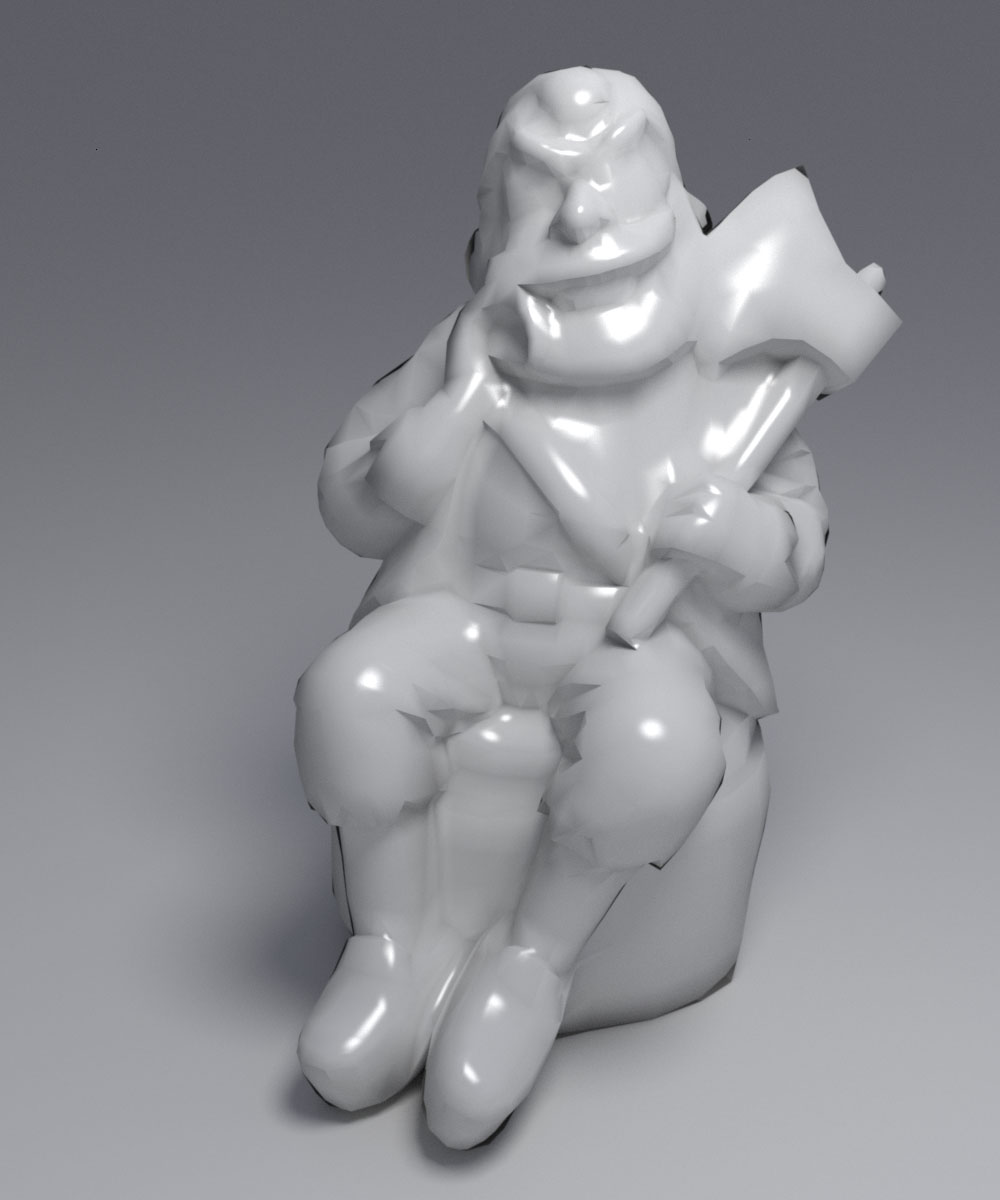}&
\includegraphics[width=0.25\linewidth]{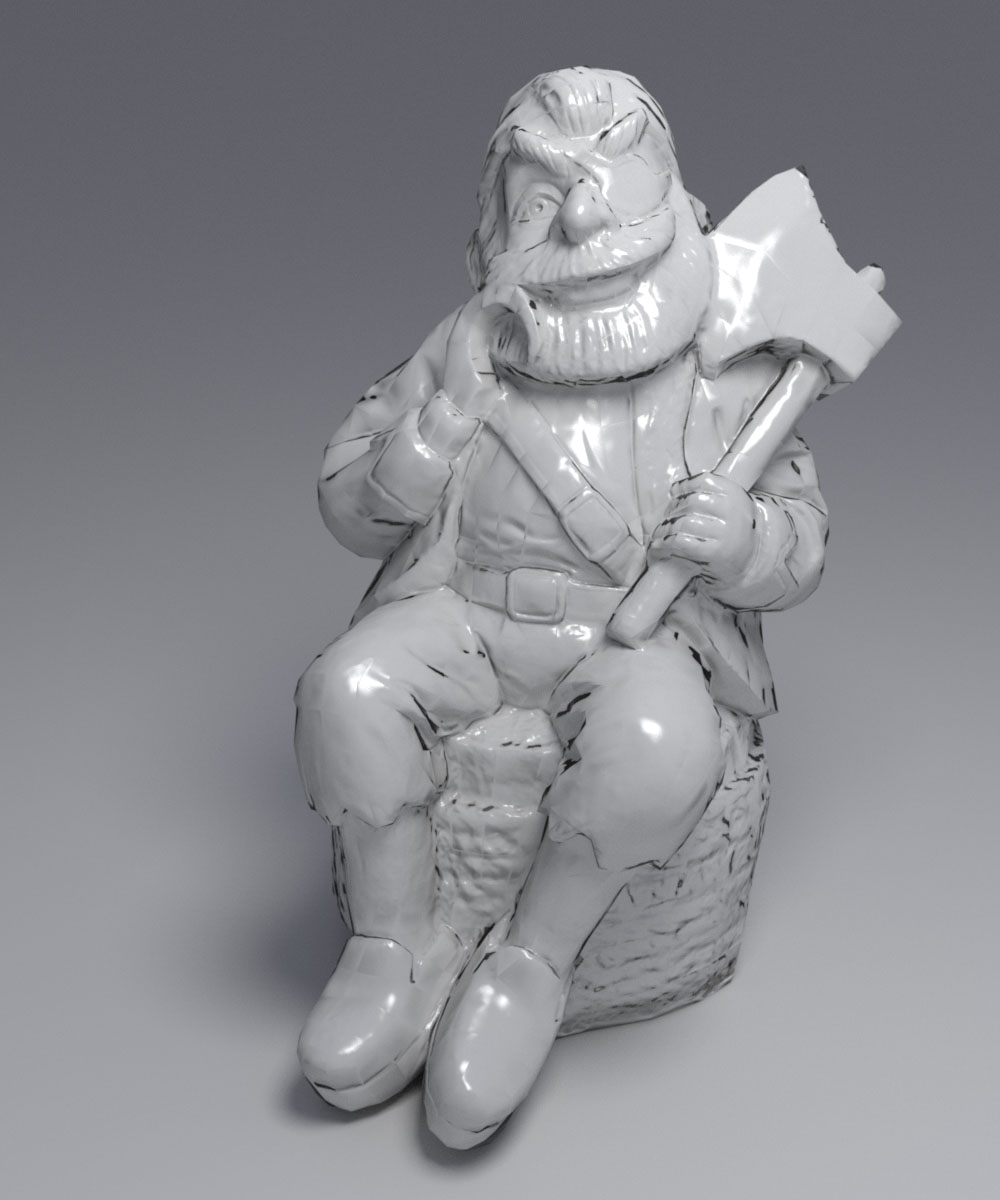}&
\includegraphics[width=0.25\linewidth]{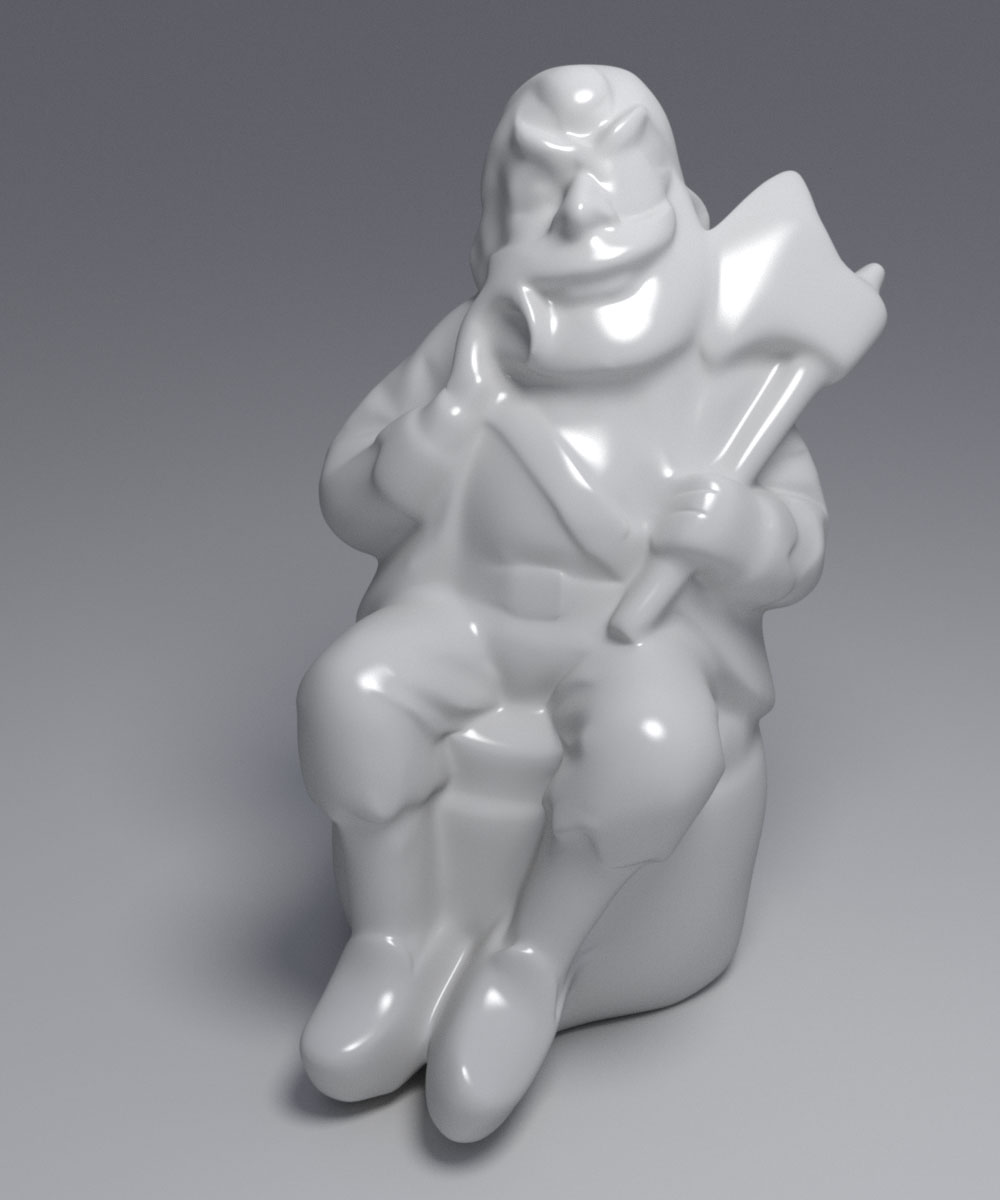}&
\includegraphics[width=0.25\linewidth]{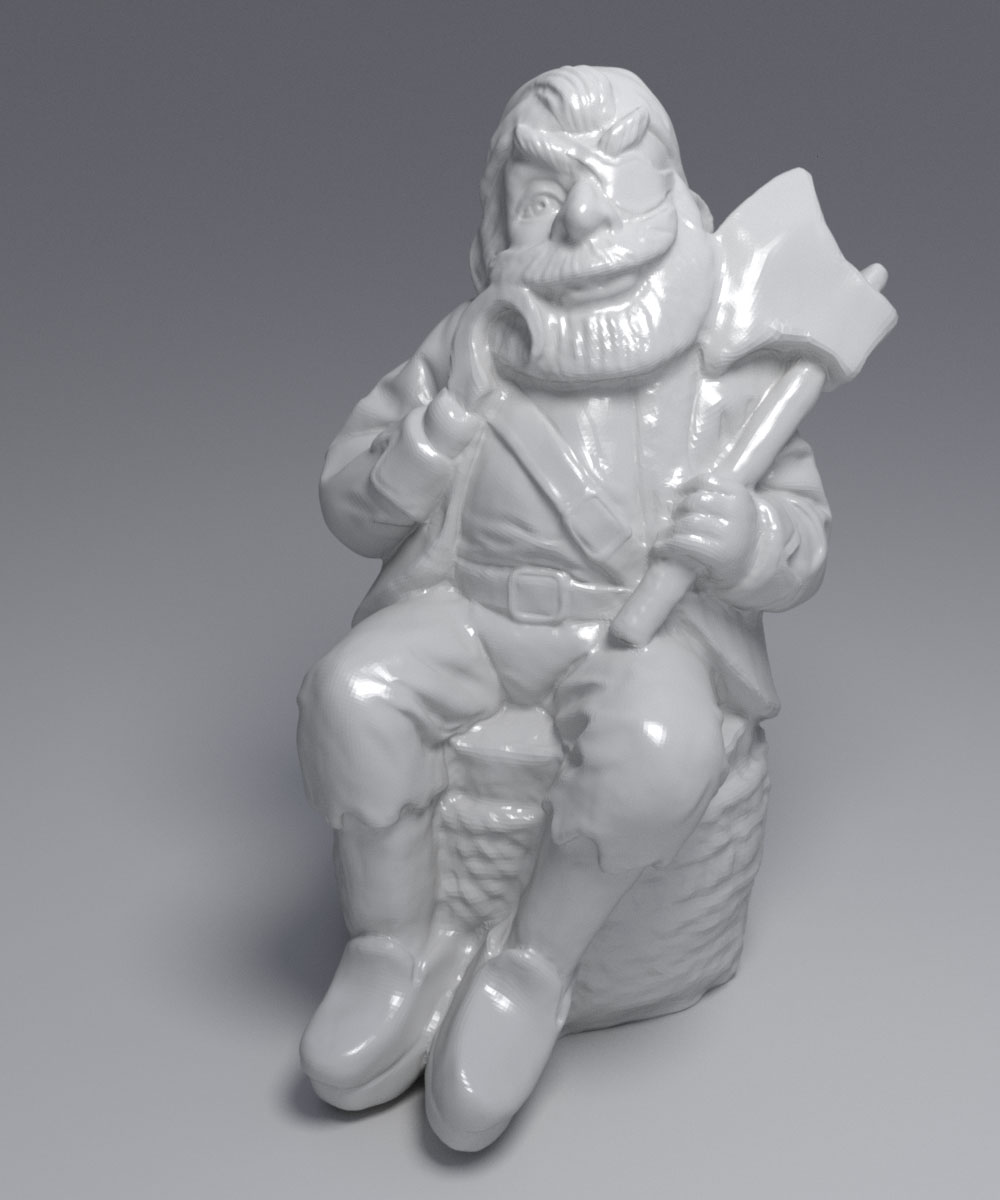}\\
\includegraphics[width=0.25\linewidth]{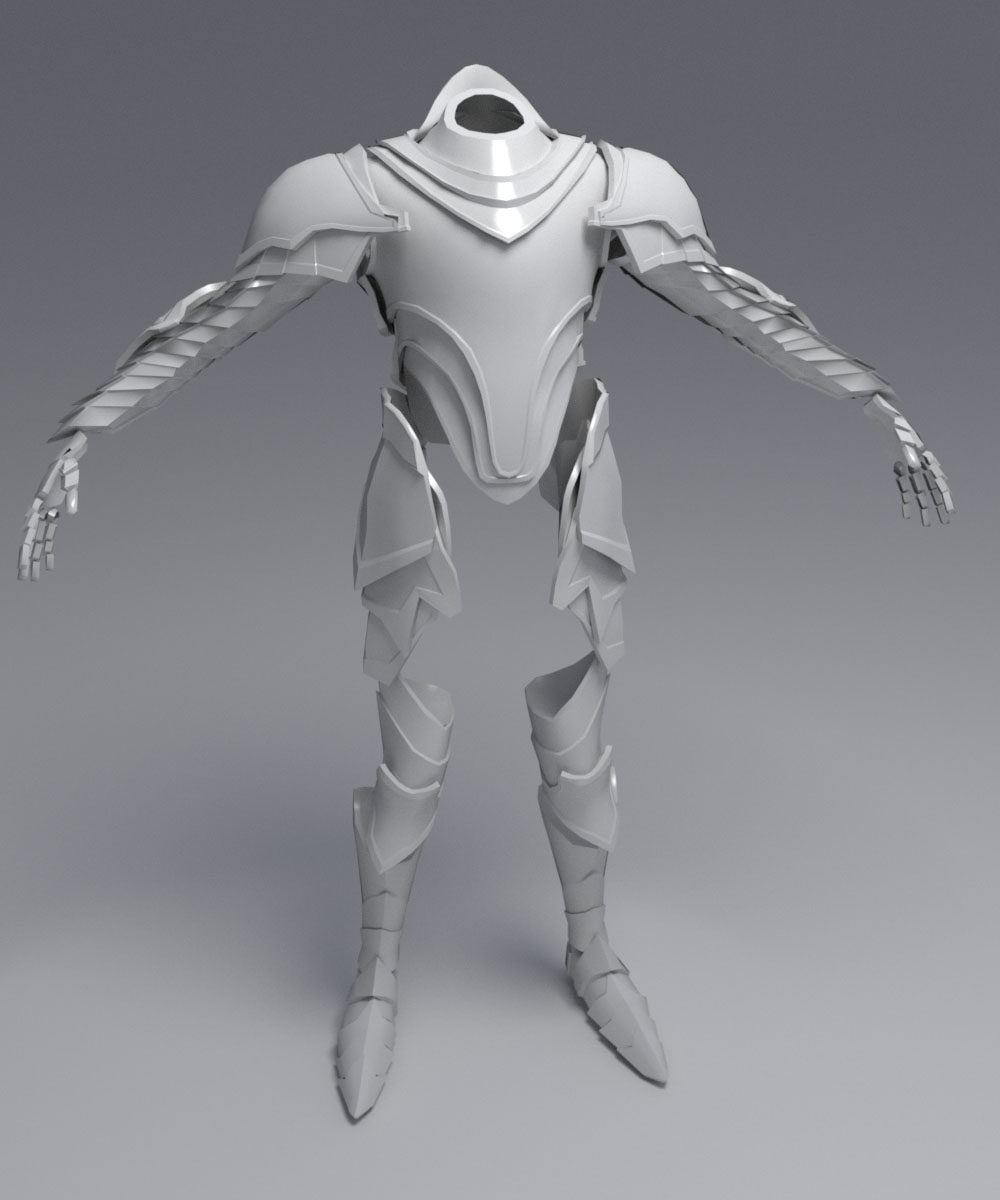}&
\includegraphics[width=0.25\linewidth]{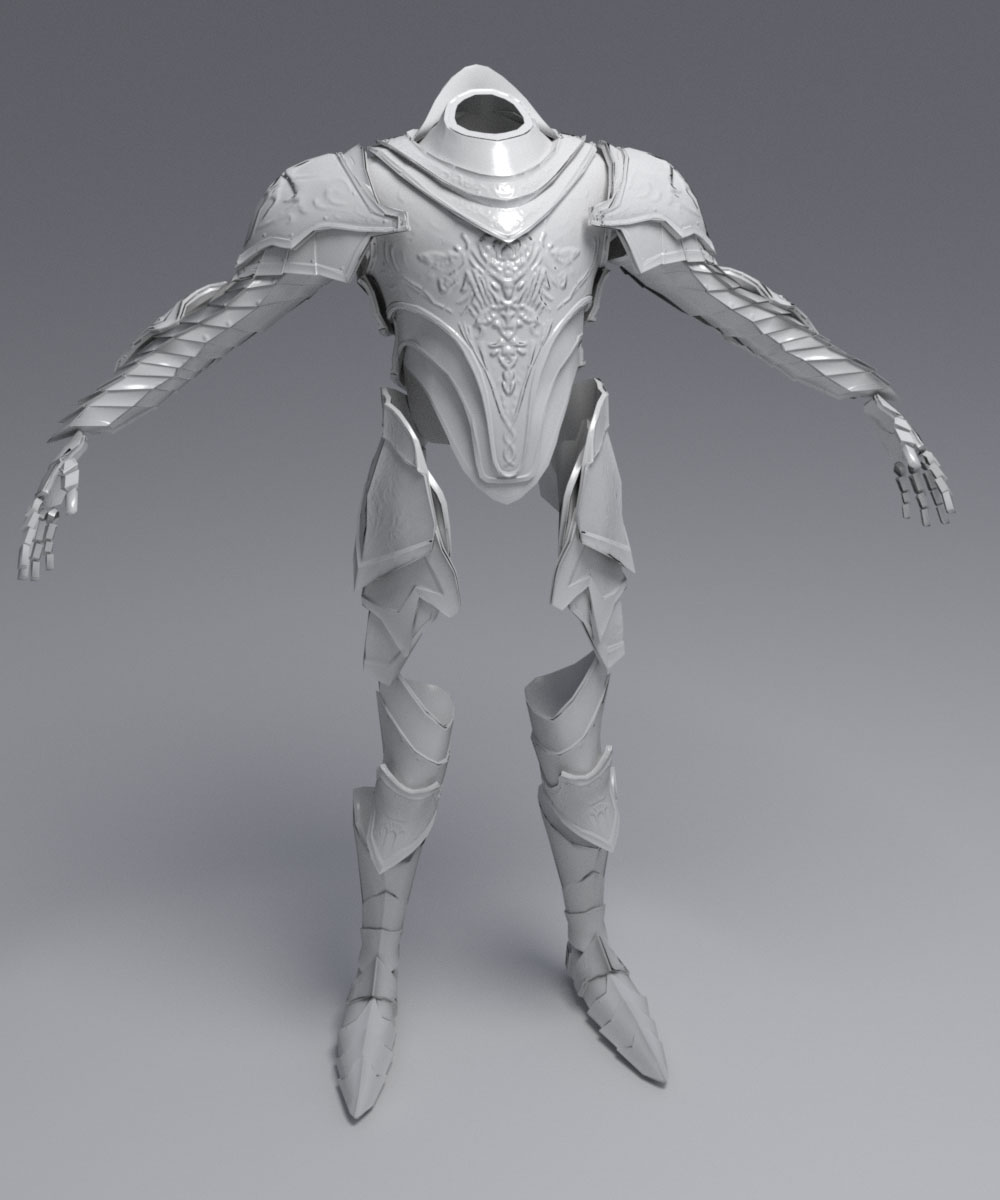}&
\includegraphics[width=0.25\linewidth]{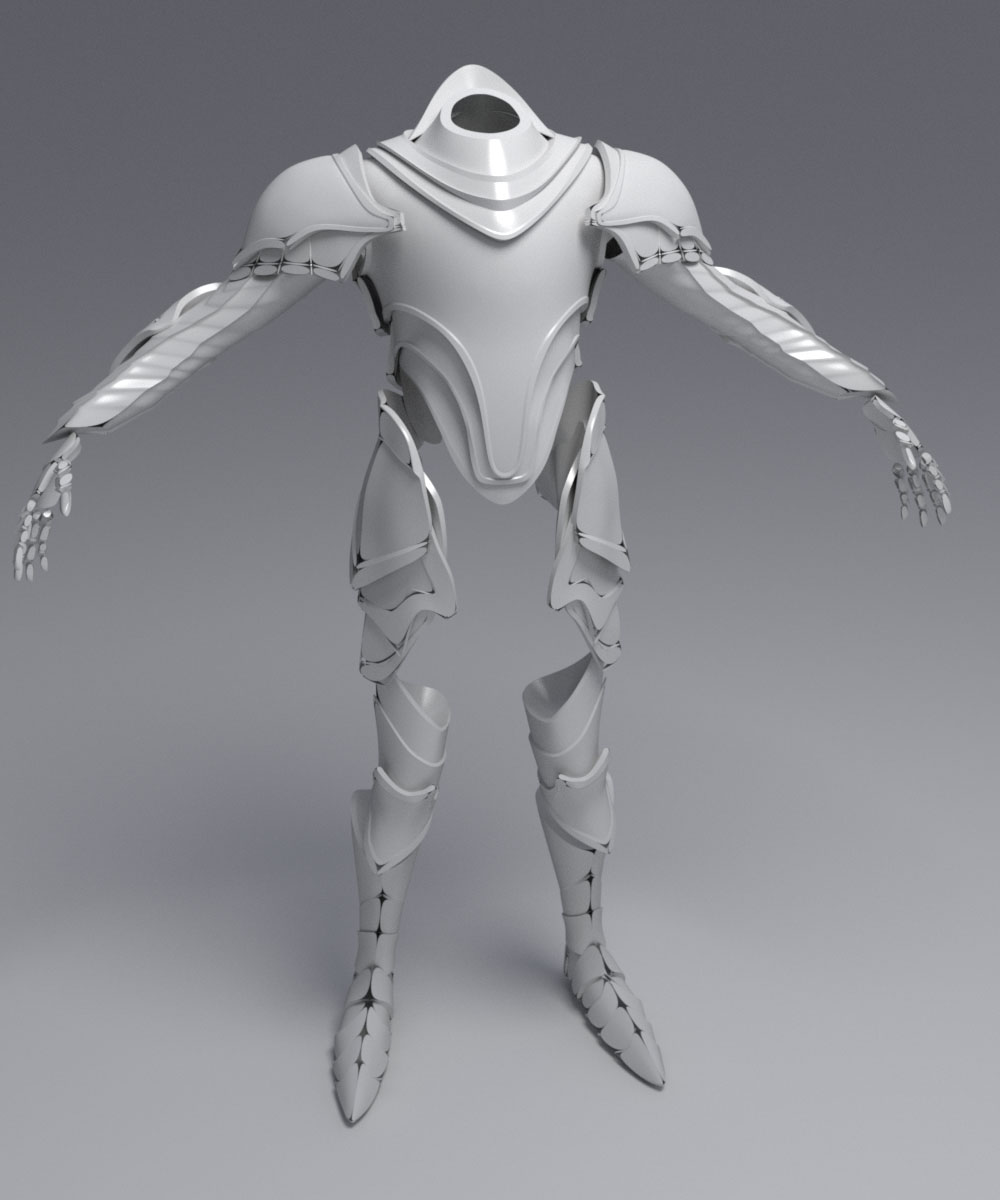}&
\includegraphics[width=0.25\linewidth]{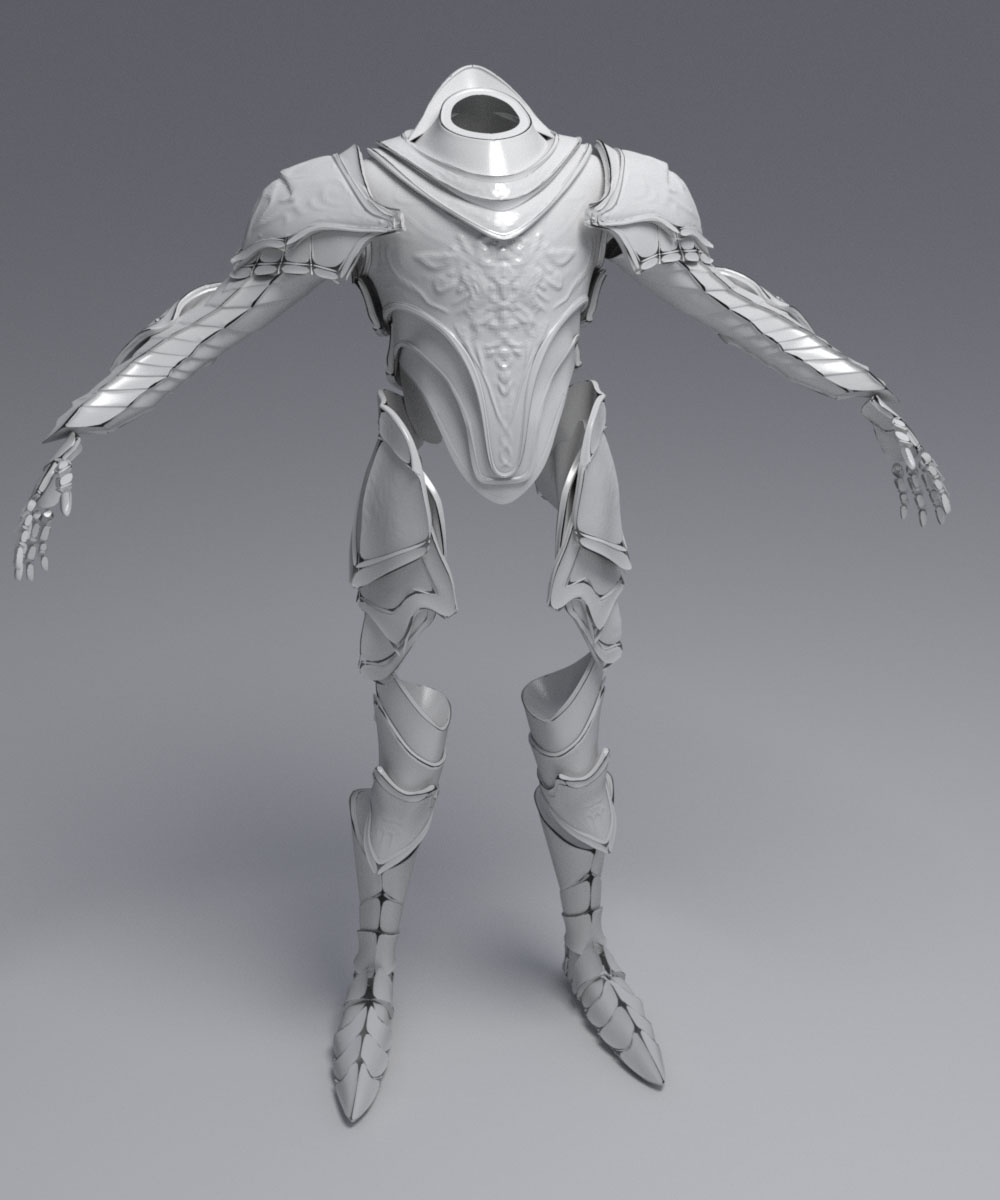}\\
\includegraphics[width=0.25\linewidth]{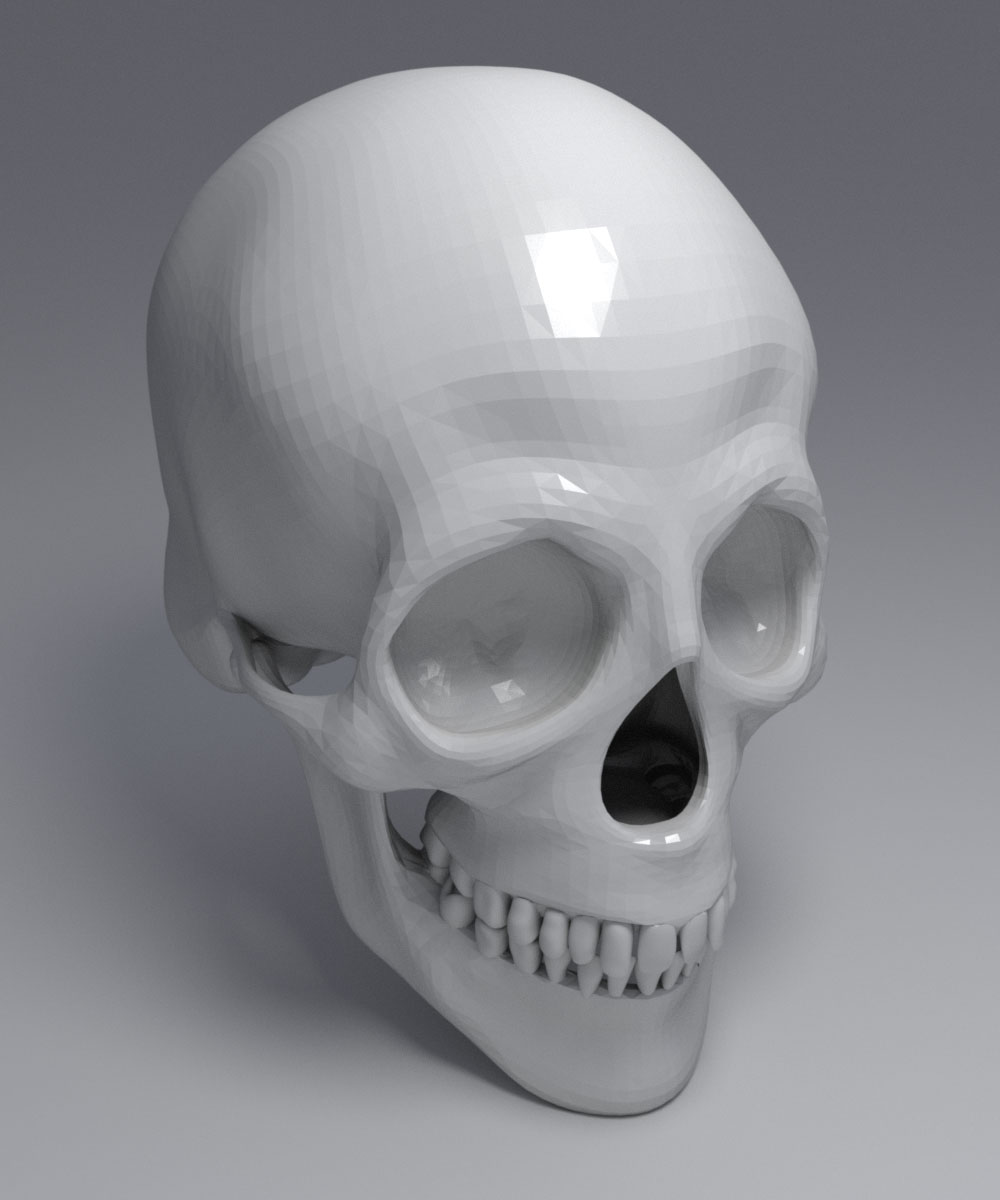}&
\includegraphics[width=0.25\linewidth]{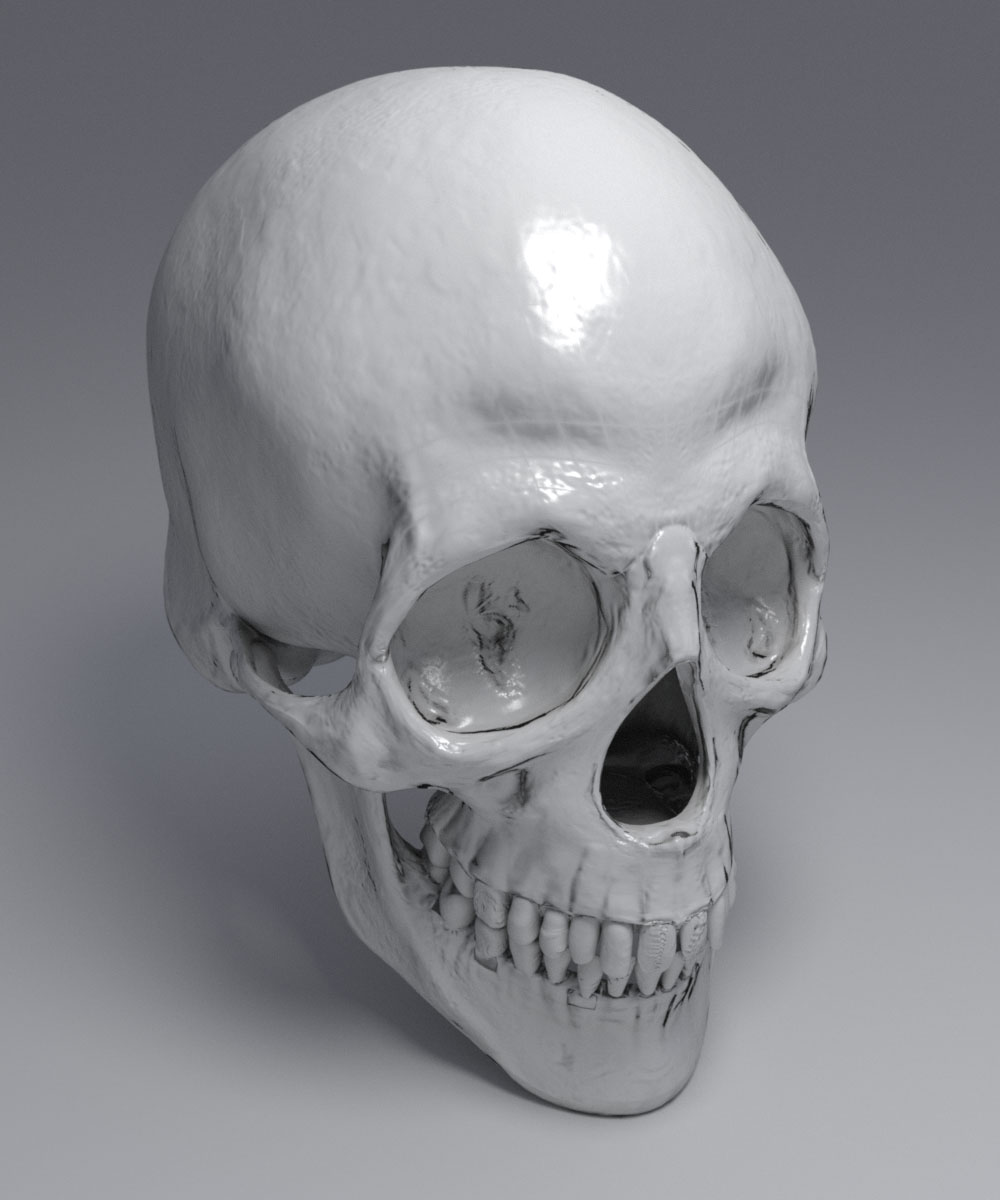}&
\includegraphics[width=0.25\linewidth]{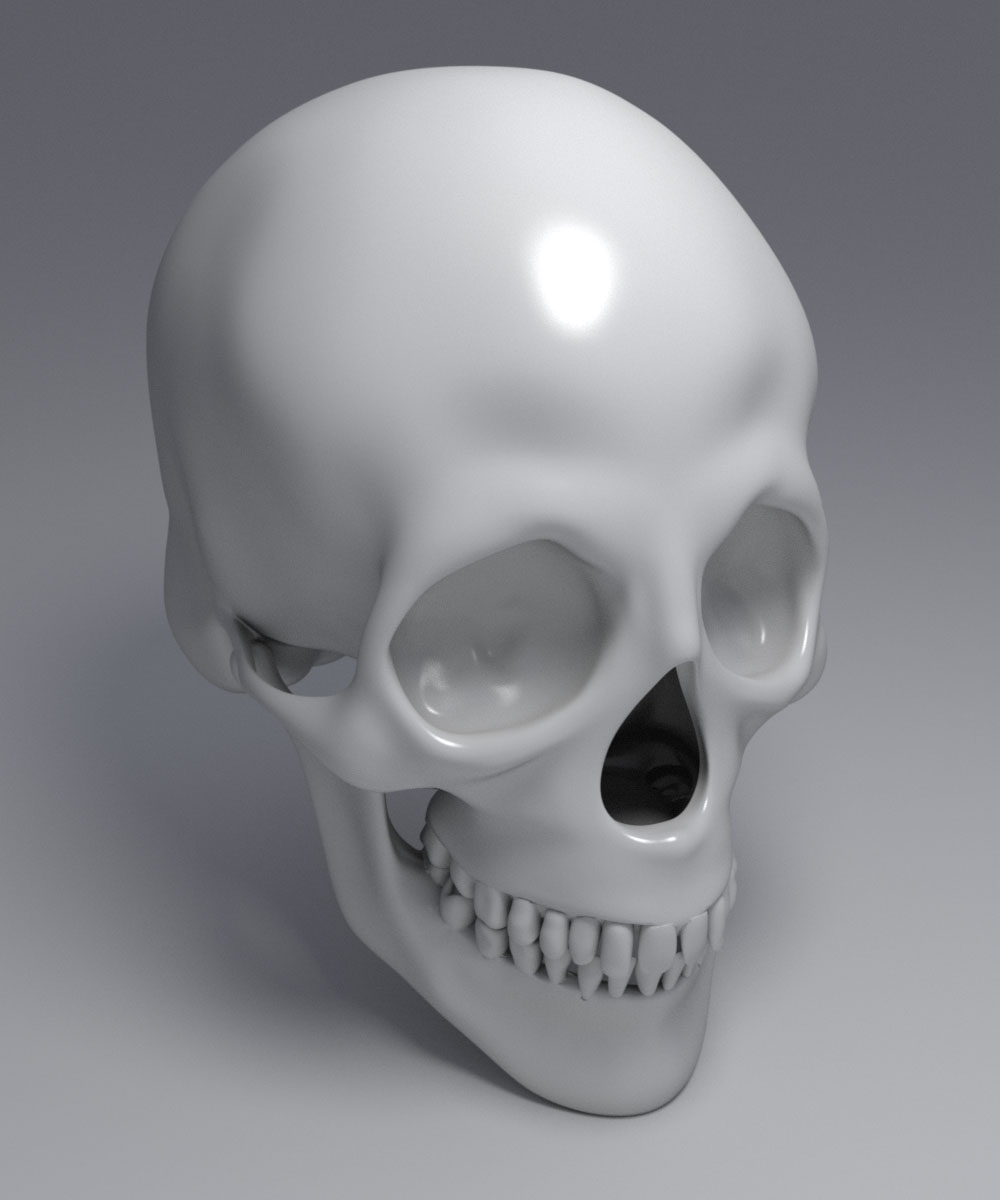}&
\includegraphics[width=0.25\linewidth]{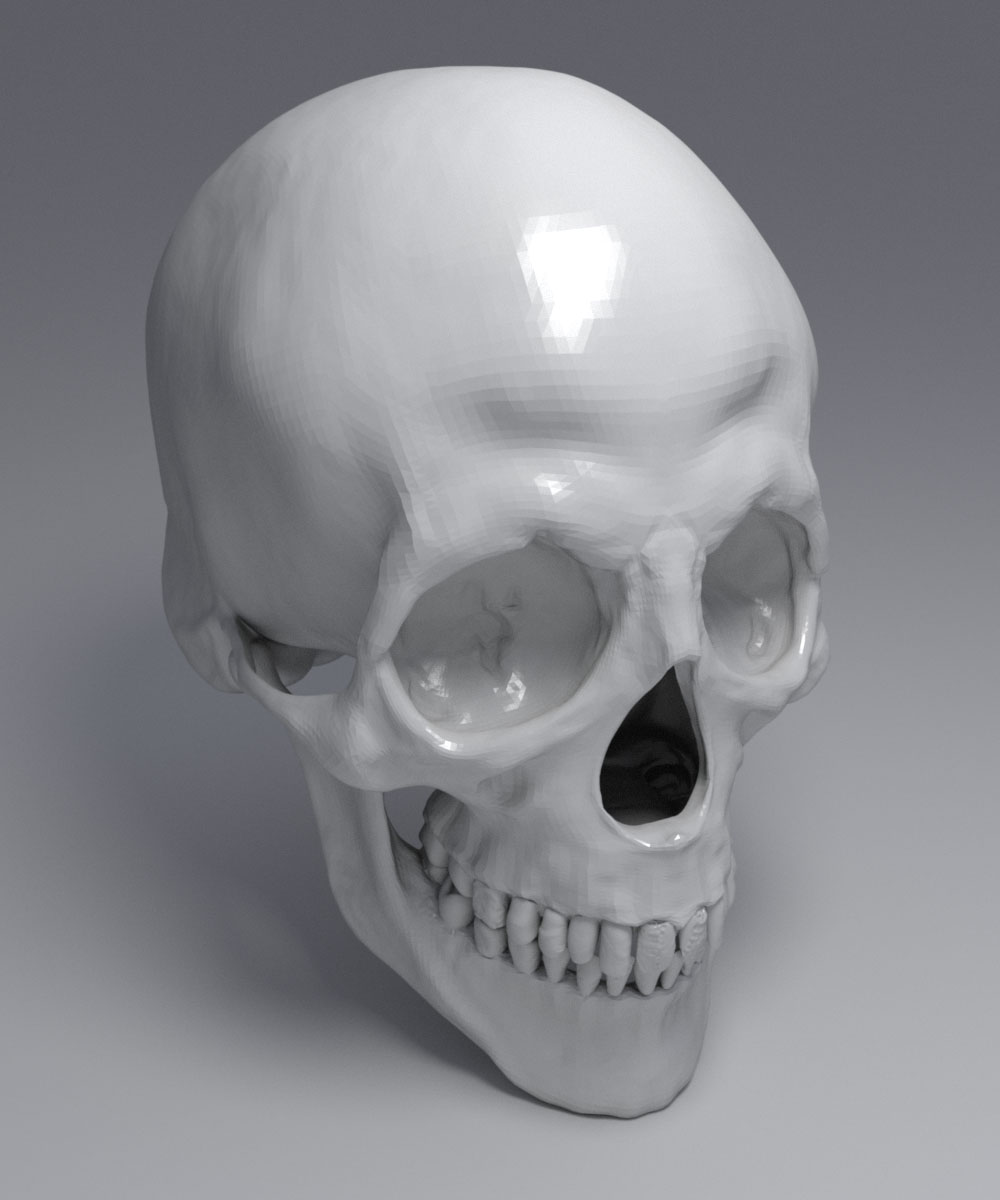}\\
\includegraphics[width=0.25\linewidth]{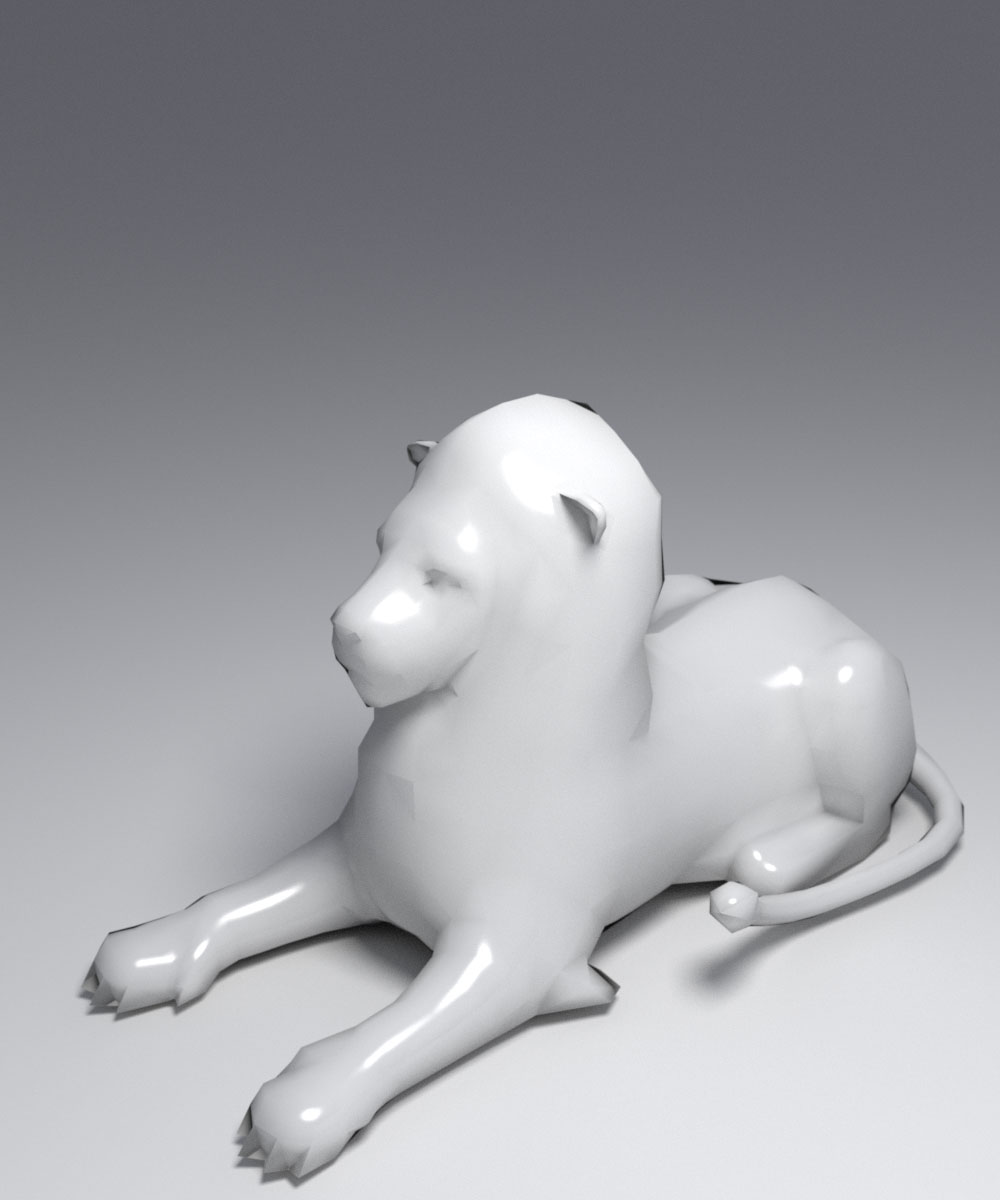}&
\includegraphics[width=0.25\linewidth]{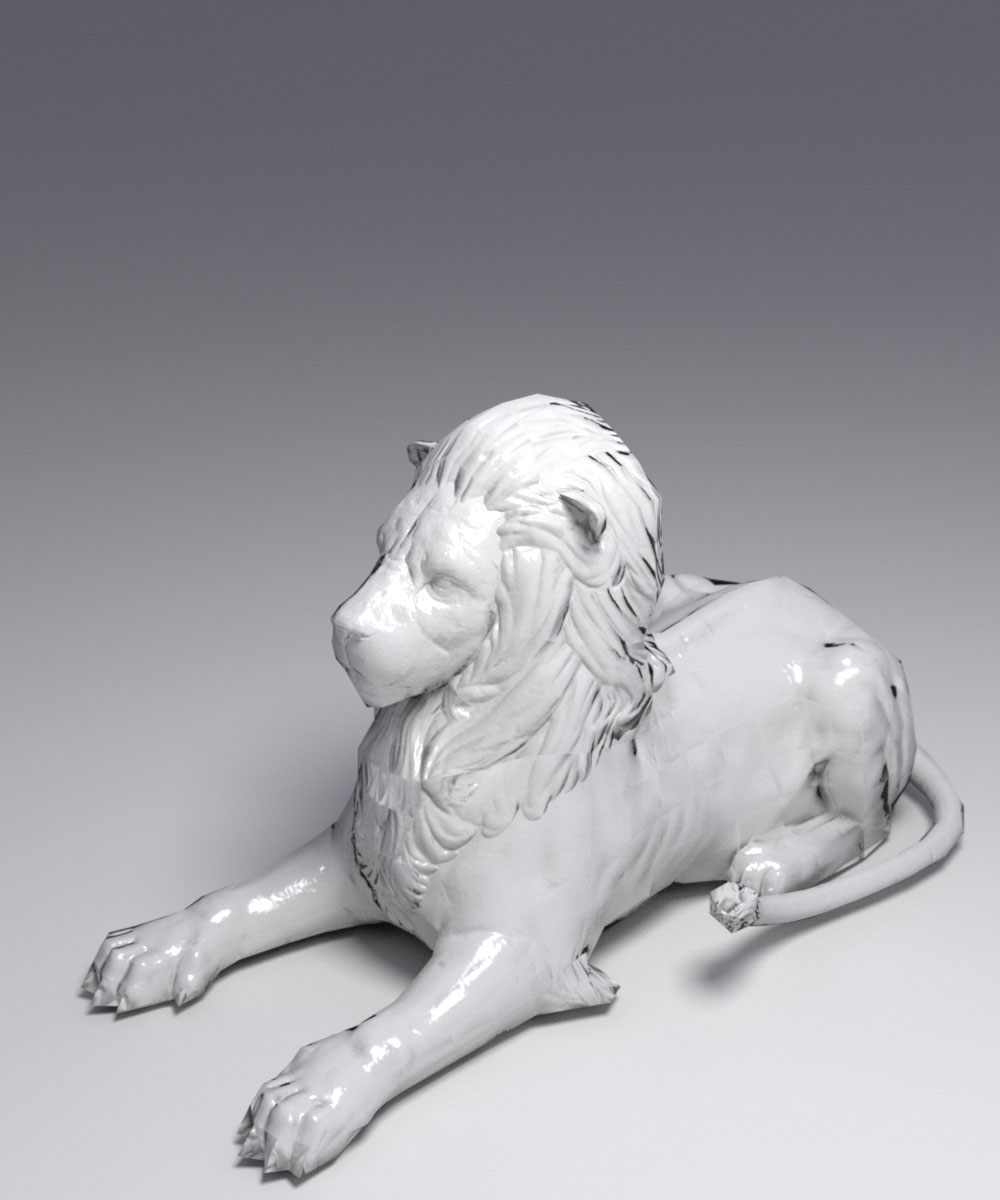}&
\includegraphics[width=0.25\linewidth]{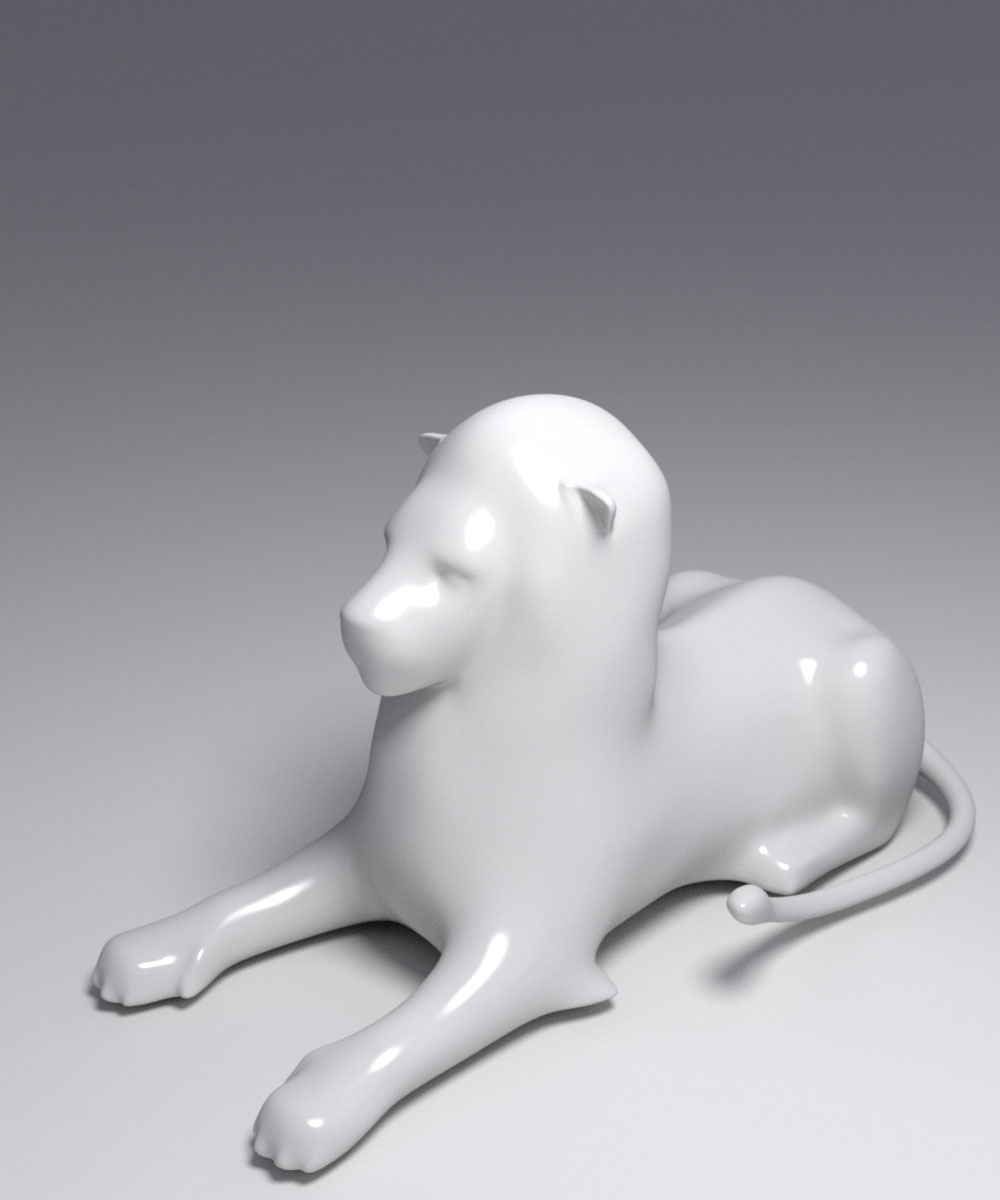}&
\includegraphics[width=0.25\linewidth]{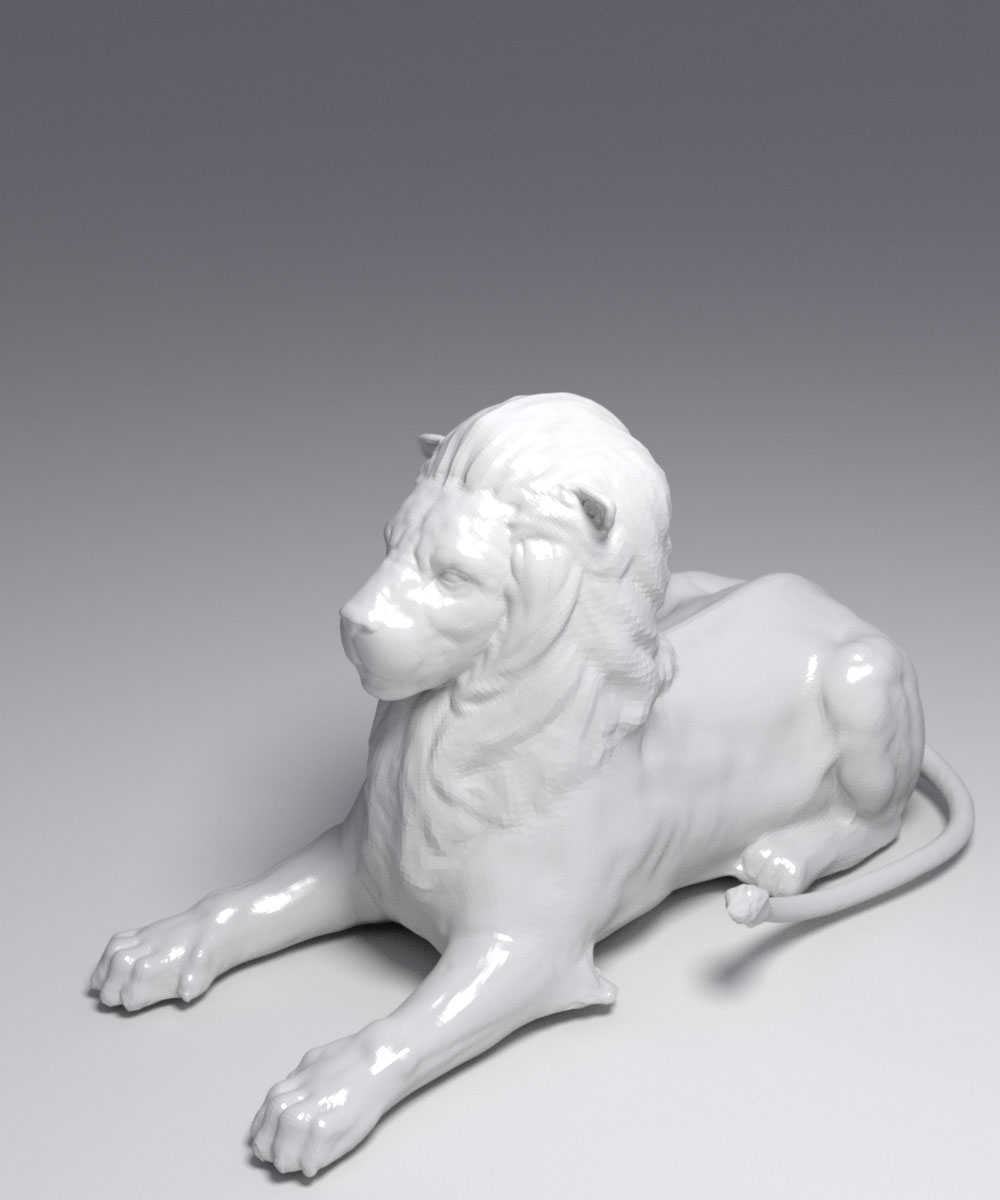}\\
(a) Original & (b) Normal map & (c) Subdivided & (d) Embossed 
\end{tabular}
\caption{Embossing results. We subdivide (c) the input model (a), and emboss the 
normal map (c) onto the subdivided model (d). When shading normals are not
  aligned with geometric normals, physically-based renderers can
  produce artifacts such as black regions (b) if not handled with particular
  care.}
\label{fig:embossing}
\end{figure}

Normal mapping is a useful technique for interactive applications, and allows for efficient low-polygon rendering by encoding geometric details as variations in normals stored in a high resolution texture. However, when it comes to realistic or physically-based rendering, this comes at the expense of several shortcomings. First, as these \textit{shading} normal variations merely fake geometric details, silhouettes remain coarse, hence precluding renderings of close-ups. Second, as shading normals are not aligned with geometric normals, this can produce artifacts in physically-based renderers such as light leaks or black regions (Fig.~\ref{fig:embossing}, (b)), or artifacts due to the lack of energy conservation if not handled with particular care~\cite{veach1998robust,2017_normalmap}.
Our projection step allows to recover finely detailed models for offline rendering from coarse assets along with their normal map. First, we increase the resolution of the input model (for subdivision surfaces, we apply a Catmull-Clark scheme). We then look up the normal value at each vertex from the normal map, and finally apply the projection step described in Sec.~\ref{sec:projection}. Several embossing results are shown in Fig.~\ref{fig:embossing}.

\section{Discussions}

\subsection{Implementation and Performance}

We implemented our method in C++ \dav{using the DEC operators within the
libDCC library~\cite{Crane:2013:DGP}. }  Our implementation is
mono-threaded, although fast parallel sparse linear solvers could be
used instead~\cite{eigenweb}. Still, our unoptimized code already
produces results in a reasonable amount of time. It processes meshes
of nearly 300k triangles in less than 15 minutes, or in less than a
minute if only the projection step is required (see
Table.~\ref{tab:performance}). Our implementation is available within 
our supplemental materials.

\begin{table}
\centering\begin{tabular}{|c|c|c|c|c|}
\hline
Model & \# vertices & \# triangles & MS (s) & Projection (s)\\
\hline
Helmet & 10,022 & 20,247 & 28.9&2.5  \\
Dragon & 13,504 & 27,224 & 203 & N/A\\
Witch & 14,255 & 28,524 & 151 & 8\\
Anvil & 25,202 & 50,400 & 273& N/A\\
Shell & 30,036 & 60,068 & 120  & 10\\
Chair & 57,933 & 115,910 & 214 & 15 \\
Lion & 60,066 & 120,128 & 248  & 14\\
Skull & 81,051 & 162,120 & 315 & 18\\
Pirate &136,770 & 273,536 & 832 & 46\\
Boar & 223,251& 444,809& 1,133 & 53\\
Gun & 392,747 & 784,785 & 2,189 & 87\\
Skull (HR) & 324,251 & 648,480 & N/A & 104 \\
Armor & 447,052 & 806,144 & N/A  & 65\\
\hline
\end{tabular}
\caption{Performance of the MS solver and vertex projection steps with respect to the model complexity, using an unoptimized monothreaded implementation.}
\label{tab:performance}
\end{table}

\subsection{Limitations}

\begin{figure}[htbp]
\centering
\begin{tabular}{@{}c@{}c@{}}
\includegraphics[width=0.3\linewidth]{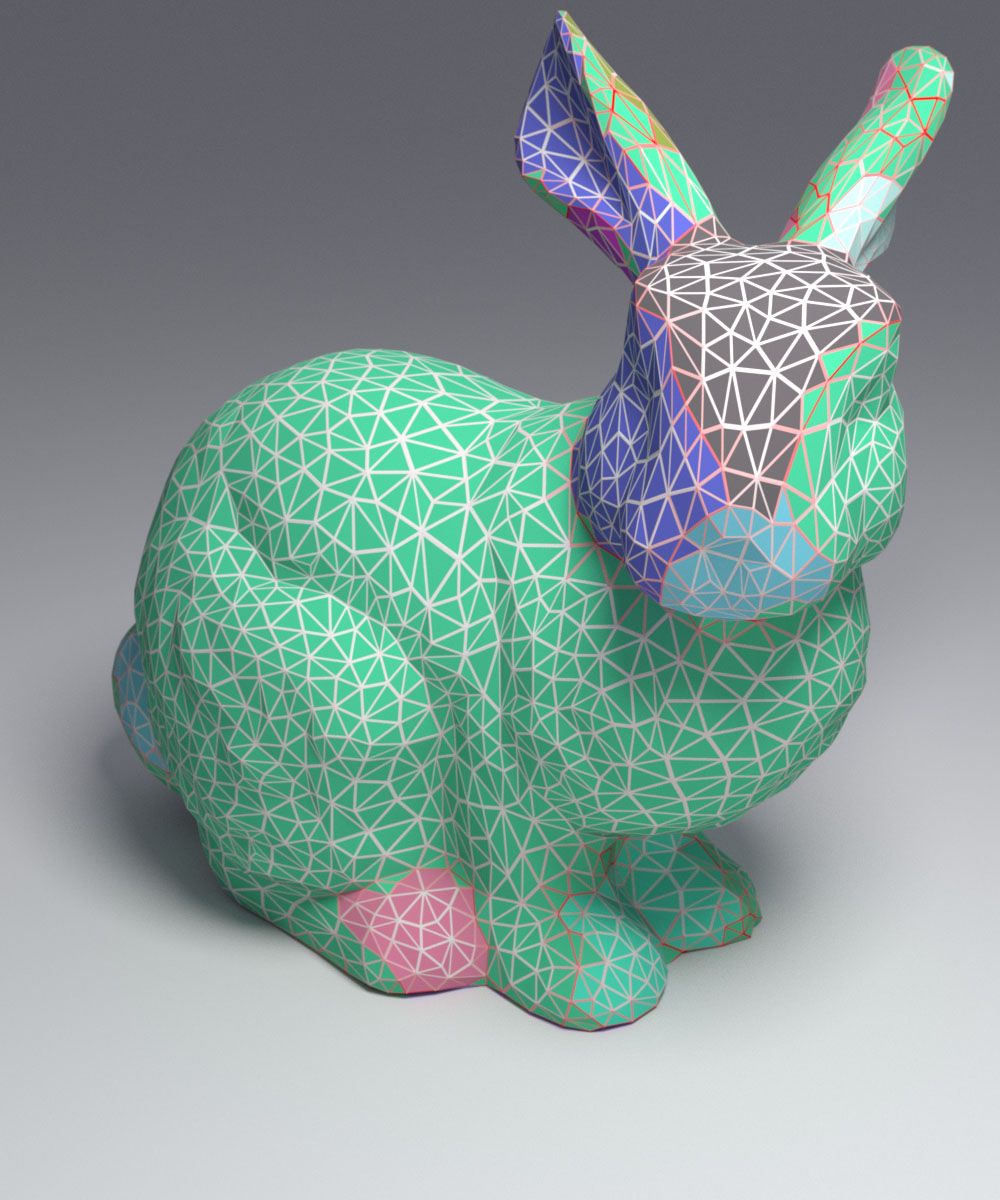} &
\includegraphics[width=0.3\linewidth]{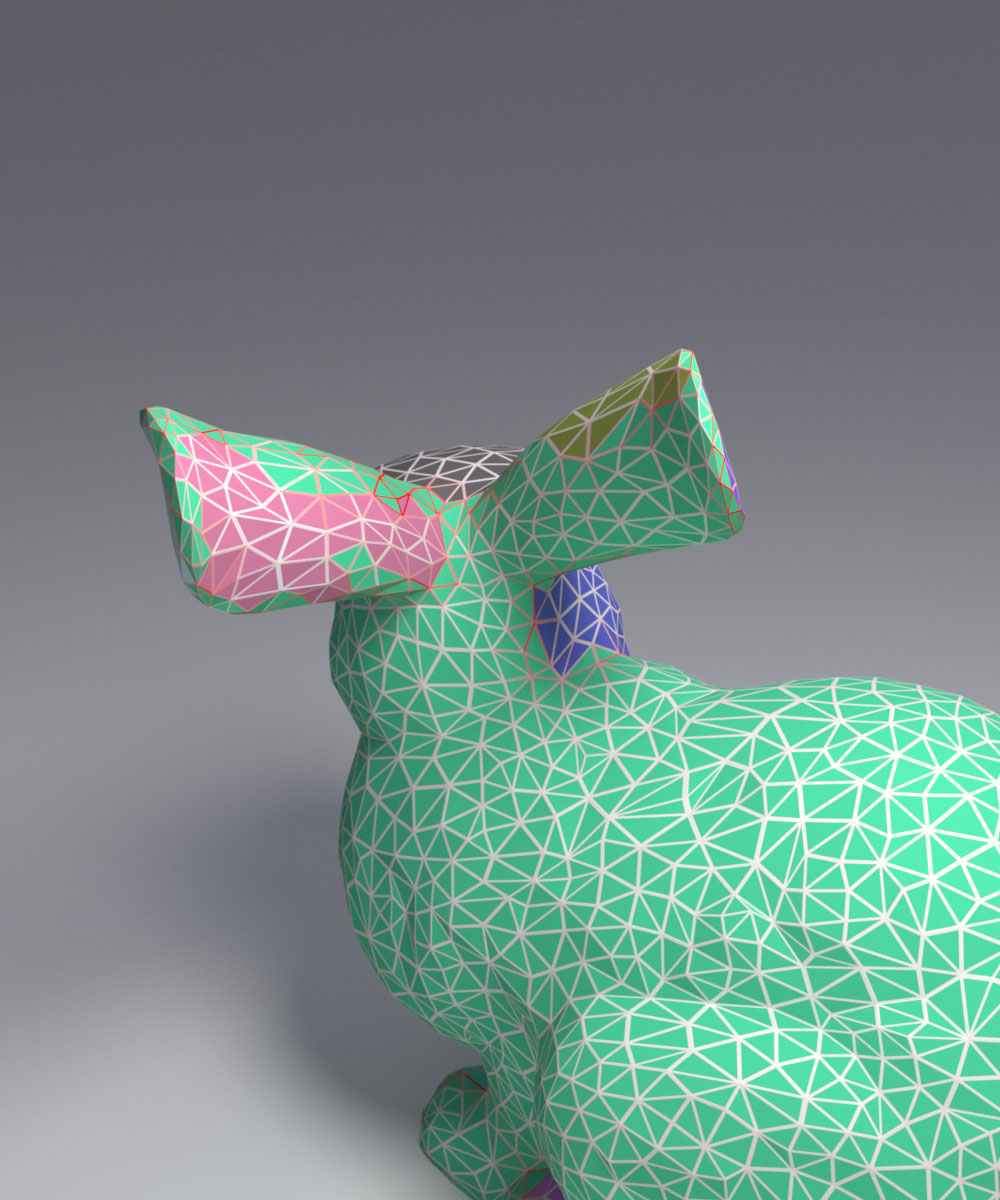} 
\end{tabular}
\begin{tabular}{@{}c@{}c@{}c@{}}
\includegraphics[width=0.3\linewidth]{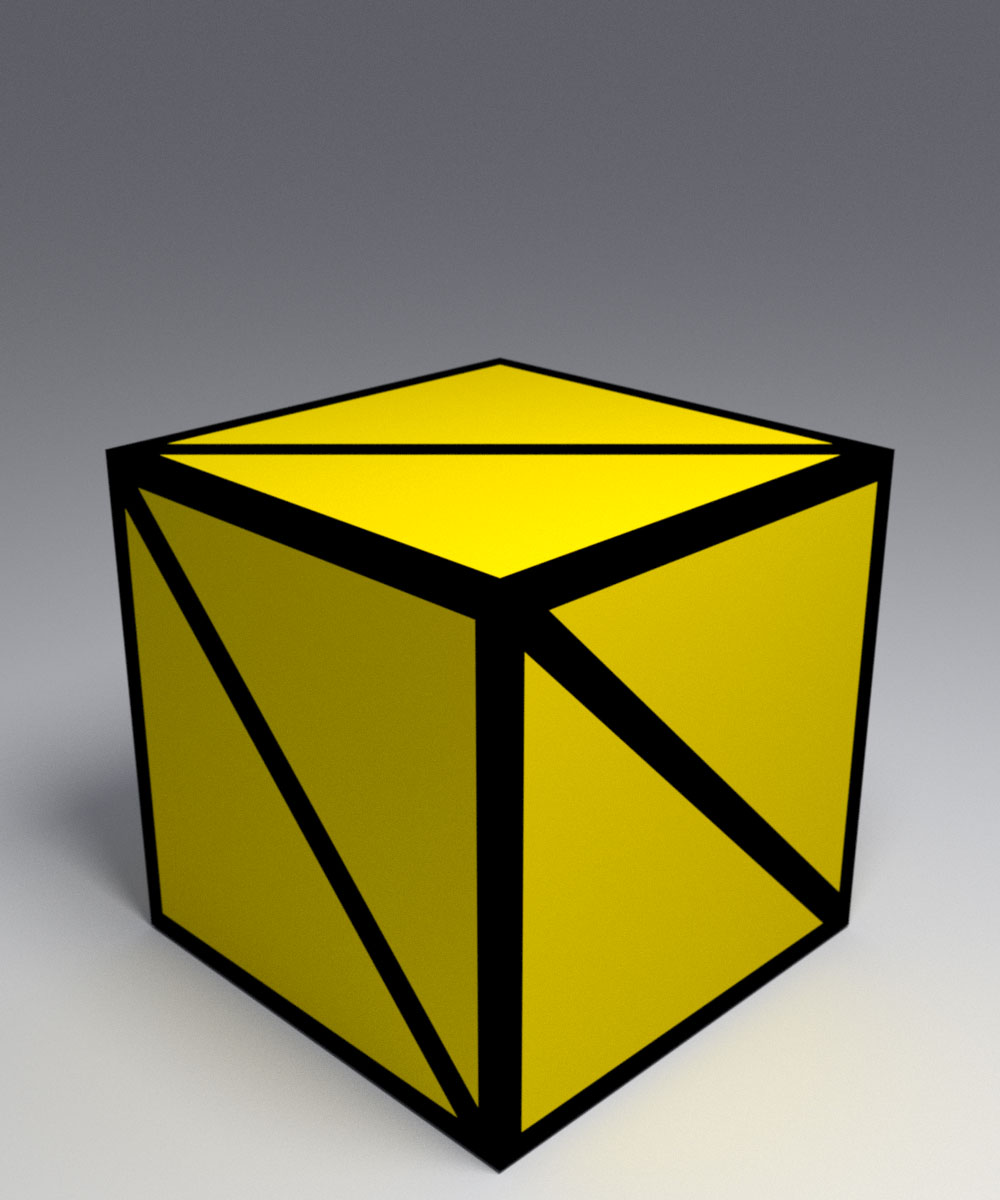}&
\includegraphics[width=0.3\linewidth]{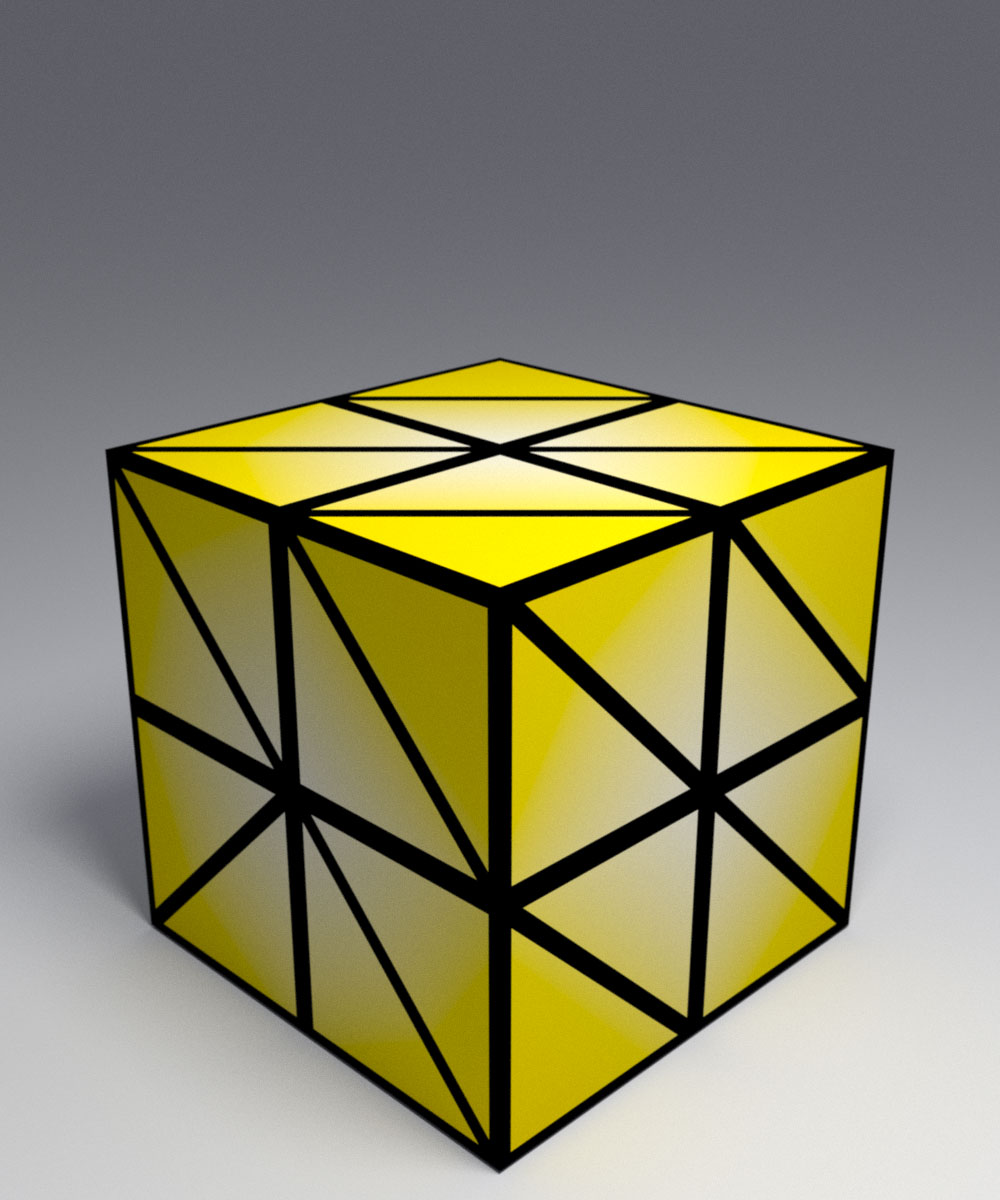} &
\includegraphics[width=0.3\linewidth]{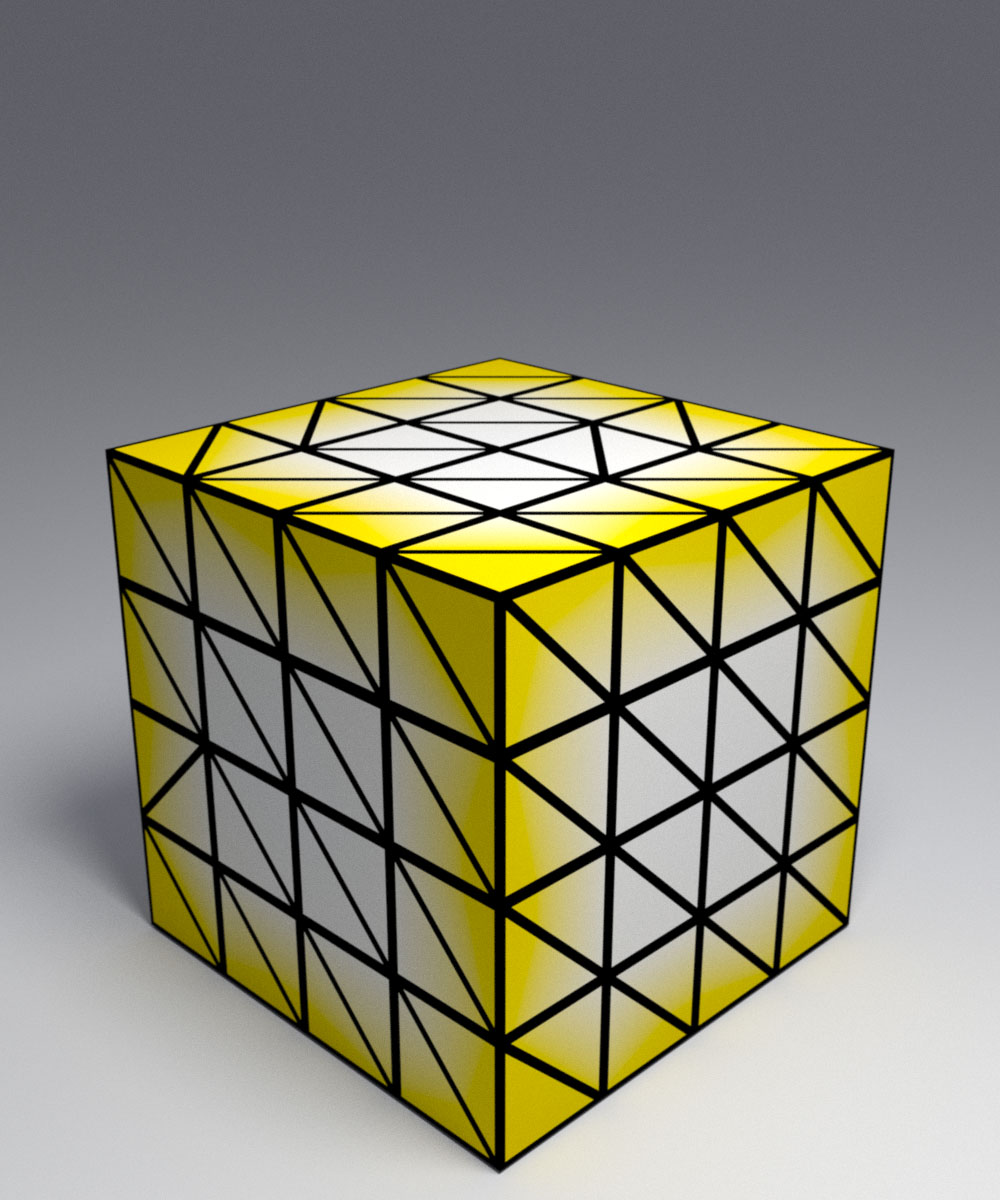} 
\end{tabular}
\caption{First row: multicut segmentation fails to capture the bunny
  ears geometry. Second row: a coarse mesh does not capture features
  (in yellow) computed per vertex accurately. As we subdivide the
  mesh, features become more accurate.}
\label{fig:cube}
\end{figure}

First of all, to better capture the geometry, our approach requires to
have meshes with enough triangles. For instance, low resolution shapes
may lead to unrelevant segmentations (see Fig.~\ref{fig:cube}, first row). For a
more extreme case, since the feature vector $v$ is represented as a
$0-$form we may highlight as feature all edges of a thin triangular
strip (see Fig.~\ref{fig:cube}, second row).

We also believe the normal map embossing would benefit from a mesh refinement strategy
that adapts to the local frequency content of the normal map. Some details
from the normal map indeed appear to be lost, and this is due to a
too coarse underlying mesh at some places. This could be
investigated in the future.

{Finally, we do not explicitly prevent triangle flips in our projection step (Sec.~\ref{sec:projection}). If flipping occurs, other work have specifically formulated their variational problem to prevent flipping~\cite{zhangstatic,wang2015rolling} and could be an inspiration for alleviating this issue.}

\subsection{Conclusions and Future Work}

We have introduced a discretization of a functional for mesh processing along with a
series of preliminary applications, which produces state-of-the-art
results in mesh denoising, segmentation, inpainting and embossing. We
leave as a future work a more extensive validation of each individual
application, and rather emphasize the broad scope of our method. In
particular, we expect the features detected by our Mumford-Shah
discretization to be useful for feature sensitive
remeshing~\cite{nivoliers2015anisotropic}. We also believe that
stitching meshes together~\cite{sharf2006snappaste} and mesh super-resolution~\cite{Tsai01} could be performed
similarly to our inpainting application. More generally, we expect the
success of the MS model in geometry processing could parallel its
success in image processing.

\subsection*{Acknowledgments}
{This work has been partly funded by COMEDIC ANR-15-CE40- 0006 research grant.}

\bibliographystyle{eg-alpha-doi}
\bibliography{paper}

\end{document}